%% file: NMSSMmass.tex
\documentclass[11pt]{article}

\usepackage{epsfig}
\usepackage{latexsym}
\usepackage{graphicx}
\usepackage{color}
\usepackage{amsmath,amssymb}
\usepackage{cite}

\makeatletter
\def\section{\@startsection {section}{1}{\z@}{-3.5ex plus -1ex minus -.2ex}{2.3ex plus .2ex}{\large\bf}}
\def\subsection{\@startsection{subsection}{2}{\z@}{-3.25ex plus -1ex
minus -.2ex}{1.5ex plus .2ex}{\normalsize\bf}}
\makeatother
\newcommand{\captionfonts}{\small}
\makeatletter  
\long\def\@makecaption#1#2{%
  \vskip\abovecaptionskip
  \sbox\@tempboxa{{\captionfonts #1: #2}}%
  \ifdim \wd\@tempboxa >\hsize
    {\captionfonts #1: #2\par}
  \else
    \hbox to\hsize{\hfil\box\@tempboxa\hfil}%
  \fi
  \vskip\belowcaptionskip}
\makeatother   

\catcode`@=11
\def\marginnote#1{}
\newcount\hour
\newcount\minute
\newtoks\amorpm
\hour=\time\divide\hour
by60
\minute=\time{\multiply\hour by60 \global\advance\minute
by-\hour}
\edef\standardtime{{\ifnum\hour<12 \global\amorpm={am}
\else\global\amorpm={pm}\advance\hour by-12 \fi
 \ifnum\hour=0
\hour=12 \fi
 \number\hour:\ifnum\minute<10
0\fi\number\minute\the\amorpm}}
\edef\militarytime{\number\hour:\ifnum\minute<10
0\fi\number\minute}
\def\draftlabel#1{{\@bsphack\if@filesw
{\let\thepage\relax
 \xdef\@gtempa{\write\@auxout{\string
\newlabel{#1}{{\@currentlabel}{\thepage}}}}}\@gtempa
 \if@nobreak
\ifvmode\nobreak\fi\fi\fi\@esphack}
\gdef\@eqnlabel{#1}}
\def\@eqnlabel{}
\def\@vacuum{}
\def\draftmarginnote#1{\marginpar{\raggedright\scriptsize\tt#1}}
\def\draft{\oddsidemargin
0.0truein
 \def\@oddfoot{\sl preliminary draft \hfil
\rm\thepage\hfil\sl\today\quad\militarytime}
 \let\@evenfoot\@oddfoot
\overfullrule 3pt
 \let\label=\draftlabel
\let\marginnote=\draftmarginnote
\def\@eqnnum{(\theequation)\rlap{\kern\marginparsep\tt\@eqnlabel}
\global\let\@eqnlabel\@vacuum}
}
\catcode`@=12

\newcommand{\beq}{\begin{eqnarray}}
\newcommand{\eeq}{\end{eqnarray}}

\def\mchi01{m_{\tilde{\chi}^0_1}}
\def\mst1{m_{\tilde{t}_1}}
\def\msc1{m_{\tilde{c}_L}}
\def\mn1{m_{\tilde{\chi}_1^0}}

\def\acdj1{a_{j1}^{\tilde{c} d_k}}
\def\atdj1{a_{j1}^{\tilde{t} d_k}}
\def\bcdj1{b_{j1}^{\tilde{c} d_k}}
\def\btdj1{b_{j1}^{\tilde{t} d_k}}

\newcommand{\gsim}{\raisebox{-0.13cm}{~\shortstack{$>$ \\[-0.07cm]
      $\sim$}}~}
\newcommand{\lsim}{\raisebox{-0.13cm}{~\shortstack{$<$ \\[-0.07cm]
      $\sim$}}~}

\newcommand{\s}{\newline \vspace*{-3.5mm}}
\newcommand{\id}{{\rm 1\kern-.12em
\rule{0.3pt}{1.5ex}\raisebox{0.0ex}{\rule{0.1em}{0.3pt}}}}
\renewcommand{\thefootnote}{\fnsymbol{footnote}}

\begin{document}

\thispagestyle{empty}

\begin{center}
\hfill CERN-PH-TH/2011-290 \\
\hfill FR-PHENO-2011-020 \\
\hfill KA-TP-33-2011 \\
\hfill SFB/CPP-11-68

\begin{center}

\vspace{1.7cm}

{\LARGE\bf Analysis of the NMSSM Higgs Boson Masses\\[0.1cm] at
  One-Loop Level} 
\end{center}

\vspace{1.4cm}

{\bf K. Ender$^{\,a}$}, {\bf T. Graf$^{\,b}$}, {\bf
  M. M\"uhlleitner$^{\,a}$} and {\bf H. Rzehak\footnote{On leave from:
  Albert-Ludwigs-Universit\"at Freiburg, Physikalisches Institut,
  Freiburg, Germany.}$^{\,c}$} \\

\vspace{1.2cm}

${}^a\!\!$
{\em {Institut f\"ur Theoretische Physik, Karlsruhe Institute of Technology, 76128 Karlsruhe, Germany}
}\\
${}^b\!\!$
{\em {IPPP, Department of Physics, University of Durham,
  Durham DH1 3LE, UK}} \\
${}^c\!\!$
{\em {CERN, Theory Division, 1211 Geneva 23, Switzerland}}

\end{center}

\vspace{0.8cm}

\centerline{\bf Abstract}
\vspace{2 mm}
\begin{quote}
\small
For a reliable prediction of the NMSSM Higgs boson signatures at
present and future high-energy colliders and a proper distinction
of the NMSSM and MSSM Higgs sector the precise knowledge
of the Higgs boson masses including higher-order corrections is
indispensable. In this paper, the one-loop corrections to the neutral
NMSSM Higgs boson masses and mixings are calculated in three different
renormalisation schemes. In addition to the $\overline{\mbox{DR}}$ 
renormalisation scheme, existing in the literature, two other
schemes are adopted. Furthermore, the dependence on the value of
the top quark mass is investigated. The resulting Higgs mass corrections have
been compared and the residual theory error due to missing
higher-order corrections can be estimated to be of the order of $10$\%.
\end{quote}

\newpage
\renewcommand{\thefootnote}{\arabic{footnote}}
\setcounter{footnote}{0}

\section{Introduction}

Supersymmetric theories \cite{susy,susyhierarchy} provide a natural
solution to the hierarchy problem \cite{susyhierarchy,hierarchy}. The
latter is related to the fact that the 
Standard Model (SM) Higgs boson mass receives radiative corrections
which are quadratic in the cut-off scale, usually chosen to be the GUT scale
$M_{GUT}=10^{16}$~GeV. To keep the Higgs mass of the order of the
electroweak (EW) scale an extreme fine-tuning of the model parameters is
necessary. Supersymmetry (SUSY) introduces a new symmetry between
fermionic and bosonic degrees of freedom, leading to extra
contributions to the radiative corrections of the Higgs boson mass due
to the new SUSY particles which cancel the dangerous quadratic
divergence of their corresponding SM counterpart. In this way the
Higgs mass is naturally kept at a phenomenologically valid level even
in the presence of high mass scales. \s

As the superpotential must be analytic in the chiral superfields, two
complex Higgs doublets have to be introduced, $H_u$ to provide masses
to the up-type fermions and $H_d$ to ensure non-zero down-type fermion
masses. In this way the theory is also kept anomaly-free. In the
superpotential of the Minimal Supersymmetric Extension of the
Standard Model (MSSM) \cite{mssm} the two Higgs doublet fields mix
through the term $\mu H_u \epsilon H_d$ which involves the higgsino
mass parameter 
$\mu$. Phenomenology requires $\mu$ to be of the order of the
electroweak scale \cite{muew}. On the other hand, $\mu$ is present in
the superpotential before the symmetry breaking and hence not a priori
of the order of the EW scale. In the Next-to-Minimal Supersymmetric
Extension of the Standard Model (NMSSM)
\cite{nmssm1,nmssm2,nmssm3,nmssm4} this $\mu$ problem is solved 
by dynamically relating the value of $\mu$ to the EW scale. The
parameter $\mu$ arises here as the vacuum expectation value of the
neutral component of an additional Higgs field $S$ which
is a singlet field with respect to the SM gauge groups. It couples to
the MSSM Higgs fields via the interaction term $\lambda S (H_u \epsilon
H_d)$. With the scalar field $S$ acquiring a non-zero vacuum
expectation value (VEV) $v_s/\sqrt{2}$ an effective $\mu$ term is generated,
naturally expected to be of the order of the EW scale, $\mu = \lambda
\langle S \rangle \equiv \lambda v_s/\sqrt{2}$. 
Furthermore, new contributions to the quartic coupling increase the
tree-level mass value of the lightest Higgs boson. 
In the MSSM the tree-level mass of the lightest Higgs boson is
predicted to be below the mass of the $Z$ boson. Large radiative
corrections involving top and stop loops are necessary to lift the
Higgs mass value beyond the lower bound from direct searches at LEP
\cite{lep}. The additional NMSSM contributions lift the mass of the
SM-like Higgs boson more easily beyond the LEP bound. \s

The extension of the Higgs sector by two more degrees
of freedom through the introduction of an additional singlet superfield
field $\hat{S}$\footnote{We denote superfields by a hat over the
  field. Fields without hat are the corresponding
  component fields.} leads to a total of 7 Higgs bosons after
electroweak symmetry breaking, three
neutral CP-even, two neutral CP-odd and two charged Higgs bosons. The
fermionic component of $\hat{S}$ mixes with the MSSM higgsinos and
neutral gauginos to
yield five neutralinos. Among the considerable phenomenological
modifications in the Higgs and neutralino sector compared to the
MSSM are possible new Higgs-to-Higgs decays, as {\it e.g.} the decay of a
SM-like scalar Higgs boson into a pair of lighter pseudoscalar Higgs
states, so that the present Tevatron and LHC search studies for
supersymmetric Higgs particles have to be revisited
\cite{htohdecs}. Such a Higgs boson 
could have escaped the LEP bounds \cite{guniondecay}.\s

From the above discussions it is clear that the precise knowledge of
the Higgs boson masses is indispensable to distinguish between MSSM
and NMSSM predictions, to properly define scenarios with new
Higgs-to-Higgs decays within the NMSSM and to correctly interpret the
experimental data. In the MSSM, where radiative corrections
are crucial to accommodate the light Higgs boson mass with the LEP
limits, the Higgs boson masses have been calculated with impressive
accuracy. After the calculation of the dominant one-loop contributions
due to top/stop loops \cite{mssmalphat} the full one-loop corrections
have been provided in \cite{fulloneloopmssm}. The leading logarithmic
two-loop effects obtained through renormalisation group equations
\cite{ll2loop} have been completed by the genuine ${\cal
  O}(\alpha_t \alpha_s)$ \cite{alsalt1,alsalt2,alsalt20,alsalt3,alsalt4}, 
${\cal O}(\alpha_t^2)$ \cite{alsalt1,alsalt3,alth2}, ${\cal
  O}(\alpha_b \alpha_s)$ \cite{alsalb} and ${\cal O} (\alpha_t\alpha_b
+ \alpha_b^2)$ \cite{altalbalth2} two-loop corrections in the limit of
zero external momentum.  
The electroweak two-loop effects
including part of the external momentum dependence have been
calculated \cite{2lew} and the leading three-loop contributions
have been evaluated in \cite{lead3l}. Also in the case of a CP-violating
MSSM a great effort has been undertaken to calculate the higher-order
corrections. 
After first investigations~\cite{firstinvest}, they have
been evaluated at one-loop order in the effective potential 
approach~\cite{effpotcp} and with the renormalisation group improved 
effective potential method through next-to-leading order \cite{reneffpot}.
In the Feynman diagrammatic approach the one-loop leading $m_t^4$
corrections have been provided in \cite{1lfd}, a full one-loop
calculation in \cite{1lfull}, and more recently the leading two-loop
contributions of ${\cal O} (\alpha_t \alpha_s)$ have been evaluated in
\cite{2lcomplex}. The corrections have been implemented in the public
computer code {\tt FeynHiggs} \cite{alsalt20,1lfull,fhcomplex}. {\tt CPsuperH}
\cite{cpsuperh}, another public code, is based on the renormalisation
group improved effective potential approach \cite{reneffpot}. The MSSM
Higgs mass spectrum for real parameters can also be obtained from the
spectrum calculators \cite{spectrum1,spectrum2,spectrum3}. \s

In the NMSSM, however, the higher-order calculations to the Higgs boson
masses have not yet reached the same level of accuracy as in the
MSSM. The leading one-loop contributions due to top/stop and
bottom/sbottom loops have been calculated in the effective potential
approach \cite{effpotnmssm}, the one-loop contributions due to
chargino, neutralino and scalar loops have been evaluated in leading
logarithmic order in Ref. \cite{llognmssm}. These corrections and the
leading logarithmic two-loop terms of ${\cal O} (\alpha_t \alpha_s)$
and ${\cal O} (\alpha_t^2)$, taken over from the MSSM results, have
been implemented in the public computer
code {\tt NMHDECAY} \cite{nmhdecay}. Furthermore, the full
one-loop contributions have been computed in the $\overline{\mbox{DR}}$
renormalisation scheme and the ${\cal O} (\alpha_t \alpha_s + \alpha_b
\alpha_s)$ corrections have been provided in the approximation of zero
external momentum \cite{slavich}. They have been
implemented in  {\tt NMHDECAY} as well as in the spectrum calculator
{\tt SPheno} \cite{spectrum3}. \s 

In this paper we complement the effort to reach a higher level of
accuracy in the computation of the NMSSM Higgs boson masses by
providing the 
full one-loop corrections in a mixed $\overline{\mbox{DR}}-$on-shell
and in an on-shell (OS) renormalisation scheme. By comparison with the
results of Ref. \cite{slavich} in the $\overline{\mbox{DR}}$ scheme, the
dependence on the renormalisation schemes can be studied. In this way
an estimate of the theoretical error due to missing higher-order corrections can
be derived. Having available the corrections in the mixed and in the
OS renormalisation scheme in addition enables a comparison of the
NMSSM results in the MSSM limit with the corresponding MSSM
corrections given in the OS scheme. \s

The paper is organised as follows. In section \ref{sec:calc} we
present the details of our calculation starting by introducing the
NMSSM Higgs sector and setting up our notation in
\ref{sec:nmssmhiggs}. After presenting the chargino and neutralino sector in
\ref{sec:gaugino} we list the parameters, which we employ, in section
\ref{sec:param}. The renormalisation is explained in detail in
\ref{sec:counter}. The explicit 
computation of the one-loop corrected Higgs boson masses and mixing
matrix elements is described in \ref{sec:oneloopmass}. Section
\ref{sec:numerical} finally is devoted to the numerical analysis. Our
results are summarised in \ref{sec:summary}.

\section{\label{sec:calc} Calculation}

\subsection{\label{sec:nmssmhiggs} The NMSSM Higgs Boson Sector}

The Higgs mass matrix is obtained from the NMSSM Higgs potential,
which is derived from the NMSSM superpotential $W_{NMSSM}$, the
corresponding soft SUSY breaking terms and the $D$ term contributions.
The NMSSM superpotential for the Higgs superfields
$\hat{H}_u,\hat{H}_d,\hat{S}$ in our conventions reads
\beq
W_{NMSSM} = W_{MSSM} - \epsilon_{ab} \lambda \hat{S} \hat{H}_d^a
\hat{H}_u^b + \frac{1}{3} \kappa \hat{S}^3
\;, \label{eq:nmssmsuperpot}
\eeq
where $a,b=1,2$ are the $SU(2)_L$ fundamental representation indices
and $\epsilon_{ab}$ denotes the totally antisymmetric tensor with
$\epsilon_{12}=\epsilon^{12}=1$. The Higgs superfield which couples to
up-type (down-type) fermion superfields is given by $\hat{H}_{u(d)}$. 
The parameters $\lambda$ and $\kappa$ are dimensionless and, working
in the CP-invariant NMSSM, are chosen to be real.
The first term in  Eq.~(\ref{eq:nmssmsuperpot}) is the MSSM superpotential,
which in terms of the quark and lepton superfields $\hat{Q},\hat{U}^c,
\hat{D}^c,\hat{L},\hat{E}^c$ reads\footnote{The superscript $c$
 denotes charge conjugation.}
\beq
W_{MSSM} = \epsilon_{ab} [y_e \hat{H}^a_d \hat{L}^b \hat{E}^c + y_d
\hat{H}_d^a \hat{Q}^b \hat{D}^c - y_u \hat{H}_u^a \hat{Q}^b \hat{U}^c]
\;, \label{eq:mssmsuperpot}
\eeq
where we have omitted colour and generation indices. Note that the
MSSM $\mu$ term in NMSSM constructions is commonly assumed to be zero
as well as terms linear and quadratic in $\hat{S}$.
The soft SUSY breaking terms in the NMSSM in terms of the component
fields $H_u, H_d, S$ are given by
\beq
\mathcal L_{soft} = {\cal L}_{soft,\, MSSM} - m_S^2 |S|^2 +
(\epsilon_{ab} \lambda 
A_\lambda S H_d^a H_u^b - \frac{1}{3} \kappa
A_\kappa S^3 + h.c.) \;,
\eeq
with the MSSM soft SUSY breaking Lagrangian
\beq
{\cal L}_{soft, \, MSSM} &=& -m_{H_d}^2 H_d^\dagger H_d - m_{H_u}^2
H_u^\dagger H_u -
m_Q^2 \tilde{Q}^\dagger \tilde{Q} - m_L^2 \tilde{L}^\dagger \tilde{L}
- m_U^2 \tilde{u}_R^* 
\tilde{u}_R - m_D^2 \tilde{d}_R^* \tilde{d}_R 
\nonumber \\\nonumber
&-& m_E^2 \tilde{e}_R^* \tilde{e}_R - (\epsilon_{ab} [y_e A_e H_d^a
\tilde{L}^b \tilde{e}_R^* + y_d
A_d H_d^a \tilde{Q}^b \tilde{d}_R^* - y_u A_u H_u^a \tilde{Q}^b
\tilde{u}_R^*] + h.c.) \\
&-&\frac{1}{2}(M_1 \tilde{B}\tilde{B} + M_2
\tilde{W}_i\tilde{W}_i + M_3 \tilde{G}\tilde{G} + h.c.) \;,
\eeq
where in the first two lines 
tilde denotes the scalar component of the corresponding quark
and lepton superfield, and $\tilde{Q}=(\tilde{u}_L,\tilde{d}_L)^T$,
$\tilde{L}=(\tilde{\nu}_L,\tilde{e}_L)^T$. Note that {\it e.g.}
$\tilde{u}_R^*$ is the scalar component of $\hat{U}^c$. In the last line the
soft SUSY breaking gaugino mass terms for the gaugino fields
$\tilde{B}$, $\tilde{W}_i$ 
($i=1,2,3$) and $\tilde{G}$ are given. All soft SUSY breaking trilinear couplings
$A_k$ ($k=\lambda,\kappa,d,u,e)$ and gaugino mass parameters $M_i$ ($i=1,2,3$) are assumed to be real, and squark and slepton mixing between the
generations is neglected. Furthermore, possible soft SUSY breaking
terms linear and quadratic in the singlet field $S$ are set to zero in
accordance with the majority of phenomenological NMSSM
constructions\footnote{In the MSSM there is an additional
  soft SUSY breaking term $m_{H_d H_u}^2 \epsilon_{ab} H^a_d
  H^b_u$ with $m_{H_d H_u}^2$ usually expressed in terms of
  the soft SUSY breaking parameter $B$ and the higgsino parameter
  $\mu$ as $m_{H_d H_u}^2 = B \mu$. This term is not
  explicitly added to the NMSSM soft SUSY breaking Lagrangian but is
  generated dynamically via the term $\epsilon_{ab} \lambda A_\lambda
  S H_d^a H_u^b$.}. \s 

The neutral components of the Higgs fields can be parametrised in
terms of CP-even and CP-odd fluctuations around their vacuum
expectation values, 
\beq
H_d^1 = \frac{1}{\sqrt{2}} (v_d + h_d + i a_d) \, , \quad
H_u^2 = \frac{1}{\sqrt{2}} (v_u + h_u + i a_u) \, , \quad
S = \frac{1}{\sqrt{2}} (v_s + h_s + i a_s) \;,
\eeq
with the VEVs $v_d,v_u,v_s$ chosen to be real and
positive. The parameter $\mu$ then arises dynamically from the
singlet field expanded about its VEV, {\it cf.}~Eq.~(\ref{eq:nmssmsuperpot}),
\beq
\mu = \frac{\lambda v_s}{\sqrt{2}} \;.
\eeq  
The minimisation conditions of the tree-level scalar
potential can be applied to replace the soft SUSY breaking Higgs
mass parameters $m_{H_u}^2, m_{H_d}^2, m_S^2$ by combinations of $\lambda,
\kappa$, the electroweak gauge couplings $g$ and $g'$, the VEVs and the
trilinear couplings $A_\lambda$, $A_\kappa$, so that the tree-level 
mass matrix $M_S^2$ of the neutral CP-even Higgs bosons 
obtained from the second derivative of the Higgs potential with
respect to the fields in the vacuum, in the basis $h^S= (h_d, h_u, h_s)^T$ can be cast into the form
\beq
M_S^2= && \phantom{LLLLLLLLLLLLLLLLLLLLLLLLLLLLLLLLLLLLLLLLLLLLLLLLLLLLLLLL} \nonumber\\
&& \hspace*{-1.7cm}
\begin{pmatrix} 
\bar{g}^2 v_d^2 + \frac{(R_\lambda + R v_s/2)  v_u v_s}{v_d} &
(\lambda^2  -\bar{g}^2 )v_u v_d - (R_\lambda + R v_s/2)
v_s &
\lambda^2 v_d v_s - (R_\lambda + R v_s) v_u
\\[0.2cm]
(\lambda^2  -\bar{g}^2 )v_u v_d - (R_\lambda + R v_s/2) v_s
&
\bar{g}^2 v_u^2 + \frac{(R_\lambda + R v_s/2)  v_d v_s}{v_u}
&
\lambda^2 v_u v_s - (R_\lambda + R v_s) v_d
\\[0.2cm]
\lambda^2 v_d v_s - (R_\lambda + R v_s) v_u
&
\lambda^2 v_u v_s - (R_\lambda + R v_s) v_d
&
2 \kappa^2 v_s^2 + \frac{R_\lambda v_u v_d }{v_s} + R_\kappa v_s
\end{pmatrix}
\eeq
where we have defined 
\beq
\bar{g}^2 = \frac{1}{4}(g^2 + g'^2) \, , \qquad
R = \lambda \kappa \, , \qquad R_\lambda = \frac{1}{\sqrt{2}} \lambda
A_\lambda \, , \qquad  R_\kappa = \frac{1}{\sqrt{2}} \kappa A_\kappa \;.
\eeq
Note, that the MSSM limit can be recovered by $\lambda,\kappa \to 0$
(with the ratio $\kappa/\lambda$ kept constant for a smooth approach) and
keeping the parameter $\mu= \lambda v_s/\sqrt{2}$ as well as the parameters
$A_\lambda$ and $A_\kappa$ fixed. In this limit we hence have $v_s \to \infty$.
\s

The CP-even mass eigenstates $H_i$ ($i=1,2,3$) are obtained by an
orthogonal transformation ${\cal R}^S$ 
(the summation over paired indices is implicit),
\beq
H_i = {\cal R}_{ij}^S h_j^S \;, \quad j=d,u,s \;. \label{eq:cpevenrot}
\eeq
They are ordered by ascending mass with $M_{H_1} \leq M_{H_2} \leq M_{H_3}$.
The CP-odd fields $(a_d,a_u,a_s)$ can be rotated first to separate a
massless Goldstone boson $G$ 
\vspace*{0.3cm}
\beq
\left( \begin{array}{c} a \\ a_s \\ G\end{array} \right) = 
\left( \begin{array}{ccc} s_{\beta_n} & c_{\beta_n} & 0 \\
 0 & 0 & 1 \\
 c_{\beta_n} & -s_{\beta_n} & 0 
\end{array} \right)
\left( \begin{array}{c} a_d \\ a_u \\ a_s \end{array} \right)  \;,
\label{eq:rotmatcpodd}
\eeq
with the abbreviations $c_x \equiv \cos x$ and $s_x \equiv \sin x$, in
the following also $t_x \equiv \tan x$. Starting from the tree-level
CP-odd mass matrix squared in the basis 
$(a_d,a_u,a_s)^T$ and using Eq.~(\ref{eq:rotmatcpodd}) results in the mass
matrix squared $M_P^2$ in the basis $h^P = (a, a_s, G)^T$, which
reads
\beq
M_P^2 = 
\begin{pmatrix}
(2R_\lambda + Rv_s) v_s \frac{c^2_{\Delta \beta}}{s_{2 \beta}} 
& 
(R_\lambda - Rv_s) v c_{\Delta \beta} &
(R_\lambda + Rv_s/2) v_s\frac{s_{2 \Delta \beta}}{s_{2
 \beta}} \\
(R_\lambda - Rv_s) v c_{\Delta \beta}
&
(2R+R_\lambda/v_s) \frac{v^2 s_{2\beta}}{2} - 3 R_\kappa v_s &
(R_\lambda - Rv_s) v s_{\Delta \beta} \\
(R_\lambda + Rv_s/2) v_s\frac{s_{2 \Delta \beta}}{s_{2
 \beta}} &
(R_\lambda - Rv_s) v s_{\Delta \beta} &
(2R_\lambda + Rv_s) v_s \frac{s^2_{\Delta \beta}}{s_{2 \beta}} 
\end{pmatrix}
\eeq
with $\Delta \beta = \beta - \beta_n$ and $v^2 \equiv v_u^2 + v_d^2$. The
angle $\beta_n$ coincides at tree-level with the angle $\beta$ defined
through the ratio of the VEVs $v_u,v_d$, $\tan \beta = v_u/v_d$, Hence
$\Delta \beta = 0$ at tree-level leading to a massless Goldstone
boson, which decouples as it should.  The first entry $(M_P^2)_{11}$
at tree-level,  
\begin{align}
(M_P^2)_{11} = \frac{(2 R_\lambda + R v_s) v_s}{s_{2\beta}} \;,
\label{eq:mah2} 
\end{align}
becomes the mass of the pseudoscalar Higgs
boson in the MSSM limit\footnote{$A_\lambda$ takes the role of the soft
 SUSY-breaking parameter $B$ in the
 MSSM.}.
Applying an orthogonal rotation ${\cal R}^P$ to $h^P=(a,
a_s, G)^T$, the CP-odd mass eigenstates $A_i \equiv A_1,A_2,G$ ($i=1,2,3$)
are obtained, 
\begin{align}
A_i = {\cal R}_{ij}^P h_j^P \;, \label{eq:cpoddrot}
\end{align}
where at tree-level ${\cal R}_{33}^P = 1$ and ${\cal R}_{3i}^P = {\cal
  R}_{i3}^P = 0$ for $i\neq 3$.
The pseudoscalar masses are ordered by ascending mass, $M_{A_1} \leq M_{A_2}$.
The CP-even and CP-odd Higgs mass values squared are given by the
eigenvalues of the respective mass matrices $M_S^2$ and $M_P^2$. 
Analytic expressions would be rather complicated if not expanded in
special parameter regions as {\it e.g.} in Ref.~\cite{zerwaseal},
where a comprehensive investigation of the NMSSM Higgs boson sector
has been performed. In our analysis the mass eigenvalues are derived
numerically. The charged Higgs boson mass on the other hand takes a
simple form after the massless charged Goldstone boson has been
separated by an orthogonal rotation with a mixing angle $\beta_c$. At
Born level we have $\beta_c = \beta$ and get
\beq
M_{H^\pm}^2 = M_W^2 +  \sqrt{2} \lambda \,\frac{v_s}{ s_{2 \beta}} (A_\lambda
+ \kappa \frac{v_s}{\sqrt{2}}) - \frac{\lambda^2 v^2}{2} 
= M_W^2 + (M_P^2)_{11} - \frac{\lambda^2 v^2}{2} \;, \label{eq:mhchar}
\eeq
with $M_W$ being the  $W$ boson mass and where we have applied the
definition given in Eq.~(\ref{eq:mah2}).

\subsection{\label{sec:gaugino} The Neutralino and Chargino Sector} 
The chargino sector remains unchanged with respect to the
MSSM. For completeness and to set up our notation we
briefly repeat the chargino system. The chargino mass matrix depends
on the wino mass parameter $M_2$, the effective Higgsino parameter
$\mu = \lambda v_s / \sqrt{2}$ and the ratio of the vacuum expectation
values $\tan\beta$. In the interaction eigenbasis 
it is given by \cite{chargino}
\beq 
M_C = \left( \begin{array}{cc} M_2 & \sqrt{2} M_W s_\beta \\
\sqrt{2} M_W c_\beta & \mu \end{array} \right) \;.
\label{eq:charmat}
\eeq
It is diagonalised by two real matrices $U$ and $V$,
\beq \label{eq:CharDiag}
U^* M_C V^{\dagger} \quad \to \quad U = {\cal O}_- \quad
\mbox{and} \quad V = \left\{ \begin{array}{cl} {\cal O}_+  & \quad
   \mbox{if } \det M_C > 0 \\ \sigma_3
   {\cal O}_+   & \quad \mbox{if } \det M_C < 0 \end{array}
\right. \; ,
\eeq
with the Pauli matrix $\sigma_3$ to render the chargino masses
positive. The rotation matrices ${\cal O}_{\pm}$ are given in terms of
the mixing angles
\beq \label{eq:Charwinkel}
\tan 2\theta_- = \frac{2\sqrt{2} M_W (M_2 c_\beta + \mu
 s_\beta)}{M_2^2-\mu^2-2M_W^2 c_\beta} \quad , \quad
\tan 2\theta_+ = \frac{2\sqrt{2} M_W (M_2 s_\beta + \mu
 c_\beta)}{M_2^2-\mu^2+2M_W^2 c_\beta} \; ,
\eeq
and the two chargino masses read
\beq \label{eq:Charmasses}
m^2_{\chi^\pm_{1,2}} = \frac{1}{2} \left\{ M_2^2 +\mu^2+2M_W^2 \mp
 [(M_2^2-\mu^2)^2+4M_W^2 (M_W^2 c_{2\beta}^2 + M_2^2 + \mu^2 + 2M_2
 \mu s_{2\beta})]^{\frac{1}{2}} \right\} \;.
\eeq 

In the neutralino sector, the mixing of the fermionic component of the
singlet superfields $\hat{S}$ with the neutral gauginos $\tilde{B},
\tilde{W}_3$ and higgsinos $\tilde{H}_d^0, \tilde{H}_u^0$ yields in
the Weyl spinor basis $\psi^0 = (\tilde{B},\tilde{W}_3,
\tilde{H}_d^0,\tilde{H}_u^0,\tilde{S})^T$ the $5\times 5$ neutralino
mass matrix
\beq
M_N = \left( \begin{array}{ccccc} M_1 & 0 & - c_\beta s_W M_Z &
   s_\beta s_W M_Z & 0 \\
0 & M_2 & c_\beta c_W M_Z & - s_\beta c_W M_Z & 0 \\
- c_\beta s_W M_Z & c_\beta c_W M_Z & 0 & -\lambda v_s / \sqrt{2} & -
\lambda v_u / \sqrt{2} \\ s_\beta s_W M_Z & - s_\beta c_W M_Z & -
\lambda v_s / \sqrt{2} & 0 & - \lambda v_d / \sqrt{2} \\ 0 & 0 & -
\lambda v_u / \sqrt{2} & - \lambda v_d / \sqrt{2} & \sqrt{2} \kappa
v_s \end{array} \right) \; . 
\label{eq:neutmat}
\eeq
We have introduced the short-hand notation $s_W \equiv \sin \theta_W$,
$c_W \equiv \cos \theta_W$ 
for the Weinberg angle $\theta_W$, and $M_Z$ denotes the $Z$ boson
mass. Diagonalisation with a unitary matrix ${\cal N}$ yields the five
neutralino mass eigenstates $\chi^0_i$ ($i=1,...,5$),
\beq
\chi^0_i = {\cal N}_{ij} \psi^0_j \;, \qquad j=1,...,5 \;.
\eeq
They are ordered by ascending mass, $m_{{\chi}_1^0}\le ... \le
  m_{{\chi}_5^0}$. With our assumption of CP 
invariance and allowing for negative neutralino mass eigenvalues,
the mass matrix ${\cal N}$ is real. 


\subsection{\label{sec:param} Parameter Basis}
The NMSSM Higgs potential depends on 12 independent parameters in the
CP-conserving case. They are given by the soft SUSY breaking mass parameters
$m_{H_u}^2, m_{H_d}^2, m_{S}^2$, the gauge couplings $g,g'$, the
vacuum expectation values $v_u,v_d,v_s$, the dimensionless parameters
$\lambda, \kappa$ and the soft SUSY breaking couplings $A_\lambda, A_\kappa$.
For the physical interpretation it is convenient to replace some of these
parameters: The minimisation of the Higgs potential $V$ requires
the terms linear in the Higgs fields to vanish in the vacuum. Hence
for the scalar fields,
\beq
\Bigg\langle \frac{\partial V}{\partial h_u} \Bigg\rangle= 
\Bigg\langle \frac{\partial V}{\partial h_d} \Bigg\rangle=
\Bigg\langle \frac{\partial V}{\partial h_s} \Bigg\rangle= 0 \; ,
\label{eq:tadpolecond} 
\eeq
where the brackets denote the vacuum. The corresponding coefficients,
which are called tadpoles, therefore have to be zero. At lowest order the
tadpole conditions for the CP-even fields read\footnote{As we work in
  the real NMSSM the derivatives of the Higgs potential with respect
  to the CP-odd fields are zero and no additional conditions have to
  be required.}
\beq
\Bigg\langle \frac{\partial V}{\partial h_d} \Bigg\rangle &\equiv& t_{h_d} = v_d
m_{H_d}^2 - R_\lambda v_u v_s + \frac{g^2+g'^2}{8}
v_d (v_d^2-v_u^2) + \frac{\lambda^2}{2} v_d (v_u^2+v_s^2) -
\frac{R}{2} v_u v_s^2 = 0  \label{eq:tadd} \\
\Bigg\langle \frac{\partial V}{\partial h_u} \Bigg\rangle &\equiv& t_{h_u} = v_u
m_{H_u}^2 - R_\lambda v_d v_s + \frac{g^2+g'^2}{8}
v_u (v_u^2-v_d^2) + \frac{\lambda^2}{2} v_u (v_d^2+v_s^2) -
\frac{R}{2} v_d v_s^2 = 0 \label{eq:tadu}\\
\Bigg\langle \frac{\partial V}{\partial h_s} \Bigg\rangle &\equiv& t_{h_s} = v_s
m_{S}^2 - R_\lambda v_d v_u + R_\kappa v_s^2 + \frac{\lambda^2}{2} v_s
(v_d^2+v_u^2) + \kappa^2 v_s^3 -R v_d v_u v_s = 0  \; . \label{eq:tads}
\eeq
Equations (\ref{eq:tadd}, \ref{eq:tadu}, \ref{eq:tads}) can be exploited to
replace $m_{H_u}^2, m_{H_d}^2, m_{S}^2$ by the tadpole parameters $t_{h_u}$,
$t_{h_d}$ and $t_{h_s}$. 
The parameters $g,g',v_u,v_d$ are replaced by the
electric charge $e$, the gauge boson masses $M_W,M_Z$ and the ratio of the
vacuum expectation values $\tan \beta \equiv t_\beta = v_u/v_d$
through the relations 
\beq \label{ggst}
g &=& \frac{e M_Z}{\sqrt{M_Z^2-M_W^2}} \qquad \qquad 
\quad g' = \frac{e M_Z}{M_W}
\\
v_u &=& \sqrt{\frac{M_Z^2-M_W^2}{1+t_\beta^2}} \frac{2 M_W t_\beta}{e M_Z}
\qquad 
v_d = \sqrt{\frac{M_Z^2-M_W^2}{1+t_\beta^2}} \frac{2 M_W}{e
 M_Z}
\;.
\eeq
Finally, $A_\lambda$ is replaced by the charged Higgs boson mass
$M_{H^\pm}^2$, 
\begin{align}
A_{\lambda}=& \frac{s_{2\beta}}{\sqrt{2} \lambda v_s}\left(\frac{v^2 \lambda^2 }{2
} - M_W^2 + \frac{M_{\text{H}^{\pm}}^2}{c_{\Delta \beta}^2} \right)
-\frac{\kappa v_s}{\sqrt{2}} 
-\frac{\sqrt{2}}{\lambda v v_s} \left(\frac{s_\beta s_{\beta_B}^2}{c_{\Delta
   \beta}^2} t_{h_d}+ 
\frac{c_\beta c_{\beta_B}^2}{c_{\Delta \beta}^2} t_{h_u}\right) 
\label{eq:alamrepl}
\end{align}
with
\begin{equation}
 v=\frac{2 M_W }{e}\sqrt{1-\frac{M_W^2}{M_Z^2}} \; , 
\end{equation}
where we have kept the dependence on the tadpole parameters and the 
mixing angles $\beta_c=\beta_n\equiv \beta_B$. At tree-level they
coincide with $\beta$. 
It is the angle $\beta$ which will be needed for the renormalisation:
According to our renormalisation schemes, presented
in section \ref{sec:counter}, the renormalisation procedure can be performed
before the transformation into mass eigenstates. This means that
the mixing matrices, also those
separating the neutral and charged Goldstone bosons from the Higgs
bosons and hence the angles $\beta_n$ and $\beta_c$, which
appear in these matrices, do not receive
counterterms. Therefore, special care has to be taken to determine the
elements of the mass matrices without inserting the tree-level
relation $\beta_B = \beta$ and to apply the renormalisation
procedure uniquely for $\tan \beta$ given by the ratio of $v_u$ and
$v_d$.  \s 

To summarise, we work with the following parameter set
\beq
t_{h_u}, \, t_{h_d}, \, t_{h_s}, \, e, \, \, M_W^2, \, M_Z^2, \,\tan\beta,
\, M_{H^\pm}^2, \, \lambda, \, \kappa, \, v_s, \, A_\kappa \:. \label{eq:parset}
\eeq

\subsection{\label{sec:counter} Renormalisation Schemes}
For the determination of the loop-corrected Higgs boson masses the Higgs
self-energies have to be calculated. They develop ultraviolet (UV)
divergences. Evaluating the self-energies in $D=4-2\epsilon$
dimensions, the divergences can be parametrised by $1/\epsilon$ 
leading to poles in $D=4$ dimensions. To get a finite result the
parameters entering the loop calculation have to be
renormalised by the introduction of appropriate counterterms absorbing
the UV divergences. Note, that at higher orders also the terms linear in
the Higgs fields get loop contributions. Therefore, also the tadpole parameters
$t_{h_{i}}$ ($i=u,d,s$) have to be renormalised, in order to fulfill 
the tadpole conditions Eq.~(\ref{eq:tadpolecond}). Before we describe
the determination of the one-loop corrected Higgs boson masses in
section \ref{sec:oneloopmass}, the renormalisation schemes which have
been adopted, shall be presented in detail in the following. \s

We start with a renormalisation scheme which is a mixture between
on-shell and $\overline{\mbox{DR}}$ renormalisation
conditions. In order to make contact and to compare to earlier results
presented in Ref.~\cite{slavich} our result will also be converted to a
pure $\overline{\mbox{DR}}$ scheme. To get an estimate of
the uncertainties due to missing higher-order corrections we study the
influence of the renormalisation scheme on the one-loop Higgs mass
corrections. We therefore also compare to a pure OS scheme. In all
three renormalisation schemes, mixed,  $\overline{\mbox{DR}}$  and OS,
we start out from the parameter set
Eq.~(\ref{eq:parset}) and replace the parameters by the renormalised ones 
and their corresponding counterterms,
\beq
\begin{array}{lcllcl}
M_Z^2 &\to& M_Z^2 + \delta M_Z^2 & \qquad \qquad t_{h_u} &\to & t_{h_u} +
\delta t_{h_u} \\
M_W^2 &\to& M_W^2 + \delta M_W^2 & \qquad \qquad t_{h_d} &\to &
t_{h_d} + \delta t_{h_d} \\
M_{H^\pm}^2 &\to& M_{H^\pm}^2 + \delta M_{H^\pm}^2 & \qquad \qquad
t_{h_s} &\to & t_{h_s} + \delta t_{h_s} \\[0cm]
e &\to & (1+\delta Z_e) e & \qquad \qquad \tan \beta &\to & \tan \beta
+\delta \tan \beta \\[0cm]
\lambda &\to & \lambda + \delta \lambda & \qquad \qquad \kappa &\to &
\kappa + \delta \kappa \\
v_s &\to & v_s + \delta v_s & \qquad \qquad A_\kappa &\to & A_\kappa +
\delta A_\kappa~. \label{eq:CTersetzung}
\end{array}  
\eeq
For the field renormalisation, the Higgs boson doublet and singlet
fields are replaced by the renormalised ones and a corresponding
single field
renormalisation constant for each doublet and the singlet, respectively,
\begin{alignat}{2}\nonumber
H_u &\to \sqrt{Z_{H_u}} \,H_u  \,&=& \, \left(1 + \frac{1}{2} \delta
  Z_{H_u}\right)  H_u\\ 
H_d &\to \sqrt{Z_{H_d}} \, H_d  \,&=& \, \left(1 + \frac{1}{2} \delta
  Z_{H_d} \right)  H_d \label{eq:fieldren}\\\nonumber
S &\to \sqrt{Z_{S}} \, S  \,&=& \, \left(1 + \frac{1}{2} \delta Z_{S} \right)  S \;.
\end{alignat}
Applying this renormalisation procedure, the renormalised 
self-energies in the basis $\{h^S_i \,|\, i=1,2,3 \} =\{ h_d,h_u,h_s \}$
for the CP-even Higgs bosons and in the
basis  $\{h^P_i \,|\, i=1,2,3 \} = \{a,a_s,G \}$ for the CP-odd ones can
be derived as ($X=S,P$) 
\begin{align}\nonumber
\hat{\Sigma}_{h^X_i h^X_j}(k^2) &= \Sigma_{h^X_i h^X_j}(k^2)  + \frac{1}{2} k^2 \bigl[\delta Z_{h^X_j h^X_i} +\delta Z_{h^X_i h^X_j}\bigr] \\&\quad\
- \frac{1}{2}\bigl[\delta Z_{h^X_k h^X_i}(M_X^2)_{kj} + (M_X^2)_{ik}\delta Z_{h^X_k h^X_j}\bigr]  - \delta (M_X^2)_{ij} \quad \text{with} \quad
i,j = 1,2,3~, \label{eq:CPevenoddrS}
\end{align}
where 
\begin{align}
\delta Z_{h^S_1 h^S_1} = \delta Z_{H_d}, \quad\ \delta Z_{h^S_2 h^S_2} = \delta Z_{H_u}, \quad\ \delta Z_{h^S_3 h^S_3} = \delta Z_{S}, 
\quad\ \delta Z_{h^S_i h^S_j} = 0 \quad\ \text{for} \quad\ i \neq j,
\end{align}
and
\beq
\delta Z_{h^P_1 h^P_1} &=& s_\beta^2 \delta Z_{H_d} + c_\beta^2 \delta
Z_{H_u}~, \qquad \delta Z_{h^P_2 h^P_2} = \delta Z_{S}~, \qquad 
\delta Z_{h^P_3 h^P_3} = c_\beta^2 \delta Z_{H_d} + s_\beta^2 \delta
Z_{H_u} \\
\delta Z_{h^P_1 h^P_3} &=& \delta Z_{h^P_3 h^P_1} = s_\beta c_\beta
(\delta Z_{H_d} - \delta Z_{H_u})~,\\ 
\delta Z_{h^P_1 h^P_2} &=& \delta Z_{h^P_2 h^P_1} =\delta Z_{h^P_2
  h^P_3} = \delta Z_{h^P_3 h^P_2} = 0~. 
\eeq
Here, $M_S^2$ and $M_P^2$ are the tree-level CP-even and CP-odd Higgs
boson mass matrices squared, respectively. For the derivation of the
counterterm matrices $\delta (M_S^2)$ and $\delta (M_P^2)$, 
the CP-even and CP-odd Higgs boson mass matrices squared are expressed
in terms of the parameter set Eq.~\eqref{eq:parset} including 
also their dependence on the tadpole parameters as well as on the
mixing angle $\beta_B$. As the expressions are quite lengthy they are given
in Appendix A. Then, the
parameters entering these matrices are replaced according to
Eqs.~(\ref{eq:CTersetzung}) and an expansion about the counterterms is
performed. The part of $M_S^2$ and $M_P^2$, respectively, which is
linear in the counterterms, corresponds to $\delta (M_S^2)$
and $\delta (M_P^2)$.\s

The renormalised self energies in Eq.~(\ref{eq:CPevenoddrS}) are
related to the ones in the basis of the mass eigenstates through, {\it
cf.}~Eqs.~(\ref{eq:cpevenrot},\ref{eq:cpoddrot}),
\beq
\hat{\Sigma}_{H_iH_j}(k^2) &=& \mathcal{R}^S_{ik}\; \mathcal{R}^S_{jl} \;\hat{\Sigma}_{h^S_k h^S_l}(k^2) \\
\hat{\Sigma}_{A_iA_j}(k^2) &=& \mathcal{R}^P_{ik} \; \mathcal{R}^P_{jl}
\; \hat{\Sigma}_{h^P_k h^P_l}(k^2) \qquad \text{with} \qquad i,j,k,l = 1,2,3~.
\eeq
The field renormalisation
constants $\delta Z_{H_d}, \delta Z_{H_d}, \delta Z_S$ are obtained from
\beq
\delta Z_{H_i H_i} = |{\cal R}^S_{i1}|^2 \delta Z_{H_d} + |{\cal
  R}^S_{i2}|^2 \delta Z_{H_u} + |{\cal R}^S_{i3}|^2 \delta Z_{S} \; ,
\qquad i =1,2,3 \; , 
\label{eq:renconst}
\eeq 
with
\beq
\delta Z_{H_i H_i} = -\left. \frac{\partial \Sigma_{H_i H_i} (k^2)}{\partial
k^2} \right|_{k^2=(M^{(0)}_{H_i} )^2}^{\scriptsize{\mbox{div}}} \; ,
\qquad i=1,2,3 \;, 
\eeq
where $(M_{H_i}^{(0)})^2$ denotes the corresponding tree-level
mass squared. The field renormalisation constants are defined in all three
renormalisation schemes via $\overline{\text{DR}}$~conditions. This 
is indicated by the superscript 'div' and means that in the field
renormalisation only the divergent part $\Delta = 2/(4 - D) - \gamma_E + \ln(4 \pi)$ is kept with $\gamma_E$ being the
Euler constant. Solving Eq.~\eqref{eq:renconst} for $\delta Z_{H_d},\delta
Z_{H_u},\delta Z_S$ results in 
\begin{align}\nonumber  
\delta Z_{H_d} &=\bigl[\bigl(|\mathcal{R}^S_{23}|^2 |\mathcal{R}^S_{32}|^2-|\mathcal{R}^S_{22}|^2
  |\mathcal{R}^S_{33}|^2\bigr)\delta Z_{H_1 H_1}
+ \bigl(|\mathcal{R}^S_{12}|^2 |\mathcal{R}^S_{33}|^2-|\mathcal{R}^S_{13}|^2
  |\mathcal{R}^S_{32}|^2\bigr)\delta Z_{H_2 H_2} \\& \quad
+ \bigl(|\mathcal{R}^S_{13}|^2 |\mathcal{R}^S_{22}|^2-|\mathcal{R}^S_{12}|^2
  |\mathcal{R}^S_{23}|^2\bigr)\delta Z_{H_3 H_3}\bigr]/R^S_r \label{eq:dZHd}\\[0.1cm]\nonumber
\delta Z_{H_u} &=\bigl[ \bigl(|\mathcal{R}^S_{21}|^2
  |\mathcal{R}^S_{33}|^2-|\mathcal{R}^S_{23}|^2
  |\mathcal{R}^S_{31}|^2\bigr)\delta Z_{H_1 H_1}
+ \bigl(|\mathcal{R}^S_{13}|^2
  |\mathcal{R}^S_{31}|^2-|\mathcal{R}^S_{11}|^2
  |\mathcal{R}^S_{33}|^2\bigr) \delta Z_{H_2 H_2}\\& \quad
+ \bigl(|\mathcal{R}^S_{11}|^2
  |\mathcal{R}^S_{23}|^2-|\mathcal{R}^S_{13}|^2
  |\mathcal{R}^S_{21}|^2\bigr) \delta Z_{H_3 H_3}\bigr]/R^S_r \label{eq:dZHu}\\[0.1cm]\nonumber
\delta Z_{S} &=\bigl[ \bigl(|\mathcal{R}^S_{22}|^2
|\mathcal{R}^S_{31}|^2-|\mathcal{R}^S_{21}|^2
|\mathcal{R}^S_{32}|^2\bigr)\delta Z_{H_1 H_1}
+ \bigl(|\mathcal{R}^S_{11}|^2
|\mathcal{R}^S_{32}|^2-|\mathcal{R}^S_{12}|^2
|\mathcal{R}^S_{31}|^2\bigr)\delta Z_{H_2 H_2}\\& \quad
+ \bigl(|\mathcal{R}^S_{12}|^2 |\mathcal{R}^S_{21}|^2-|\mathcal{R}^S_{11}|^2
  |\mathcal{R}^S_{22}|^2\bigr)\delta Z_{H_3 H_3}\bigr]/R^S_r \label{eq:dZS}
\end{align} 
where
\begin{align}
R^S_r &= -|\mathcal{R}^S_{11}|^2 |\mathcal{R}^S_{22}|^2 |\mathcal{R}^S_{33}|^2 \nonumber
        +|\mathcal{R}^S_{11}|^2 |\mathcal{R}^S_{23}|^2 |\mathcal{R}^S_{32}|^2
        +|\mathcal{R}^S_{12}|^2 |\mathcal{R}^S_{21}|^2 |\mathcal{R}^S_{33}|^2\\&\quad
        -|\mathcal{R}^S_{12}|^2 |\mathcal{R}^S_{23}|^2 |\mathcal{R}^S_{31}|^2
        -|\mathcal{R}^S_{13}|^2 |\mathcal{R}^S_{21}|^2 |\mathcal{R}^S_{32}|^2
        +|\mathcal{R}^S_{13}|^2 |\mathcal{R}^S_{22}|^2 |\mathcal{R}^S_{31}|^2~.
\end{align}
\s

Contrary to the field renormalisation constants the renormalisation
conditions for the remaining parameters are different in the three
chosen renormalisation schemes, as will be described in the
following. \s

\noindent
\underline{\bfseries Mixed renormalisation scheme}

In the mixed renormalisation scheme we divide the parameters into parameters
defined through on-shell conditions\footnote{In slight abuse of the
  language we also call the renormalisation conditions for the tadpole
  parameters on-shell.}  
and into parameters defined via $\overline{\mbox{DR}}$~conditions:
\beq
\underbrace{M_Z,M_W,M_{H^\pm},t_{h_u},t_{h_d},t_{h_s},e}_{\mbox{on-shell
 scheme}}, 
\underbrace{\tan \beta, \lambda, v_s, \kappa, 
A_\kappa}_{\overline{\mbox{DR}} \mbox{ scheme}} \;.
\eeq
In the following, the various counterterms shall be specified in more
detail. \s

\noindent \underline{\it (i,ii)  Gauge boson masses} 

The gauge boson masses are defined through on-shell conditions,
\beq
\mbox{Re} \hat{\Sigma}^T_{ZZ} (M_Z^2) = 0, \quad
\mbox{Re} \hat{\Sigma}^T_{WW} (M_W^2) = 0 \;,
\eeq
where $T$ denotes the transverse part of the respective
self-energy. For the mass counterterms this yields
\beq
\delta M_Z^2 = \mbox{Re} \Sigma_{ZZ}^T (M_Z^2), \quad
\delta M_W^2 = \mbox{Re} \Sigma_{WW}^T (M_W^2) \;.
\eeq
Note, that the NMSSM gauge boson self-energies differ from the MSSM
case due to the introduction of an additional
superfield $\hat{S}$.
\s

\noindent \underline{\it (iii)  Mass of the charged Higgs boson} 

The mass of the charged Higgs boson is determined through the on-shell
condition, 
\beq
\mbox{Re} \hat{\Sigma}_{H^\pm H^\mp} (M_{H^\pm}^2) = 0 \;,
\eeq
resulting in the corresponding counterterm
\beq
\delta M_{H^\pm}^2 = \mbox{Re} \Sigma_{H^\pm H^\mp} (M_{H^\pm}^2) \;.
\eeq
\s

\noindent \underline{\it (iv-vi)  Tadpole parameters} 

The tadpole coefficients are required to vanish also at one-loop
order, yielding
\beq
t^{(1)}_{h_i} - \delta t_{h_i} = 0 \, , \quad i=d,u,s \;, \label{eq:tadpolecond1}
\eeq
where $t^{(1)}_{h_i}$ stands for the contributions coming from the 
corresponding genuine Higgs boson tadpole graphs. As the tadpole graphs
are calculated in the mass eigenstate basis, they have to be
transformed to the interaction basis. Applying
Eq.~(\ref{eq:cpevenrot}) we have 
\beq
\delta t_{h_i} = {\cal R}^S_{ji} \; t^{(1)}_{H_j} \quad, \qquad \quad
i=d,u,s, \quad j=1,2,3 \;.
\eeq
\s

\noindent \underline{\it (vii)  Electric charge} 

The electric charge is defined to
be the full electron-positron photon coupling for
on-shell external particles in the Thomson limit, so that all
corrections to this vertex vanish on-shell and for zero momentum
transfer. The counterterm for
the electric charge is then given in terms of the transverse part of the
photon-photon and photon-$Z$ self-energies
\cite{Denner:1991kt}\footnote{Note that the sign of the second term in
Eq.~\eqref{eq:delze} differs from the one in \cite{Denner:1991kt} due
to our conventions in the Feynman rules.},
\beq
\delta Z_e = \frac{1}{2} \left.\frac{\partial \Sigma^T_{\gamma\gamma}
   (k^2)}{\partial k^2}\right|_{k^2=0} +
\frac{s_W}{c_W} \frac{\Sigma^T_{\gamma Z} (0)}{M_Z^2} \;.
\label{eq:delze}
\eeq 
\s

\noindent \underline{\it (viii) $\tan\beta$}

For the renormalisation of $\tan \beta$ we adopt the
$\overline{\mbox{DR}}$ scheme. Applying Eq.~(\ref{eq:fieldren})
and
\beq
v_i \to v_i + \delta v_i \qquad \qquad i = u,d
\eeq
results in
\beq
\delta \tan \beta = \tan \beta \left[ \frac{1}{2} (\delta Z_{H_u} -
  \delta Z_{H_d})+ \left( \frac{\delta v_u}{v_u} - \frac{\delta
      v_d}{v_d} \right) \right]_{\mbox{\scriptsize{div}}} = 
\left[ \frac{\tan \beta }{2} (\delta Z_{H_u} - \delta Z_{H_d})
\right]_{\mbox{\scriptsize{div}}} \;,
\eeq
where we have used in the last step $\delta
v_u/v_u|_{\mbox{\scriptsize{div}}} = \delta v_d
/v_d|_{\mbox{\scriptsize{div}}}$ \cite{vuvd}. 
The field renormalisation constants $\delta Z_{H_d}$ and $\delta
Z_{H_u}$ are  given in Eqs.~(\ref{eq:dZHd}, \ref{eq:dZHu}).
\s

\noindent\underline{\it  (ix) Coupling $\lambda$} 

In the mixed renormalisation scheme $\lambda$ is defined as a
$\overline{\text{DR}}$ parameter.  The counterterm is determined 
via the renormalised self-energy $\hat{\Sigma}_{h_1^P h_1^P}$, see
Eq.~(\ref{eq:CPevenoddrS}), using that
\begin{align}\label{eq:findingdlambda}
\hat{\Sigma}_{h_1^P h_1^P} \left.\Bigl((M_P^2)_{11}\Bigr)
\right|_{\scriptsize{\mbox{div}}} = 0 \quad \Longleftrightarrow \quad
\delta (M_P^2)_{11} = \left. \Sigma_{h_1^P h_1^P} \Bigl((M_P^2)_{11}\Bigr)
\right|_{\scriptsize{\mbox{div}}} \;.
\end{align}
As $\delta (M_P^2)_{11}$ contains the counterterm $\delta
\lambda$\footnote{Compare with Eq.~(\ref{eq:mah2}).},
Eq.~\eqref{eq:findingdlambda} can be solved for $\delta \lambda$
resulting in,
\begin{align}
\nonumber
\delta \lambda = \frac{e^2}{4 \lambda M_W^2 s_W^2}\Bigl[&\Sigma_{h_1^P h_1^P}
 ((M_P^2)_{11})  - \delta M_{H^\pm}^2  + \delta M_W^2 \Bigl(1 + \frac{2 \lambda^2
   (c_W^2-s_W^2)}{e^2} \Bigr) 
\\&
- \frac{2 \lambda^2 c_W^4}{e^2} \delta M_Z^2
+ \frac{4 \lambda^2 M_W^2 s_W^2}{e^2} \delta Z_e 
\Bigr]_{\scriptsize{\mbox{div}}} \label{eq:dellambda} \quad .
\end{align}
The self-energy $\Sigma_{h_1^P h_1^P}$ is obtained from the
self-energies in the mass eigenstate basis $\Sigma_{A_iA_j}$
($i,j=1,2,3$) through
\beq
\Sigma_{h_1^P h_1^P} = {\cal R}^P_{i1} \, \Sigma_{A_i A_j} \, {\cal R}^P_{j1} \quad \;.
\eeq
\s

\noindent
\underline{\it (x)  Singlet vacuum expectation value $v_s$} 

The vacuum expectation value $v_s$ of the singlet field is renormalised in the
$\overline{\mbox{DR}}$ scheme. The counterterm is derived by exploiting the
chargino sector. In fact, the lower right entry of the chargino mass
matrix $M_C$ in the interaction basis reads, {\it
  cf.}~Eq.\eqref{eq:charmat},
\beq
(M_C)_{22} = \frac{\lambda v_s}{\sqrt{2}} \;.
\eeq
Applying Eq.~\eqref{eq:CTersetzung}, expanding around the counterterms
and extracting the terms linear in the counterterms yields
\beq
\delta v_s = \left[\frac{\sqrt{2}}{\lambda} \delta (M_C)_{22} 
- v_s \frac{\delta \lambda}{\lambda}
\right]_{\mbox{\scriptsize{div}}} \;,
\label{eq:delvs}
\eeq
with $\delta \lambda$ given by Eq.~(\ref{eq:dellambda}). The
counterterm $\delta (M_C)_{22}$ is obtained from the renormalised
chargino self-energies in the following way. Defining the
general structure of a fermionic self-energy\footnote{The
  decomposition can be applied both for
unrenormalised self-energies $\Sigma$ and renormalised self-energies
$\hat{\Sigma}$.} as,
\beq
\Sigma_{ij} (k^2) = \slash{\!\!\! k} \Sigma^L_{ij} (k^2) {\cal P}_L + 
\slash{\!\!\! k} \Sigma^R_{ij} (k^2) {\cal P}_R + 
 \Sigma^{Ls}_{ij} (k^2) {\cal P}_L + \Sigma^{Rs}_{ij} (k^2) {\cal P}_R \; ,
\label{eq:strucself}
\eeq
where ${\cal P}_{L,R} = (\id \mp \gamma_5)/2$ are the left- and
right-handed projectors, we use the condition
\beq
\left[ (M_C)_{22} [V^\dagger \hat{\Sigma}^L_{\chi^\pm} (k^2) V +
  U^T \hat{\Sigma}^R_{\chi^\pm} (k^2) U^*]_{22} +  [U^T
  \hat{\Sigma}^{Ls}_{\chi^\pm} (k^2) V +
  V^\dagger \hat{\Sigma}^{Rs}_{\chi^\pm} (k^2)
  U^*]_{22}\right]_{\scriptsize \mbox{div}} = 0 \;.
\label{eq:vsrencond}
\eeq
Note that $\hat{\Sigma}_{\chi^\pm}$ is a $2\times 2$ matrix with the
entries given by the renormalised self-energies of the charginos in
the mass eigenbasis. The renormalised self-energies in terms of the
unrenormalised ones, the mass counterterms and field renormalisation
constants are given 
in Appendix B, Eqs.~\eqref{eq:cren1}-\eqref{eq:cren4}. The structure of
the condition 
Eq.~\eqref{eq:vsrencond} has been chosen such that the (divergent)
contributions of the chargino field renormalisation constants drop
out. Replacing the renormalised self-energy by the relations
\eqref{eq:cren1}-\eqref{eq:cren4} leads to the counterterm $\delta
(M_C)_{22}$, 
\beq 
\delta (M_C)_{22}  = \frac{1}{2} \left[ (M_C)_{22} 
[V^\dagger \Sigma^L_{\chi^\pm} (k^2) V + U^T \Sigma^R_{\chi^\pm} (k^2)
U^*]_{22} + [U^T \Sigma^{Ls}_{\chi^\pm} (k^2) V +
  V^\dagger \Sigma^{Rs}_{\chi^\pm} (k^2) U^*]_{22}\right]_{\scriptsize
  \mbox{div}} 
\label{eqdelmchi}
\eeq
Note that the $k^2$ dependence in Eq.~\eqref{eqdelmchi} drops out as
only the divergent part is taken. \s 

\noindent 
\underline{\it (xi)  Coupling $\kappa$} 

The counterterm for the $\overline{\mbox{DR}}$ renormalised parameter
$\kappa$ is derived from the neutralino sector in an analogous
procedure as the chargino sector was exploited to determine $\delta
v_s$. The lower right entry  $(M_N)_{55}$ of the neutralino mass matrix
Eq.~(\ref{eq:neutmat}) reads
\beq
(M_N)_{55} = \sqrt{2} \kappa v_s \;,
\eeq
leading to the counterterm $\delta \kappa$,
\beq
\delta \kappa = \frac{1}{\sqrt{2} v_s} \delta (M_N)_{55} - \kappa
\frac{\delta v_s}{v_s} \;.
\label{eq:delkap}
\eeq
The determination of $\delta v_s$ has been described in the previous
paragraph. For the determination of $\delta (M_N)_{55}$ we use the condition
\beq
\left[ (M_N)_{55} [{\cal N}^T (\hat{\Sigma}^L_{\chi^0} (k^2)+
  \hat{\Sigma}^R_{\chi^0} (k^2)) {\cal N}]_{55}
+ [{\cal N}^T (\hat{\Sigma}^{Ls}_{\chi^0} (k^2) + 
  \hat{\Sigma}^{Rs}_{\chi^0} (k^2)) {\cal N}]_{55}
\right]_{\scriptsize \mbox{div}}  =  0 \;,
\label{eq:neutcond}
\eeq
where the fermionic self-energy structure Eq.~\eqref{eq:strucself} has
been applied for the decomposition of the renormalised neutralino
$5\times 5$ self-energy matrix $\hat{\Sigma}_{\chi^0}$. Once again the
condition has been chosen such that the divergent parts of the field
renormalisation constants cancel in
Eq.~\eqref{eq:neutcond}. Rewriting the equation in terms of the
unrenormalised self-energies, {\it
  cf.}~Eqs.~\eqref{eq:nren1}-\eqref{eq:nren4}, yields 
\beq
\delta (M_N)_{55} = \frac{1}{2} \left[ (M_N)_{55} 
[{\cal N}^T (\Sigma^L_{\chi^0} (k^2) + \Sigma^R_{\chi^0} (k^2) ) {\cal N}]_{55}
+ [{\cal N}^T (\Sigma^{Ls}_{\chi^0}  (k^2) + \Sigma^{Rs}_{\chi^0} (k^2))
{\cal N}]_{55}  \right]_{\scriptsize \mbox{div}} \;,
\eeq
which is inserted in Eq.~\eqref{eq:delkap} to determine $\delta \kappa$.
\s

\newpage 
\noindent
\underline{\it (xii)  Trilinear coupling $A_\kappa$}

The trilinear coupling $A_\kappa$ is also defined as a $\overline{\mbox{DR}}$~parameter. For the derivation of the counterterm we use that
\begin{align}\label{eq:findingAkappa}
\hat{\Sigma}_{h_2^P h_2^P} \Bigl((M_P^2)_{22}\Bigr)
\Big|_{\scriptsize{\mbox{div}}} = 0 \;.
\end{align}
Equation~\eqref{eq:findingAkappa} depends on $\delta A_\kappa$ via the
mass matrix squared counterterm $\delta (M_P^2)_{22}$. Solving for  
$\delta A_\kappa$ we have
\begin{align}
\delta A_\kappa = \Bigl[- \frac{\sqrt{2}}{3 \kappa v_s}\bigl[\Sigma_{h_2^P h_2^P} \Bigl((M_P^2)_{22}\Bigr) - \delta f\bigr]
- A_\kappa \bigl[\frac{\delta \kappa}{\kappa} + \frac{\delta
  v_s}{v_s}\bigr]  \Bigr]_{\scriptsize{\mbox{div}}} \;.
\end{align}
The counterterm $\delta f$ is derived from 
\begin{align} \nonumber
f &= \frac{t_{h_s}}{v_s} 
-\frac{2 M_W s_W s_{\beta} c^2_{\beta} c^2_{\beta_B}}{e v_s^2
 c_{\Delta \beta}^2} [t_{h_u}+ t_{h_d} t_{\beta} t^2_{\beta_B}]
+ \frac{M_W^2 s_W^2 s^2_{2\beta} }{e^2 v_s^2  c_{\Delta \beta}^2}
[M^2_{H^\pm} - M_W^2 c_{\Delta \beta}^2]
\\& \quad\
+ \frac{\lambda M_W^2 s_W^2 s_{2\beta}}{e^4 v_s^2} [2\lambda M_W^2
s_W^2 s_{2\beta} + 3 \kappa e^2 v_s^2] \;,
\end{align}
with $\Delta \beta =  \beta - \beta_B$. This is done by replacing the
parameters $t_{h_u}$, $ t_{h_d}$, $ t_{h_s}$, $ e$, $ M_Z$, $ M_W$, $ M_{H^\pm}$, $
\tan\beta$, $ \lambda$, $\kappa$, and $v_s$ according  to
Eq.~\eqref{eq:CTersetzung}, performing an expansion about the
counterterms, extracting the part linear in the counterterms and
finally applying the tree-level relations for the tadpole parameters
$t_{h_u}= t_{h_d}= t_{h_s} = 0$ and for the mixing angle, $\beta_B =
\beta$. \s 
 
The self-energy $\Sigma_{h_2^P h_2^P}$ in terms of
the corresponding self-energies in the mass eigenbasis is given by 
\beq
\Sigma_{h_2^P h_2^P}= {\cal R}_{i 2}^P \Sigma_{A_iA_j} {\cal R}_{j 2}^P
\;, \qquad i,j=1,2,3 \;.
\eeq

Alternatively, we could have derived the counterterms $\delta v_s$ and
$\delta \kappa$ from the Higgs sector instead of resorting to the
chargino and neutralino sector. We have explicitly verified that this
leads to the same results for the one-loop corrected Higgs boson masses. Our
choice of renormalisation allows a non-trivial cross-check of the
renormalisation procedure. Moreover, it paves the way for an extension
of the one-loop corrections to the Higgs decays into charginos and
neutralinos. \s

\noindent
\underline{\bfseries On-shell renormalisation scheme} 

In this renormalisation scheme we keep the conditions
(i)-(viii) ($\tan\beta$ is still renormalised in the
$\overline{\mbox{DR}}$ scheme)
but the parameters $v_s,\lambda,\kappa$ and $A_\kappa$ are determined
via on-shell renormalisation conditions. 
This is done by applying on-shell renormalisation conditions
on the two CP-odd Higgs boson mass eigenstates $A_1,A_2$, on the mass
eigenstates of the two charginos and on the mass eigenstates of the
two lightest neutralinos. As OS conditions are 
imposed on the mass eigenstates not only single elements of the
counterterm mass matrices occur in the conditions, in 
contrast to the $\overline{\mbox{DR}}$ conditions exploited within the 
mixed scheme. Thus, the equations get more involved and
include also the parameters $M_1,M_2$. This is why six
renormalisation conditions are required, although in the end the
counterterms for $M_1,M_2$ are not needed. \s

Requiring on-shell masses for the CP-odd Higgs bosons as well as for the two charginos and for the lightest and 
next-to-lightest neutralinos leads to the following relations,
\begin{equation}
\begin{alignedat}{2}
\Big( \mathcal{R}^P\;\delta
(M^2_P)\;(\mathcal{R}^P)^\mathrm{T}\Big)\Big|_{11}&=\text{Re}
\Sigma_{A_1A_1}((M_{A_1}^{(0)})^2)\,,
 \quad
\Big( \mathcal{R}^P\;\delta
(M_P^2)\;(\mathcal{R}^P)^\mathrm{T}\Big)\Big|_{22}=\text{Re}
\Sigma_{A_2A_2}((M_{A_2}^{(0)})^2)\,,&&\\
\Big(U^*\;\delta (M_C) \;V^\dagger\Big)\Big|_{11}&=  \frac{1}{2}\text{Re}
\Bigl[m_{\chi^\pm_1}\bigl(\Sigma^L_{\chi^\pm_{11}}(m^2_{\chi^\pm_1})
+ \Sigma^R_{\chi^\pm_{11}}(m^2_{\chi^\pm_1})\bigr) + \Sigma^{Ls}_{\chi^\pm_{11}}(m^2_{\chi^\pm_1}) 
+ \Sigma^{Rs}_{\chi^\pm_{11}}(m^2_{\chi^\pm_1}) \Bigr]
\,
, &&\\
\Big(U^*\;\delta (M_C)\;V^\dagger\Big)\Big|_{22}&= \frac{1}{2}\text{Re}
\Bigl[m_{\chi^\pm_2}\bigl(\Sigma^L_{\chi^\pm_{22}}(m^2_{\chi^\pm_2})
+ \Sigma^R_{\chi^\pm_{22}}(m^2_{\chi^\pm_2})\bigr) + \Sigma^{Ls}_{\chi^\pm_{22}}(m^2_{\chi^\pm_2}) + 
\Sigma^{Rs}_{\chi^\pm_{22}}(m^2_{\chi^\pm_2}) \Bigr]\,,&&\\
\Big(\mathcal{N}\;\delta (M_N)\;\mathcal{N}^T\Big)\Big|_{11}&=\frac{1}{2}\text{Re}
\Bigl[m_{\chi^0_1}\bigl(\Sigma^L_{\chi^0_{11}}(m^2_{\chi^0_1})
+ \Sigma^R_{\chi^0_{11}}(m^2_{\chi^0_1})\bigr) + \Sigma^{Ls}_{\chi^0_{11}}(m^2_{\chi^0_1}) 
+ \Sigma^{Rs}_{\chi^0_{11}}(m^2_{\chi^0_1}) \Bigr]
\,,&&\\
 \Big(\mathcal{N}\;\delta (M_N)\;\mathcal{N}^T\Big)\Big|_{22}&= \frac{1}{2}\text{Re}
\Bigl[m_{\chi^0_2}\bigl(\Sigma^L_{\chi^0_{22}}(m^2_{\chi^0_2})
+ \Sigma^R_{\chi^0_{22}}(m^2_{\chi^0_2})\bigr) + \Sigma^{Ls}_{\chi^0_{22}}(m^2_{\chi^0_2}) + 
\Sigma^{Rs}_{\chi^0_{22}}(m^2_{\chi^0_2}) \Bigr]
\,,
\label{eq:eqsystem}
\end{alignedat}
\end{equation}
where $(M_{A_{1,2}}^{(0)})^2$ are the tree-level masses squared of the
pseudoscalar mass eigenstates $A_1,A_2$. The counterterm matrix
$\delta (M_P^2)$ is derived as described above.
The entries $\delta (M_C)_{ij}$ ($i,j=1,2$) of the counterterm matrix
$\delta (M_C)$ for the charginos read
\beq
\delta M_{C_{11}}&=& \delta M_2\\
\delta M_{C_{12}}&=& \delta M_W^2\frac{s_{\beta } }{\sqrt{2} M_W}+ \delta\!\tan\!\beta\,\sqrt{2} c_{\beta }^3 M_W\\
\delta M_{C_{21}}&=&\delta M_W^2\frac{c_{\beta } }{\sqrt{2} M_W}-\delta\!\tan\!\beta \,\sqrt{2}  c_{\beta }^2 M_W s_{\beta }\\
\delta M_{C_{22}}&=&\delta \lambda\frac{  v_s}{\sqrt{2}}+\delta
v_s\frac{\lambda  }{\sqrt{2}} \;.
\eeq
And finally, the entries $\delta (M_N)_{ij}= \delta (M_N)_{ji}$ ($i,j=1...5$) of the
counterterm matrix $\delta (M_N)$ for the neutralinos can be cast into
the form
\begin{align}
\delta M_{N_{11}}=&\delta M_1\displaybreak[3]\\
\delta M_{N_{13}}=& \left(\delta M_W^2-\delta
  M_Z^2\right)\frac{c_{\beta } }{2 M_Z s_W}+\delta\!\tan\!\beta\, \frac{c_{\beta } M_Z s_W  s_{2 \beta }}{2} \displaybreak[3] \\
\delta M_{N_{14}}=&  \left(\delta M_Z^2-\delta
  M_W^2\right)\frac{s_{\beta } }{2 M_Z s_W}+\delta\!\tan\!\beta
\,\frac{ c_{\beta }^3 M_W s_W}{c_W}\displaybreak[3]\\
\delta M_{N_{22}}=&\delta M_2\displaybreak[3]\\
\delta M_{N_{23}}=&  \delta M_W^2\frac{c_{\beta } }{2 M_W}-\delta\!\tan\!\beta\, c_{\beta }^2 M_W s_{\beta }\displaybreak[3]\\
\delta M_{N_{24}}=&  -\delta M_W^2\frac{s_{\beta }}{2 M_W}-\delta\!\tan\!\beta\, c_{\beta }^3 M_W\displaybreak[3]\\
\delta M_{N_{34}}=& -\delta \lambda\frac{  v_s}{\sqrt{2}}-\delta
v_s\frac{\lambda  }{\sqrt{2}} \displaybreak[3]\\
\delta M_{N_{35}}=&  -\left(c_W^4\delta M_Z^2-2
  c_W^2 \delta 
  M_W^2+\delta M_W^2\right)\frac{\lambda  s_{\beta } }{\sqrt{2} e M_W
  s_W}-\delta\!\tan\!\beta\, \frac{\sqrt{2}  \lambda  c_{\beta }^3 M_W s_W}{e}+\nonumber\\
&+\left(\lambda  \delta Z_e-\delta \lambda\right)\frac{\sqrt{2} M_W s_W s_{\beta } }{e}\displaybreak[3]\\
\delta M_{N_{45}}=& -\left(c_W^4 \delta M_Z^2-2
  c_W^2 \delta 
  M_W^2+\delta M_W^2\right)\frac{\lambda  c_{\beta } }{\sqrt{2} e M_W
  s_W}+\delta\!\tan\!\beta\, \frac{ \lambda  c_{\beta } M_W s_W s_{2 \beta }}{\sqrt{2} e}+\nonumber\\
&+\left(\lambda  \delta Z_e-\delta \lambda\right)\frac{\sqrt{2} c_{\beta } M_W s_W}{e}\displaybreak[3]\\
\delta M_{N_{55}}=&\sqrt{2} \delta \kappa  v_s+\sqrt{2} \kappa  \delta
v_s \displaybreak[3]\\ 
\delta M_{N_{12}}=& \delta M_{N_{15}}= \delta M_{N_{25}}= \delta
M_{N_{33}}= \delta M_{N_{44}}= 0 \;.
\end{align}
With these ingredients the explicit expressions for the equations
\eqref{eq:eqsystem} can be derived. In our calculation, the system of
equations \eqref{eq:eqsystem} is solved numerically, keeping, however,
the dependence on the divergent part $\Delta$ explicitly.
\s

\vspace*{0.2cm}
\noindent
\underline{\bfseries \mbox{$\overline{\mbox{DR}}$} renormalisation scheme}

The $\overline{\mbox{DR}}$ renormalisation scheme differs 
in the conditions (i)-(vii) from the mixed one, which means in the conditions for $M_Z,M_W,M_{H^\pm},t_{h_u},t_{h_d},t_{h_s},e$.
Note that a change of the  renormalisation condition for $M_{H^\pm}$ can be interpreted as a change of the condition for
$A_\lambda$, {\it cf.} Eq.~(\ref{eq:mhchar}). For these parameters instead of
OS conditions $\overline{\mbox{DR}}$ renormalisation conditions are
adopted now.  The other renormalisation conditions do not change.

\subsection{\label{sec:oneloopmass} Loop Corrected Higgs Boson Masses and Mixing Matrix Elements}
The one-loop corrected scalar Higgs boson masses
squared are extracted numerically as the zeroes of the determinant of the
two-point vertex functions $\hat{\Gamma}^S$,
\beq
\hat{\Gamma}^S (k^2) = && \phantom{LLLLLLLLLLLLLLLLLLLLLLLLLLLLLLLLLLLLLLLLLLLLLLLLLLLL} \nonumber\\
&& \hspace*{-2cm} i \left( \begin{array}{ccc} k^2 - (M_{H_1}^{(0)})^2 +
    \hat{\Sigma}_{H_1 H_1} (k^2) & \hat{\Sigma}_{H_1 H_2} (k^2) &
    \hat{\Sigma}_{H_1 H_3} (k^2) \\
\hat{\Sigma}_{H_2 H_1} (k^2) & k^2 - (M_{H_2}^{(0)})^2 +
\hat{\Sigma}_{H_2 H_2} (k^2) & \hat{\Sigma}_{H_2 H_3} (k^2) \\
\hat{\Sigma}_{H_3 H_1} (k^2) & \hat{\Sigma}_{H_3 H_2} (k^2) &
k^2 - (M_{H_3}^{(0)})^2 +\hat{\Sigma}_{H_3 H_3} (k^2)
\end{array} \right)  
\hspace*{-2cm }\label{eq:cpevenoneloop}
\eeq
In the same way the pseudoscalar masses squared are obtained from 
$\hat{\Gamma}^P$,
\beq
 \hat{\Gamma}^P (k^2) =  i \left( \begin{array}{cc} k^2 - (M_{A_1}^{(0)})^2 +
    \hat{\Sigma}_{A_1 A_1} (k^2) & \hat{\Sigma}_{A_1 A_2} (k^2) \\
\hat{\Sigma}_{A_2 A_1} (k^2) & k^2 - (M_{A_2}^{(0)})^2 +
\hat{\Sigma}_{A_2 A_2} (k^2) 
\end{array} \right) \;. 
\label{eq:cpoddoneloop} 
\eeq
The superscript $(0)$ denotes the tree-level values of the masses squared.
It should be noted that in Eq.~\eqref{eq:cpoddoneloop} the mixing
with the Goldstone bosons is not taken into account.  We have checked
explicitly that the numerical effect is negligible. 
The unrenormalised self-energy and tadpole contributions that occur implicitly in Eqs.~\eqref{eq:cpevenoneloop} and 
\eqref{eq:cpoddoneloop} are evaluated at one-loop
order. They contain fermion, Goldstone and Higgs boson,
gauge boson and ghost loops as well as loops from the corresponding
superpartners {\it i.e.} sfermions, charginos and neutralinos. \s

The mass eigenvalues are obtained iteratively. In order to obtain the
lightest scalar Higgs boson mass {\it e.g.}, in the first iteration
the external momentum squared  $k^2$ in the renormalised self-energies
$\hat{\Sigma}_{H_i H_j}$ is set equal to the lightest scalar
tree-level mass squared. Then, the mass matrix part of
$\hat{\Gamma}^S$, meaning  $(i \hat{\Gamma}^S + k^2 \id)$, is
diagonalised and the resulting mass eigenvalues squared are used in
the next iteration where $k^2$ is set equal to 
the lightest of the obtained mass eigenvalues. Once again the
mass eigenvalues are obtained. The 
procedure is repeated until the deviation between the lightest
eigenvalue and the one of the previous iteration is less than
$10^{-9}$. The other Higgs mass eigenvalues are derived accordingly \s

Due to the radiative corrections, not only the masses of the particles receive contributions but also
the fields are affected.  
To take these effects into account, new matrices, ${\cal R}^{S,1l}$,
${\cal R}^{P,1l}$, are introduced which transform the fields $h_u, h_d, h_s$ and $a, a_s$ into the corresponding one-loop mass eigenstates, 
respectively. These matrices are no physical observables
and beyond lowest order they depend on the external momentum in the
self-energies. For the derivation of the radiatively corrected
matrices, ${\cal R}^{S,1l}$,
${\cal R}^{P,1l}$, we follow the procedure applied in Ref.~\cite{1lfull}. 
It ensures the
correct on-shell properties for the external particle in processes with
external on-shell Higgs bosons at higher orders and thus accounts also for the mixing between the Higgs bosons. This leads to
finite wave function correction factors. In the
scalar case {\it e.g.} we have to apply the additional factor ${\bf Z}_f^S$ to the tree-level matrix ${\cal R}^S$,
which rotates the interaction states $(h_d,h_u,h_s)^T$ to the mass
eigenstates $(H_1,H_2,H_3)^T$,  to get the
one-loop matrix elements,
\beq
{\cal R}^{S,1l}_{il} = ({\bf Z}_f^S)_{ij} {\cal R}^S_{jl} \;, \qquad
i,j = H_1,H_2,H_3\;, \quad l =  h_d,h_u,h_s \;.
\eeq
The correction factor is given by
\beq
({\bf Z}_f^S)_{ij} = \sqrt{\hat{Z}^S_i} \hat{Z}^S_{ij} \;, 
\eeq
with
\beq
\hat{Z}^S_i = \frac{1}{1+ (\mbox{Re} \hat{\Sigma}^{\mbox{\scriptsize eff}}_{ii})'
  (M_{H_i})^2} \;.
\eeq
The prime denotes the derivative with respect to $k^2$, and
$M_{H_i}^2$ is the one-loop corrected Higgs boson mass squared. The effective
self-energy $\hat{\Sigma}^{\mbox{\scriptsize eff}}$ appears in the
diagonal Higgs boson propagators 
\beq
\Delta_{ii} (k^2) = - \bigl[\bigl(\hat{\Gamma}^S (k^2)\bigr)^{-1}\bigr]_{ii} = 
\frac{i}{k^2- (M_{H_i}^{(0)})^2 + \hat{\Sigma}^{\mbox{\scriptsize
    eff}}_{ii} (k^2) } \;.
\eeq
It is given by (no summation over $i,j,l$)
\beq
\hat{\Sigma}^{\mbox{\scriptsize eff}}_{ii} (k^2) = \hat{\Sigma}_{ii} - i \frac{2
  \hat{\Gamma}^S_{ij} (k^2) \hat{\Gamma}^S_{jl} (k^2) \hat{\Gamma}^S_{li}
  (k^2) - \hat{\Gamma}^{S\, 2}_{li} (k^2) \hat{\Gamma}^S_{jj} (k^2) -
  \hat{\Gamma}^{S\, 2}_{ij} (k^2) \hat{\Gamma}^S_{ll} (k^2)}{
  \hat{\Gamma}^S_{jj} (k^2) \hat{\Gamma}^S_{ll} (k^2) -
  \hat{\Gamma}^{S\, 2}_{jl} (k^2)} \;.
\eeq
The off-diagonal Higgs boson propagator $\Delta_{ij}$ ($i\ne j$, no
summation over $i,j,l$) reads
\beq
\Delta_{ij} (k^2)= \frac{\hat{\Gamma}^S_{ij} \hat{\Gamma}^S_{ll} -
  \hat{\Gamma}^S_{jl} \hat{\Gamma}^S_{li}}{\hat{\Gamma}^S_{ii}
  \hat{\Gamma}^S_{jj} \hat{\Gamma}^S_{ll} + 2\hat{\Gamma}^S_{ij}
  \hat{\Gamma}^S_{jl} \hat{\Gamma}^S_{li} - \hat{\Gamma}^S_{ii}
  \hat{\Gamma}^{2\, S}_{jl} - \hat{\Gamma}^S_{jj} \hat{\Gamma}^{S\,
    2}_{li}  - \hat{\Gamma}^S_{ll} \hat{\Gamma}^{S\, 2}_{ij} } \;. 
\eeq
The argument $k^2$ in $\hat{\Gamma}^S_{ij}$ has been dropped for
better readability.
For $\hat{Z}^S_{ij}$ we have in terms of the propagators (again no summation over the indices)
\beq
\hat{Z}^S_{ij} &=& \left.\frac{\Delta_{ij} (k^2)}{\Delta_{ii} (k^2)}
\right|_{k^2=M_{H_i}^2} \nonumber \\
&\stackrel{i\ne j}{=}& \frac{\hat{\Sigma}_{ij} (M_{H_i}^2)
  \left(M_{H_i}^2- (M_{H_l}^{(0)})^2
  +\hat{\Sigma}_{ll} (M_{H_i}^2) \right) - \hat{\Sigma}_{jl} (M_{H_i}^2)
  \hat{\Sigma}_{li} (M_{H_i}^2)}{\hat{\Sigma}_{jl}^2 (M_{H_i}^2) -
\left(M_{H_i}^2- (M_{H_j}^{(0)})^2
  +\hat{\Sigma}_{jj} (M_{H_i}^2) \right)
\left(M_{H_i}^2- (M_{H_l}^{(0)})^2
  +\hat{\Sigma}_{ll} (M_{H_i}^2) \right)} \\
\hat{Z}^S_{ii} &=& 1 \;.
\eeq
In case of the pseudoscalar $2\times 2$ mixing matrix ${\cal
  R}^{P,1l}_{il}$ the effective self-energy reduces to
\beq
\hat{\Sigma}^{\mbox{\scriptsize eff}}_{ii} (k^2) = \hat{\Sigma}_{ii} +
i \frac{\hat{\Gamma}^{P\, 2}_{ij} (k^2)}{\hat{\Gamma}^P_{jj} (k^2)} \;, \qquad 
i,j=A_1,A_2 
\eeq
and the off-diagonal propagator reads
\beq
\Delta_{ij} (k^2) = \frac{\hat{\Gamma}_{ij}^P}{\hat{\Gamma}_{ii}^P
  \hat{\Gamma}_{jj}^P - \hat{\Gamma}_{ij}^{P\, 2} } \;.
\eeq 
The thus derived mixing matrix elements include the full momentum
dependence and imaginary parts of the Higgs boson
self-energies. Alternatively we could have set $k^2=0$ which
corresponds to the result in the effective potential
approximation and yields a unitary mixing matrix. For our parameter sets used in the numerical analysis
we found that the differences in the two approaches are
negligible. Furthermore, the imaginary parts of the mixing matrix
elements are small compared to the real parts. 

\section{\label{sec:numerical} Numerical Analysis} 
The calculation of the Higgs and gauge boson self-energies, of the
tadpoles and the counterterms has been performed numerically in two
different calculations. In the first calculation all necessary Feynman
rules have been derived from the NMSSM Lagrangian and implemented in a
{\tt FeynArts} model file \cite{Hahn:2000kx}. In the second calculation the
Feynman rules have been obtained with the Ma\-the\-ma\-tica
package {\tt SARAH} \cite{sarah}. The Feynman rules in the
two approaches have been cross-checked against each other and also
against the rules given in Ref.~\cite{nmssm3}. Subsequently, in both
calculations {\tt FormCalc} \cite{formcalc} was used to evaluate the
self-energy and tadpole diagrams in the 't Hooft-Feynman gauge, in which
the Goldstone bosons and the ghost fields have the same masses as the
corresponding gauge bosons.  The divergent integrals are regularised
applying the constrained differential renormalisation scheme
\cite{constrained} which has been shown to be equivalent \cite{equiv}
to the SUSY conserving dimensional reduction scheme \cite{dimred}. The
numerical  computation of the integrals has been performed with {\tt
  LoopTools} \cite{formcalc}. Two Mathematica programs have been
written to evaluate the counterterms, diagonalise numerically the
one-loop corrected Higgs boson mass matrices and extract the mass
eigenvalues. \s

For our numerical analysis we follow the SUSY Les Houches Accord
(SLHA) \cite{slha} and use as input values the Fermi
constant $G_F=1.16637 \cdot 10^{-5}$~GeV$^{-2}$ and the $Z$ boson
mass $M_Z=91.187$~GeV.  For the electroweak coupling we set $\alpha =
1/137$. From these input values we derive the parameters of our
input set defined in Eq.~\eqref{eq:parset}. The top quark pole mass is given
by $M_t = 173.3$~GeV. As we cross-check our results against the ones of
Ref.~\cite{slavich} which uses the running $\overline{\mbox{DR}}$
quark  masses, we need to calculate these as well. In order to obtain
the $\overline{\mbox{DR}}$ top quark 
mass we convert $M_t$ at the scale $Q_t=M_t$ in the
corresponding running mass. The SM renormalisation group equations are
then used to evolve the top mass up to a common scale $Q$ chosen to be
of the order of the SUSY breaking scale, where the gluino corrections
are added. We denote the running top mass by $m_t$ in the
following. The same procedure is applied to the bottom mass starting
from the SLHA input value $m_b (m_b)^{\overline{{\scriptsize
      \mbox{MS}}}}$ set equal to 4.19~GeV.
The masses of the light quarks are chosen as $m_u=2.5$ MeV, $m_c=1.27$
GeV, $m_d=4.95$ MeV, $m_s=101$~MeV \cite{pdg}.   The $\tau$ mass has
been set to $m_\tau = 1.777$~GeV. \s

In the following we will discuss the results for various scenarios
which exemplify different effects of the higher-order corrections. For
our scenarios, we took care not to violate unitarity bounds by
choosing $\lambda,\kappa$ such that $\sqrt{\lambda^2+\kappa^2} \lsim
0.7$. Furthermore, $v_s$ has been chosen to be of the order of the
vacuum expectation value $v$. As the one-loop corrections to the
pseudoscalar masses are small, we mostly show plots for the scalar
masses and comment briefly on the pseudoscalar masses. 
Furthermore, the corrections for the
heaviest scalar Higgs boson are not shown, as in all cases they are
negligible. \s

\begin{figure}[b]
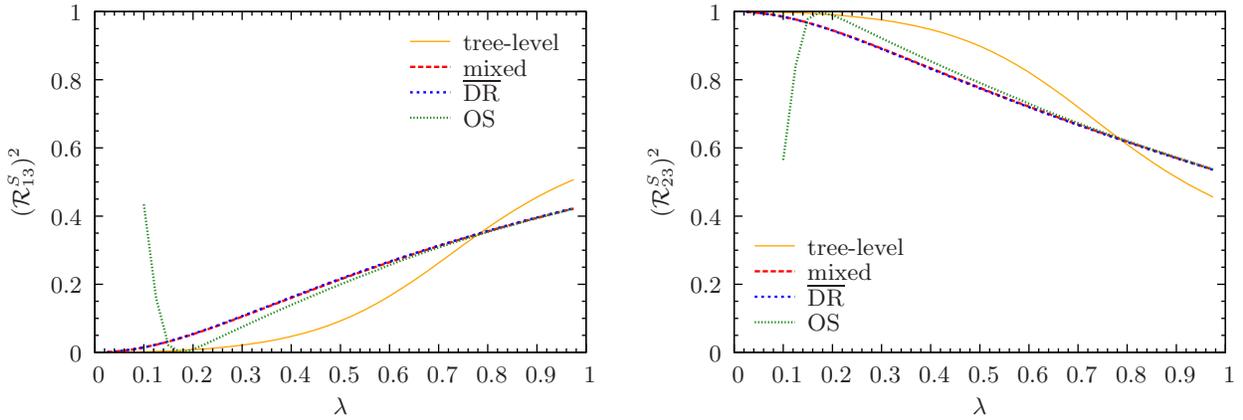

\begin{center}
\parbox{8.5cm}{\include{Fig1a}}\parbox{8.5cm}{\include{Fig1b}}
\caption{\label{fig:ri3} The matrix element squared $({\cal R}^S_{13})^2$
  (left), $({\cal R}^S_{23})^2$ (right)
  for the two lightest CP-even Higgs bosons as
  function of $\lambda$ at tree-level (yellow/full) and at 
  one-loop level adopting a $\overline{\mbox{DR}}$ (blue/dotted), a mixed
  (red/dashed) and an on-shell (green/small dotted) renormalisation scheme.}
\end{center}
\end{figure}
\noindent
\underline{\bf Variation of \boldmath{$\lambda$}}
We first verify that we reproduce the results of
Ref.~\cite{slavich} by adopting a $\overline{\mbox{DR}}$ renormalisation scheme
and choosing the same parameter set
\beq
\kappa = \lambda/5 \; , \quad \tan\beta= 2 \; , \quad A_\lambda =
500\;\mbox{GeV} \; , \quad A_\kappa = - 10\; \mbox{GeV} \; , \quad \mu
= 250 \; \mbox{GeV }
\eeq
with $\lambda$ being a free parameter and $v_s$ given by $v_s =
\sqrt{2} \mu/\lambda$. For the squarks and sleptons a common soft SUSY
breaking mass $M_S=300$~GeV has been adopted, and the remaining soft
SUSY breaking parameters have been chosen as
\beq
A_t = A_b = A_\tau = -1.5 M_S \; , \qquad M_1 = M_S/3 \; , \qquad M_2
= 2/3 M_S \; , \qquad M_3 = 2 M_S \; . 
\eeq
The renormalisation scale has been set equal to $Q_0=M_S$ and for the
top and bottom quark mass the running $\overline{\mbox{DR}}$ mass at
the scale $Q_0$ has 
been used. Furthermore, the light quark masses have been set to zero
as in Ref.~\cite{slavich}. 
We find agreement for the one-loop corrected Higgs boson
masses. Starting from this scenario in the $\overline{\mbox{DR}}$ scheme, 
in order to investigate the effect of different
renormalisation schemes, the $\overline{\mbox{DR}}$ input 
values have been converted to the mixed renormalisation scheme as defined
above, as well as to a pure on-shell scheme, and the Higgs boson masses and the matrix ${\cal R}^{S}$
are evaluated accordingly. For the definitions of
the schemes, see section \ref{sec:counter}. 
In Fig.~\ref{fig:ri3} we show the matrix elements squared $({\cal
  R}_{13}^{S})^2$ and $({\cal R}_{23}^{S})^2$, respectively, of the
mixing matrix ${\cal R}^S$. These matrix 
elements are a measure of the strength of the singlet component of the
two lightest Higgs bosons. They are shown for the three different
renormalisation schemes compared to the tree-level result as a
function of $\lambda$. In order to match the result of
Ref.~\cite{slavich} they have been evaluated at vanishing external
momentum $k^2 = 0$. At tree-level, for small values of $\lambda$, the
lightest CP-even Higgs boson is dominantly MSSM-like and the
next-to-lightest is dominantly singlet-like. With increasing $\lambda$
the mixing increases and at tree-level there is a
cross-over at $\lambda \approx 0.94$: 
the next-to-lightest Higgs boson is now more MSSM-like
than the lightest Higgs boson. The higher-order corrections change the
amount of the singlet component and for this parameter set there is no
cross-over below $\lambda=1$. Furthermore, we see that for $\lambda
\gsim 0.2$ the amount of the singlet component is hardly affected by the
renormalisation scheme. The curves for the three renormalisation
schemes lie on top of each other for 
$\lambda \gsim 0.65$. For smaller values of $\lambda$, however,
after having crossed $({\cal
  R}^S_{13})^2 = 0$ and $ ({\cal R}^S_{23})^2 = 1$, respectively, $({\cal
  R}^S_{13})^2, ({\cal R}^S_{23})^2$ in the OS-scheme start to differ largely
from the corresponding values in the two other schemes. 
The reason is that the finite parts of the counterterms involved in the conversion from the
$\overline{\mbox{DR}}$ parameters to on-shell parameters contain a
division by  $\lambda$, so that these finite counterterm contributions blow up in
the limit $\lambda \to 0$. The $H_1$ mass squared even turns negative for
$\lambda \lsim 0.1$. The matrix elements and masses in the OS scheme
are therefore not plotted any more for $\lambda$ values below this value. 
\s

\begin{figure}[t]
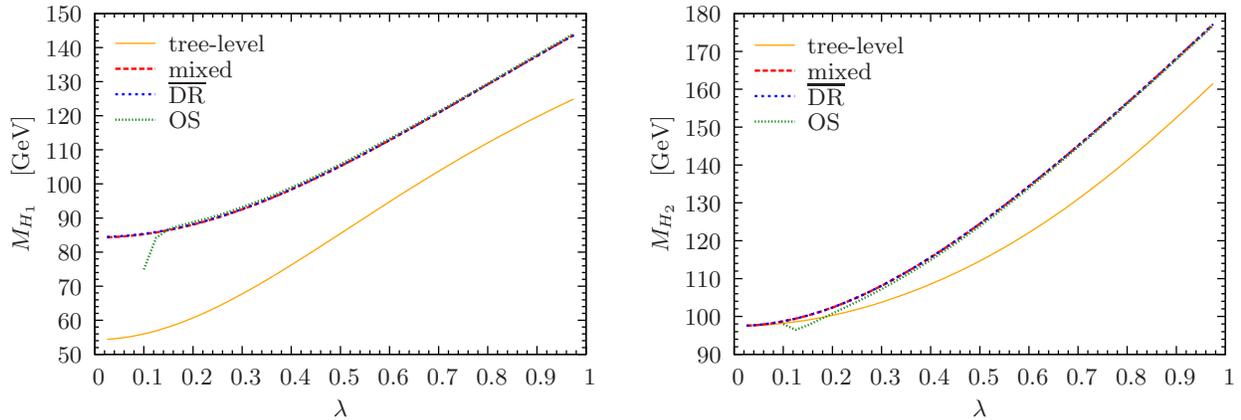

\begin{center}
\parbox{8.5cm}{\include{Fig2a}}\parbox{8.5cm}{\include{Fig2b}}
\caption{\label{fig:comp} The mass $M_{H_1}$ of the lightest (left) and
  $M_{H_2}$  of the next-to-lightest (right) CP-even Higgs boson as
  function of $\lambda$ at tree-level (yellow/full) and at 
  one-loop level adopting a $\overline{\mbox{DR}}$ (blue/dotted), a mixed
  (red/dashed) and an on-shell (green/small dotted) renormalisation scheme.}
\vspace*{-0.4cm}
\end{center}
\end{figure}
Figure~\ref{fig:comp} shows the one-loop corrected mass $M_{H_1}$ of
the lightest (left) and $M_{H_2}$ of the next-to-lightest (right) CP-even Higgs
boson as a function of $\lambda$ in the three different renormalisation
schemes compared to the tree-level result. The tree-level masses
increase with rising $\lambda$ due to the NMSSM contribution
$\sim \lambda^2 \sin^2 2 \beta$ from the Higgs quartic coupling. As 
$H_1$ is dominantly MSSM-like its one-loop corrections are much more
important than for the singlet dominated $H_2$. For the latter they are
negligible at small $\lambda$ and more important for large $\lambda$
where $H_2$ is more MSSM-like. As can be inferred from the figures
the deviations between the different schemes is negligible. The curves
for all three schemes lie on top of each other. Apart from
small values of $\lambda$ where the one-loop corrected $H_1$ and $H_2$
mass in the on-shell scheme start to differ from the
$\overline{\mbox{DR}}$ and the mixed scheme.
\s

\begin{figure}[t]
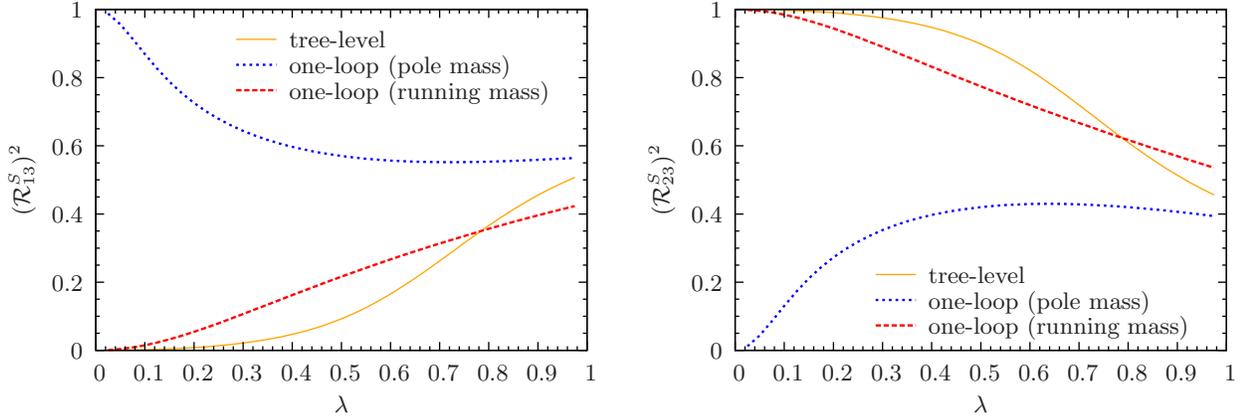

\begin{center}
\parbox{8.5cm}{\include{Fig3a}}\parbox{8.5cm}{\include{Fig3b}}
\caption{\label{fig:mixmtopvar} The matrix element squared $({\cal
    R}^S_{13})^2$ (left), $({\cal R}^S_{23})^2$ (right)
  for the two lightest CP-even Higgs bosons as
  function of $\lambda$ at tree-level (yellow/full), at one-loop level
  with the top quark pole mass (blue/dotted) and with the running
  $\overline{\mbox{DR}}$ top quark mass (red/dashed).}
\vspace*{-0.4cm}
\end{center}
\end{figure}
\begin{figure}[b]
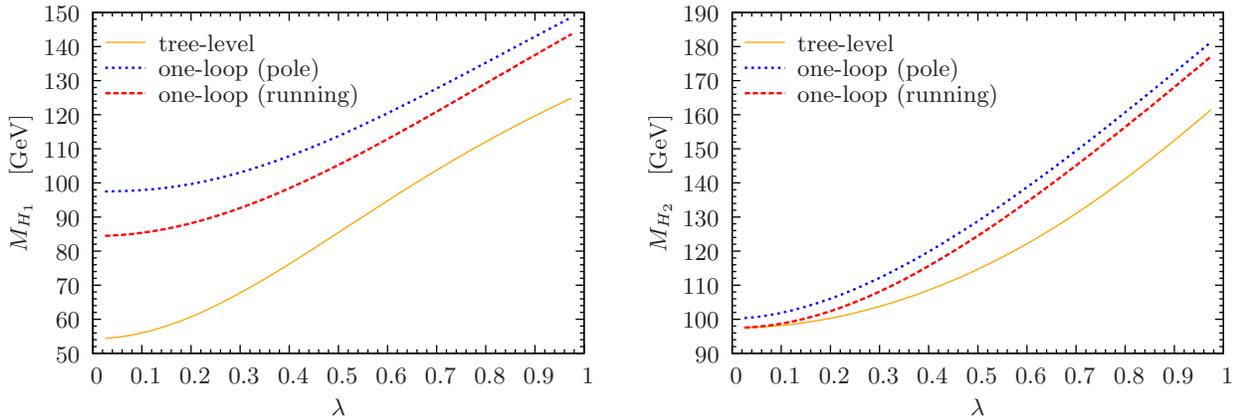

\begin{center}
\parbox{8.5cm}{\include{Fig4a}}\parbox{8.5cm}{\include{Fig4b}}
\caption{\label{fig:massmtopvar} The mass $M_{H_1}$ of the lightest (left) and
  $M_{H_2}$  of the next-to-lightest (right) CP-even Higgs boson as
  function of $\lambda$ at tree-level (yellow/full), at one-loop level
  with the top quark pole mass (blue/dotted) and with the running
  $\overline{\mbox{DR}}$ top quark mass (red/dashed).}
\end{center}
\end{figure}
The mass corrections and hence also the strengths of the singlet
and MSSM component, respectively, strongly
depend on the value of the top quark mass, which reflects the fact
that the main part of the higher-order corrections stems from the top
sector. Figure~\ref{fig:mixmtopvar}
shows the mixing matrix elements squared $({\cal R}^S_{13})^2, ({\cal
  R}^S_{23})^2$ as functions of $\lambda$ at one-loop
level, calculated in the $\overline{\mbox{DR}}$ scheme, with the
top quark mass taken as the running $\overline{\mbox{DR}}$ mass
$m_t=150.6$~GeV in one case and as the pole mass $M_t=173.3$~GeV in
the other case. For comparison the tree-level values are shown as
well. Figure~\ref{fig:massmtopvar} displays the corresponding
Higgs masses. In contrast to the case where the top quark mass has
been set to the $\overline{\mbox{DR}}$ value, using the
top pole mass leads to a one-loop corrected lightest Higgs boson which
is singlet-like and to a next-to-lightest Higgs boson being
MSSM-like. Furthermore, the Higgs mass corrections are more important
for a higher top quark mass value. Defining the relative correction $\Delta
M_{H}/M_{H}$ as
\beq
\frac{\Delta M_H}{M_H} = \frac{|M_H - M_H^{(0)}|}{M_H^{(0)}} \;,
\eeq
where $M_H$ ($M_H^{(0)}$) denotes the 1-loop corrected (tree-level)
Higgs boson mass, the relative correction for $H_1$ amounts to maximally
55\% for the running mass and 79\% for the pole mass. The large
corrections also  
explain the change of the one-loop corrected $H_1$ from a MSSM-like to
singlet-like Higgs boson. In fact, due to the large corrections the
one-loop corrected 
$H_1$ mass gets shifted above the one-loop corrected mass of
$H_2$. Due to our convention to label by ascending indices the Higgs
bosons with increasing mass, the $H_1$ one-loop corrected mass is
assigned to $H_2$ and vice versa, so that $H_1$ and $H_2$ interchange
their roles and $H_1$ becomes singlet-like at 1-loop whereas $H_2$ becomes
MSSM-like. \s  

\begin{figure}[t]
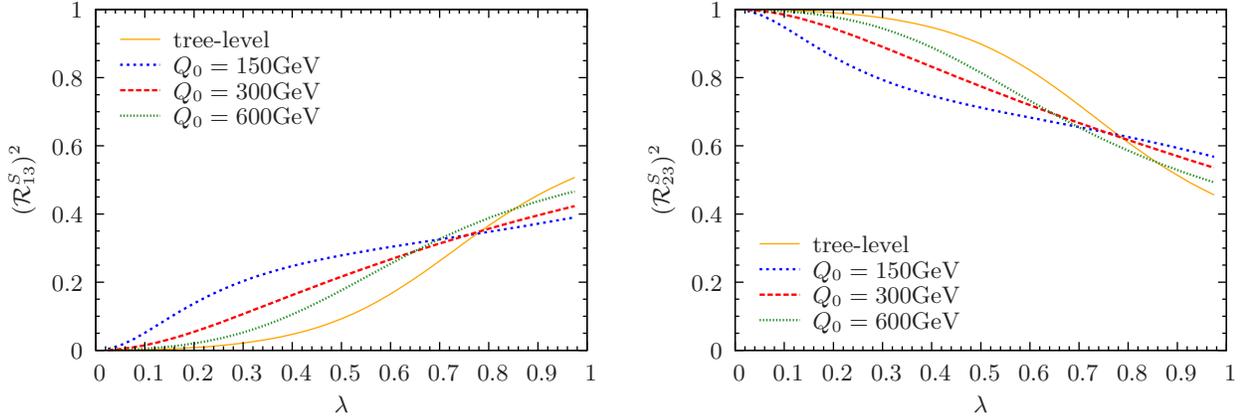

\begin{center}
\parbox{8.5cm}{\include{Fig5a}}\parbox{8.5cm}{\include{Fig5b}}
\caption{\label{fig:mixq0var} The matrix element squared $({\cal R}^S_{13})^2$
  (left), $({\cal R}^S_{23})^2$ (right)
  for the two lightest CP-even Higgs bosons as
  function of $\lambda$ at tree-level (yellow/full), at one-loop level
  at the renormalisation scale $Q_0=150$~GeV (blue/dotted), 
  300~GeV (red/dashed) and 600~GeV (green/small dotted).}
\vspace*{-0.5cm}
\end{center}
\end{figure}
\begin{figure}[t]
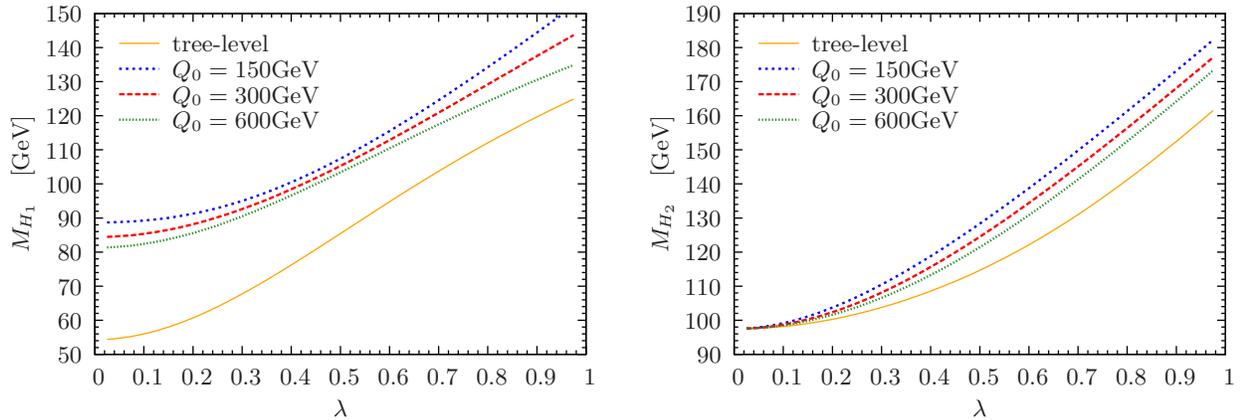

\begin{center}
\parbox{8.5cm}{\include{Fig6a}}\parbox{8.5cm}{\include{Fig6b}}
\caption{\label{fig:massq0var} The mass $M_{H_1}$ of the lightest (left) and
  $M_{H_2}$  of the next-to-lightest (right) CP-even Higgs boson as
  function of $\lambda$ at tree-level (yellow/full), at one-loop level
  at the renormalisation scale $Q_0=150$~GeV (blue/dotted), 
  300~GeV (red/dashed) and 600~GeV (green/small dotted).}
\end{center}
\end{figure}
In order to get an estimate of the missing higher-order corrections we
investigate the influence of the renormalisation scale $Q_0$. The
results are shown (in the $\overline{\mbox{DR}}$ scheme) 
in Fig.~\ref{fig:mixq0var} for the mixing matrix
elements squared. They are plotted as a function of $\lambda$ at
tree-level and at one-loop level for three different values $Q_0=150,
300$ and 600~GeV. Note that the scale $Q_0$ also changes the value of the
running $b$ and $t$ quark masses. The corresponding plots for the
masses of the two lightest Higgs bosons are depicted in
Fig.~\ref{fig:massq0var}. Whereas the change of the renormalisation
scale alters the mixing matrix elements considerably,
the effect on the Higgs boson masses is less pronounced. With rising $Q_0$
the running top quark mass decreases, leading to smaller one-loop
Higgs boson mass corrections. The relative correction for the lightest
Higgs boson mass is 63\% for $Q_0=150$~GeV and 50\% for
$Q_0=600$~GeV. The relative correction of the next-to-lightest Higgs
boson mass is changed less, with $\Delta M_{H_2}/M_{H_2}= 15\%$ and 8\%
for $Q_0=150$ and 600~GeV. This is due to the higher tree-level mass
value. The residual theoretical uncertainties due to missing
higher-order corrections can thus be estimated to ${\cal O}(10\%)$.  \s

\begin{figure}[b]
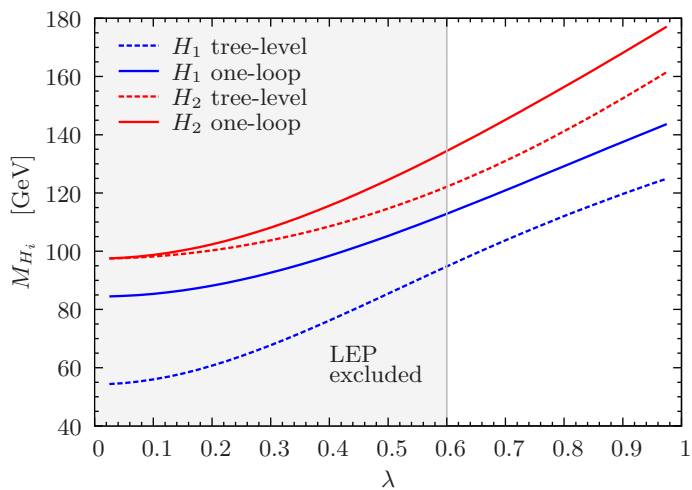

\begin{center}
\parbox{8.5cm}{\include{Fig7}}
\caption{\label{fig:lambdaexcl} The Higgs boson masses at tree-level
  (dashed) and 
  one-loop (full) for $H_1$ (blue/dark grey) and $H_2$ (red/light grey) in the
  $\overline{\mbox{DR}}$ scheme with the exclusion limits set by LEP.}
\vspace*{-0.5cm}
\end{center}
\end{figure}
The amount of singlet component of the Higgs bosons strongly affects
their couplings to fermions and gauge bosons and hence their
phenomenology. In particular, with the small $H_1$ and
$H_2$ mass values between $\sim 90$ and $\sim 180$~GeV, the question
arises if the scenario has been excluded by LEP, Tevatron or LHC. 
We have explicitly verified that there are still
regions in $\lambda$ which have not been excluded. The allowed and
excluded regions are shown in Fig.~\ref{fig:lambdaexcl}. The exclusion
limit for $\lambda \le 0.6$ where $M_{H_1} \lsim 113$~GeV is due to
the LEP exclusion\footnote{The exclusion regions always
  apply to the one-loop corrected Higgs boson masses.} of $M_{H_1}$ in
the channel $e^+e^- \to ZH \to Z b\bar{b}$  
\cite{LEPHbb}. The Tevatron \cite{TEVtautau} and the
present LHC results \cite{lhcexcl,lhcnewexcl} do not constrain the
scenario. Note, that our limits 
represent a rough estimate and cannot replace a sophisticated
study of exclusion limits set by a combination of the experimental
results. We have cross-checked though our exclusion limits against those
obtained with {\tt HiggsBounds} \cite{higgsbounds} and have found
agreement. \s

\begin{figure}[t]
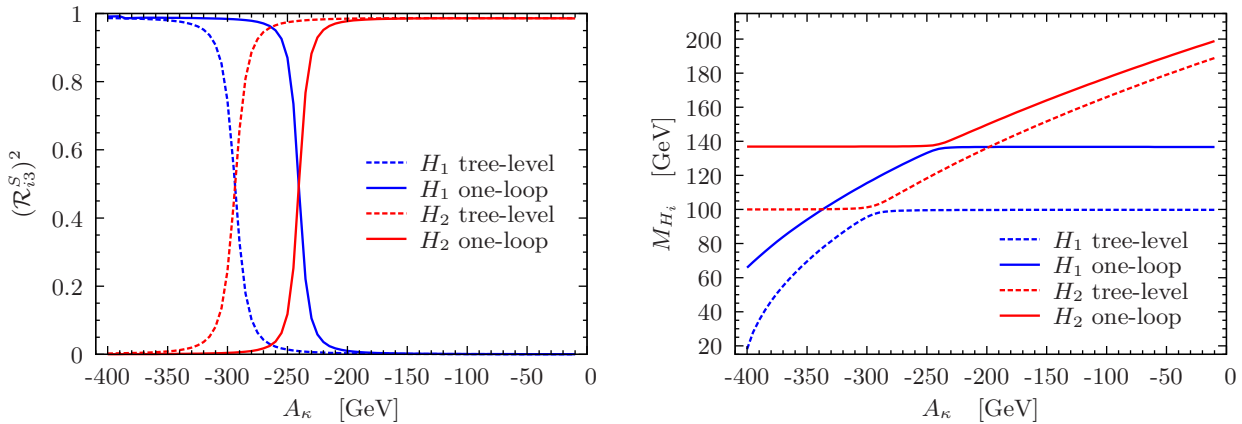

\begin{center}
\parbox{8.5cm}{\include{Fig8a}}\parbox{8.5cm}{\include{Fig8b}}
\caption{\label{fig:Akappavar} Left: The matrix element squared
  $({\cal R}^S_{i3})^2$ ($i=1,2$)
 as function of $A_\kappa$ at tree-level (dashed) and one-loop (full)
 for $H_1$ (blue/dark grey) and $H_2$ (red/light grey). Right: The
 corresponding tree-level and one-loop corrected masses.}
\vspace*{-0.5cm}
\end{center}
\end{figure}

Concerning the pseudoscalar Higgs bosons, the lighter state $A_1$ is
singlet-like, whereas the heavier one is MSSM-like, and the
corrections are small. They are negative and in this 
scenario the relative corrections are below 2\%. The masses increase
with rising $\lambda$. 
The one-loop corrected mass $M_{A_1}$ of the lighter pseudoscalar
Higgs boson ranges between 40 and 140~GeV, and $M_{A_2}$ takes values
of 586 to 598~GeV. \s

\noindent \underline{\bf Variation of \boldmath{$A_\kappa$}} In the
following the variation of the soft SUSY breaking coupling parameter
$A_\kappa$ is investigated. We have chosen a common soft SUSY breaking
squark mass $m_0=1.1$ TeV and $M_S=600$~GeV, large enough to fulfill
the present exclusion limits on the squark masses of the first two
generations and the gluino mass set by the LHC experiments
\cite{exclsquark}. The full parameter set is given by
\beq
\lambda &=& \hspace*{-0.1cm} 0.6 \;, \qquad \kappa = \lambda/3 \; ,
\quad \tan\beta= 2 \; , 
\quad A_\lambda = 500\;\mbox{GeV} \; , \quad \mu= 275 \; \mbox{GeV }
\nonumber \\ 
A_t &=& \hspace*{-0.1cm} A_b = A_\tau = -1.5 M_S \; , \qquad M_1 =
M_S/3 \; , \qquad M_2 = 2/3 M_S \; , \qquad M_3 = 2 M_S \;
. \label{eq:akappvar} 
\eeq
The top and bottom quark masses have been chosen to be the pole
masses, with $m_b^{\scriptsize \mbox{pole}}=4.88$~GeV. The
renormalisation scheme is 
the mixed scheme as defined in section \ref{sec:counter}. The mixing
matrix elements squared quantifying the amount of the singlet component of
$H_1$ and $H_2$ are presented in Fig.~\ref{fig:Akappavar} (left). 
They have been obtained by the
procedure described in section \ref{sec:oneloopmass}. We have explicitly
verified that the difference to the values obtained by setting $k^2=0$
is negligible. The tree-level and one-loop corrected $H_1$
and $H_2$ masses are shown in Fig.~\ref{fig:Akappavar} (right). As can
be inferred from the figure, for large negative values of $A_\kappa$
the lightest Higgs boson is mostly singlet-like and the heavier one
MSSM-like. With increasing $A_\kappa$ the mixing increases developing a
rapid cross-over at $A_\kappa\approx -294$~GeV. The one-loop
corrections shift 
the cross-over to a larger value $A_\kappa \approx -240$~GeV. This behaviour
is also reflected in the Higgs boson masses. Below (above) the cross-over the 
lightest (next-to-lightest) Higgs boson is mostly singlet-like and
exhibits a strong dependence on $A_\kappa$, whereas in the MSSM-like
case the Higgs bosons hardly depend on $A_\kappa$. This behaviour
results from the fact that the soft SUSY breaking term which contains
$A_\kappa$ is a cubic coupling in the singlet field $S$. A variation
of $A_\kappa$ can hence only be communicated through a singlet
contribution of the Higgs boson fields. Since the mixing is small, only
fields which are mainly singlet-like are affected. Despite the singlet
character of $H_1$ for small $A_\kappa$ the one-loop corrections
can be important and more than triple the mass. This is mainly due to
the small tree-level value. In case of the next-to-lightest Higgs
boson $H_2$ the
one-loop corrections in the singlet-like case are less pronounced and
of ${\cal O}(10\;\mbox{GeV})$. The corrections to the MSSM-like Higgs
bosons increase the masses by about 37~GeV. No Higgs boson mass values
$M_{H_{1,2}}$ have been excluded so far by the experiments. With a
slightly higher value of $\lambda$, however, the MSSM-like Higgs boson
mass gets shifted above 141 GeV, so that the scenario would be
excluded by the new LHC exclusion limits \cite{lhcnewexcl}. \s 

For the pseudoscalar masses the relative corrections are very small,
at the 6\% level for the singlet-like $A_1$ and below 0.2\% for the
heavier $A_2$. The one-loop masses decrease with decreasing absolute
value of $A_\kappa$ and are 
$110 \mbox{~GeV} \lsim M_{A_1} \lsim 345$~GeV and $639 \mbox{~GeV}
\lsim M_{A_2} \lsim 640$~GeV for $0 \mbox{~GeV} \ge A_\kappa \ge
-400$~GeV. \s

\begin{figure}[t]
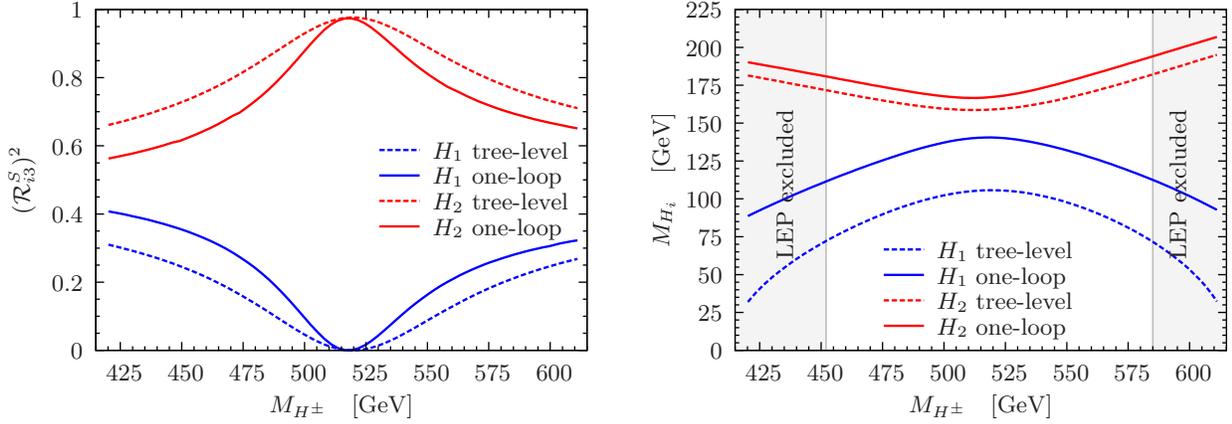

\begin{center}
\parbox{8.5cm}{\include{Fig9a}}\parbox{8.5cm}{\include{Fig9b}}
\caption{\label{fig:Alambdavar} Left: The matrix element squared
  $({\cal R}^S_{i3})^2$ ($i=1,2$) as function of $M_{H^\pm}$ at tree-level (dashed)
  and one-loop (full) for $H_1$ (blue/dark grey) and $H_2$ (red/light
  grey). Right: The 
  corresponding tree-level and one-loop corrected masses with the
  exclusion limits.}
\end{center}
\end{figure}

\noindent \underline{\bf Variation of \boldmath{$M_{H^\pm}$}} The
variation of $A_\lambda$ corresponds to a variation of the charged
Higgs boson mass, {\it cf.} Eq.~(\ref{eq:mhchar}). The other parameters
are kept fixed and chosen as
\beq
\lambda &=& \hspace*{-0.1cm} 0.65 \;, \qquad \kappa = \lambda/3 \; ,
\quad \tan\beta= 2 \; , 
\quad A_\kappa = -10\;\mbox{GeV} \; , \quad \mu= 225 \; \mbox{GeV }
\nonumber \\ 
A_t &=& \hspace*{-0.1cm} A_b = A_\tau = -1.5 M_S \; , \qquad M_1 =
M_S/3 \; , \qquad M_2 = 2/3 M_S \; , \qquad M_3 = 2 M_S \;
. \label{eq:Mhpmvar} 
\eeq
The common soft SUSY breaking squark mass is once again taken to be 
$m_0=1.1$~TeV and $M_S=600$~GeV. The parameter $A_\lambda$ is varied
between 250 and 600~GeV, which corresponds to $M_{H^\pm}\approx
420...610$~GeV. Outside this parameter range the $H_{1}$ mass squared
becomes 
negative\footnote{See also Ref.~\cite{zerwaseal} for a discussion of
  the Higgs boson masses and their dependence on the NMSSM
  parameters.}. 
The results for the singlet components and Higgs boson masses at
tree-level and one-loop are shown in Fig.~\ref{fig:Alambdavar}. The
lightest Higgs boson $H_1$ is dominantly MSSM-like with a 100\% MSSM
component at $M_{H^\pm}\approx 520$~GeV. Here $H_2$ is maximally
singlet-like, though not completely. The heaviest scalar Higgs boson $H_3$
takes over the remaining singlet component (not shown here), so that
$\sum_{i=1..3} ({\cal R}^S_{i3})^2=1$ as demanded by unitarity of the mixing
matrix\footnote{It should be noted that this is only approximately true for the matrix obtained via the 
procedure described in Sect.~\ref{sec:oneloopmass}. In general this matrix, in contrast to the one obtained 
using $k^2 =0$,  is not unitary. As already mentioned the difference between the two approaches is small 
for our scenarios.} . The one-loop corrections increase the mixing between $H_1$ and
$H_2$ away from the maximum and minimum singlet values. They lead to a
more pronounced maximum and minimum in the singlet component and 
slightly shift their positions to smaller $H^\pm$ masses, $M_{H^\pm}
\approx 518$~GeV. The $H_1$ and $H_2$ masses are
maximal and minimal, respectively, at the position of minimal
mixing. Due to its MSSM nature the lightest Higgs boson receives large
one-loop corrections. The correction is $\sim 35$~GeV for $M_{H_1}$ at its
maximum value, which is taken at $M_{H^\pm} \approx 515$~GeV,
and can even triple the tree-level mass at the borders of the
$M_{H^\pm}$ range. The latter is mostly the effect of an already very small
tree-level mass of $\sim 30$~GeV. The singlet-like $H_2$ on the other
hand receives smaller corrections which can nevertheless reach 10
GeV. The investigated parameter range has been partially excluded as
indicated in Fig.~\ref{fig:Alambdavar} (right). For $M_{H^\pm} \le
452$~GeV light Higgs boson masses $M_{H_1} \lsim 112$~GeV have been
excluded by the LEP searches in $ZH \to Zb\bar{b}$ \cite{LEPHbb}. The right
exclusion limit $M_{H^\pm} \ge 585$~GeV is also due to the
LEP exclusion of $M_{H_1} \lsim 113.5$~GeV in the $ZH \to Zb\bar{b}$
channel. The Tevatron results do not exclude Higgs mass values. The
exclusion limits obtained with {\tt HiggsBounds} agree 
with ours. The maximum mass value taken by the MSSM-like lightest
scalar Higgs boson is 140.5 GeV. It is just below the value excluded by the
newest LHC limits \cite{lhcnewexcl}. In view of the uncertainties
associated with the 1-loop Higgs mass corrections and also taking into
account the fact that our exclusion limits are only a rough estimate,
the scenario might be excluded for charged Higgs mass values
around 515~GeV. \s

\begin{figure}[b]
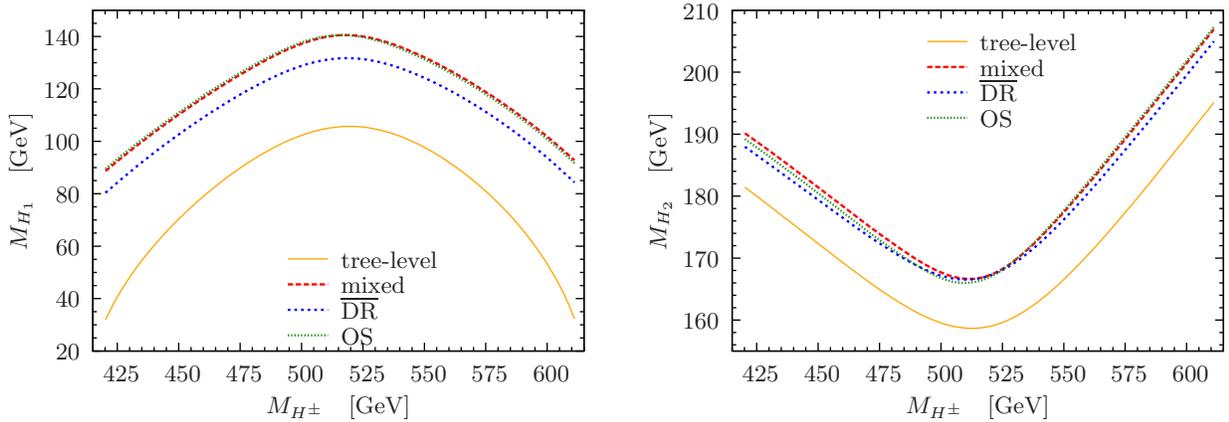

\begin{center}
\parbox{8.5cm}{\include{Fig10a}}\parbox{8.5cm}{\include{Fig10b}}
\caption{\label{fig:Alambdaschemes} The mass $M_{H_1}$ of the lightest
  (left) and $M_{H_2}$  of the next-to-lightest (right) CP-even Higgs boson as
  function of $M_{H^\pm}$ at tree-level (yellow/full) and at 
  one-loop level adopting a $\overline{\mbox{DR}}$ (blue/dotted), a mixed
  (red/dashed) and an on-shell (green/small dotted) renormalisation scheme.}
\end{center}
\end{figure}
The influence of the renormalisation scheme on the one-loop corrected
$H_{1,2}$ masses is shown in Fig.~\ref{fig:Alambdaschemes}. They are
shown as functions of $M_{H^\pm}$ in the OS, mixed and pure
$\overline{\mbox{DR}}$ scheme and compared to the tree-level
result. For the chosen parameter set the differences in the mixed and OS
scheme are negligible in contrast to the $\overline{\mbox{DR}}$
scheme. 
This is to be expected as both in the 
mixed and the OS scheme $A_\lambda$, respectively $M_{H^\pm}$, are
renormalised on-shell, whereas in the $\overline{\mbox{DR}}$ scheme
only the divergent part is included in the counterterm. The
differences in the renormalisation schemes are, however, of maximally ${\cal
  O}(10\%)$ for the masses of the light Higgs boson. For the heavier scalar particles they
are only $\sim 1$\% of the one-loop corrected mass. \s

\begin{figure}[h]
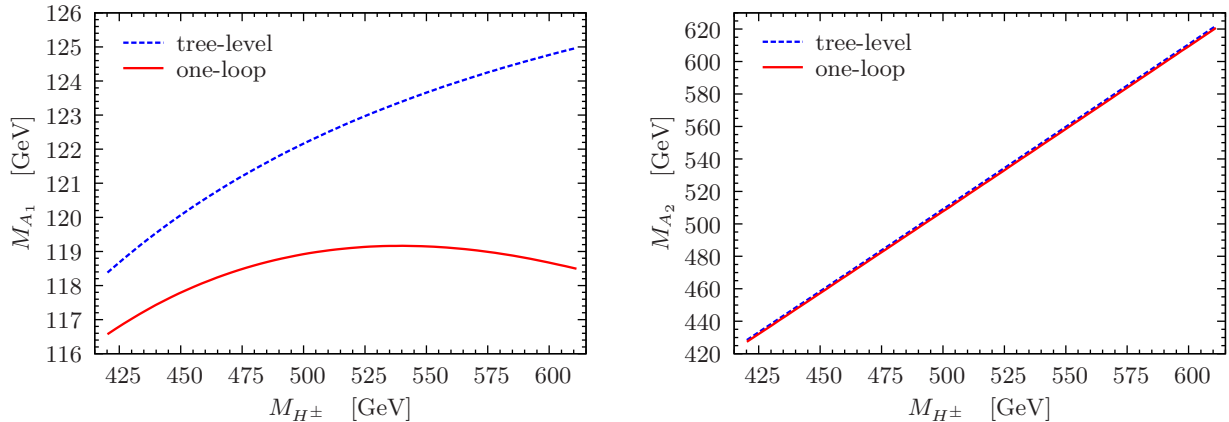

\begin{center}
\parbox{8.5cm}{\include{Fig16a}}\parbox{8.5cm}{\include{Fig16b}}
\caption{\label{fig:Mpmpseudo} The mass $M_{A_1}$  of the lighter
  (left) and $M_{A_2}$ of the heavier (right) pseudoscalar Higgs boson
 as a function of $M_{H^\pm}$ at tree-level (dashed/blue) and at one-loop
 (full/red).}
\end{center}
\end{figure}

In Figure \ref{fig:Mpmpseudo} we show the tree-level and one-loop
corrected masses of the pseudoscalar Higgs bosons as functions of
$M_{H^\pm}$. Whereas for the heavier CP-odd boson the corrections are
negligible they can be of ${\cal O} (5\%)$ for the lighter one at
large values of $M_{H^\pm}$. Nevertheless they are much smaller than the
corrections for the lightest scalar Higgs boson. \s


\section{\label{sec:summary} Summary and Conclusions} 
In summary, we have calculated the one-loop corrections to the NMSSM
Higgs masses in a renormalisation scheme which mixes on-shell and
$\overline{\mbox{DR}}$ conditions, by applying the latter solely to
the parameters $\tan\beta, v_s, \lambda, \kappa, A_\kappa$. We have
compared our result to a pure $\overline{\mbox{DR}}$ and a pure
OS-scheme. Apart from special parameter regions which unphysically
blow up the counterterms, the results differ by at most 10\% for the
masses of the lightest scalar Higgs boson. Another estimate of the
effect of the missing 
higher-order corrections is given by the variation of the
renormalisation scale. We found the effect on the Higgs mass
corrections to be ${\cal O}(10\%)$ or less. Altogether, the residual
theoretical uncertainty at one-loop level can be estimated
to be of the order of 10\%. \s

The bulk of the one-loop corrections stems from the top quark
sector as it is known from the MSSM. This is reflected in the
difference of the relative corrections for the lightest Higgs boson mass.
The relative correction in the scenario, which we investigated, is
equal to 55\% when adopting the running $\overline{\mbox{DR}}$ mass
and 79\% in case of the top quark pole mass. Furthermore, the higher-order
corrections can shift the point of 
cross-over between singlet-like and MSSM-like behaviour of the Higgs
bosons. This dictates also the amount of coupling to the gauge bosons
and fermions and hence influences the Higgs boson phenomenology. The
precise knowledge of the higher-order corrections is therefore
indispensable for a proper investigation of specific production and decay
scenarios and a proper interpretation of the experimental results. 

\section*{Appendix}
\section*{A The CP-even and CP-odd Mass Matrices}
In the following we display the mass matrices in terms of the
parameters for which we apply our renormalisation conditions. We repeat
them here for completeness, {\it cf.}~also Eq.~\eqref{eq:parset}, 
\beq
t_{h_u}, \, t_{h_d}, \, t_{h_s}, \, e, \, \, M_W^2, \, M_Z^2, \,\tan\beta,
\, M_{H^\pm}^2, \, \lambda, \, \kappa, \, v_s, \, A_\kappa
\:. \label{eq:parset1} 
\eeq
Note, that in the mass matrices we paid attention to keep the distinction
in the angle $\beta$ and the angle $\beta_B$. The difference of these two 
angles is denoted as 
\beq
\Delta \beta = \beta - \beta_B \;. 
\eeq
Here, $\beta$ is defined by the ratio of the vacuum expectation values
of the neutral components of the two Higgs doublets $H_u$ and $H_d$,
$\tan\beta = v_u/v_d$. The angle $\beta_B$ on the other hand performs
the rotation from the basis $(a_d,a_u,a_s)$ to the basis $(a,a_s,G)$
to separate a massless Goldstone boson. It is the angle $\beta$ which
receives a counterterm. The angle $\beta_B$ also enters the scalar mass
matrix. This is due to the replacement Eq.~\eqref{eq:alamrepl} of
$A_\lambda$ in terms of the parameter set Eq.~\eqref{eq:parset1}. \s

The scalar $3\times 3$ mass matrix $M_S^2$ in the basis
$h^S=(h_d,h_u,h_s)^T$ is given by the entries $M^2_{S_{ij}}=
M^2_{S_{ji}}$ ($i,j=1,2,3$), with 
\begin{align} \nonumber
M^2_{S_{11}} &= \frac{e c_\beta c^2_{\beta_B} }{2 M_W s_W
  c^2_{\Delta\beta}} [t_{h_d} (2 t_\beta t_{\beta_B} +1) -t_{h_u} t_\beta]
+ \frac{s^2_{\beta} }{c^2_{\Delta\beta}} [M_{H^\pm}^2+(M_Z^2 t^2_\beta-M_W^2)
  c^2_{\Delta\beta}]
\\& \quad\
+ \frac{2 \lambda^2 M_W^2 s_W^2s^2_\beta}{e^2} 
\displaybreak[3]\\
M^2_{S_{12}} &= 
 \frac{e c_\beta c^2_{\beta_B} }{2 M_W  s_W c^2_{\Delta\beta}}
 [t_{h_d} t_{\beta} t^2_{\beta_B} + t_{h_u}] 
- \frac{s_{\beta} c_{\beta} }{c^2_{\Delta\beta}} [ M_{H^\pm}^2 +(M_Z^2-M_W^2)
  c^2_{\Delta\beta} ]
+ \frac{\lambda^2 M_W^2 s_W^2 s_{2 \beta}}{e^2}
\displaybreak[3]\\\nonumber
M^2_{S_{13}} &= \frac{s_\beta c_\beta c^2_{\beta_B} }{v_s
  c^2_{\Delta\beta}} [t_{h_d} t_{\beta} t^2_{\beta_B} + t_{h_u}]
+ \frac{2 M_W s_W s_\beta^2 c_\beta }{e v_s c^2_{\Delta\beta}} [M_W^2
c^2_{\Delta\beta} -M_{H^\pm}^2] 
+ \frac{\lambda M_W s_W c_\beta v_s }{e} [2 \lambda - \kappa t_\beta]
\\& \quad\
+\frac{-4 \lambda^2 M_W^3 s_W^3 s^2_{\beta} c_{\beta}}{e^3 v_s}  
\displaybreak[3] \\ \nonumber
M^2_{S_{22}} &=\frac{e c_{\beta} c_{\beta_B}}{2 M_W s_W
  c^2_{\Delta\beta}}  [-t_{h_d} s_{\beta_B} t_{\beta_B} + t_{h_u}
s_{\beta_B} (t_{\beta} t_{\beta_B} + 2)] 
+ \frac{c_\beta^2 }{c^2_{\Delta\beta}} [M_{H^\pm}^2+(M_Z^2 t_\beta^2 -M_W^2)
  c^2_{\Delta\beta}]\\& \quad\ 
+ \frac{2 \lambda^2 M_W^2 s_W^2 c^2_{\beta}}{e^2} 
\displaybreak[3] \\
M^2_{S_{23}} &=  \nonumber
\frac{c^2_{\beta} c^2_{\beta_B} }{v_s c^2_{\Delta\beta}} [ t_{h_d} t_{\beta} t^2_{\beta_B} +
  t_{h_u} ]
+ \frac{2 M_W s_W s_\beta c^2_\beta }{e v_s c^2_{\Delta\beta}} [M_W^2
c^2_{\Delta\beta} -M_{H^\pm}^2]
+\frac{\lambda M_W s_W c_\beta v_s }{e} [2 \lambda t_\beta - \kappa ]
\\& \quad\
+ \frac{-4 \lambda^2 M_W^3 s_W^3 s_\beta c^2_{\beta}}{e^3 v_s}
\displaybreak[3]\\
M^2_{S_{33}} &=  \nonumber
\kappa A_\kappa \frac{v_s}{\sqrt{2}} +2 \kappa^2 v_s^2+\frac{t_{h_s}}{v_s}
+ \frac{2M_W s_W s_\beta c_\beta^2 }{e^2 v_s^2 c^2_{\Delta\beta}}
[2 M_{H^\pm}^2 M_W s_W s_\beta - e (t_{h_d}  t_\beta s^2_{\beta_B} +
t_{h_u}  c^2_{\beta_B})] 
\\& \quad\
+ \frac{M_W^2 s_W^2 s_{2\beta}}{e^4 v_s^2} [2\lambda^2 M_W^2 s_W^2
s_{2\beta} - \kappa \lambda e^2 v_s^2 - M_W^2 e^2 s_{2\beta}] \quad .
\end{align}

The entries $M^2_{P_{ij}}= M^2_{P_{ji}}$ ($i,j=1,2,3$) of the
pseudoscalar $3\times 3$ mass matrix $M_P^2$  in the basis
$h^P=(a,a_s,G)^T$ read
\begin{align}
M^2_{P_{11}} &=\frac{2 \lambda^2 M_W^2 s_W^2 c^2_{\Delta \beta}}{e^2}
+M_{H^\pm}^2 -M_W^2 c^2_{\Delta \beta}
\displaybreak[3]\\\nonumber
M^2_{P_{12}} &= 
\frac{M_W  s_W s_{2 \beta}}{e v_s c_{\Delta \beta}} [M_{H^\pm}^2 -
M_W^2 c_{\Delta\beta}^2] 
-\frac{c_{\beta} c^2_{\beta_B}}{v_s c_{\Delta \beta}} [t_{h_u} +
t_{h_d} t_{\beta} t^2_{\beta_B}] 
\\& \quad\
+ \frac{\lambda M_W s_W c_{\Delta \beta}}{e^3 v_s} [2\lambda M_W^2
s_W^2 s_{2\beta} - 3\kappa e^2 v_s^2]
\\
M^2_{P_{13}} &= M_{H^\pm}^2 t_{\Delta \beta}
+\frac{M_W^2 s_{2 \Delta \beta}}{2e^2} [2 \lambda^2 s_W^2 - e^2] 
+\frac{e c_{\beta_B}}{2 M_W s_W c_{\Delta \beta}} [t_{h_d} t_{\beta_B}
-t_{h_u}]\displaybreak[3]\\
\nonumber
M^2_{P_{22}} &=-3 A_\kappa \kappa \frac{v_s}{\sqrt{2}}
+\frac{t_{h_s}}{v_s}
- \frac{2M_W s_W s_\beta c_\beta^2 c^2_{\beta_B}}{e v_s^2 c_{\Delta \beta}^2} [t_{h_u}+ t_{h_d} t_\beta t^2_{\beta_B}]
+ \frac{M_W^2 s_W^2 s_{2\beta}^2}{e v_s^2 c_{\Delta\beta}^2}
[M_{H^\pm}^2-M_W^2 c_{\Delta \beta}^2]
\\& \quad\
+ \frac{\lambda M^2_W s^2_W s_{2\beta}}{e^4 v_s^2} [2\lambda M_W^2
s_W^2 s_{2\beta} + 3\kappa e^2 v_s^2]
\displaybreak[3]\\ \nonumber
M^2_{P_{23}} &= \frac{M_W s_W s_{2\beta}}{2ev_s c_{\Delta \beta}} [2
M_{H^\pm}^2 t_{\Delta \beta} - M_W^2 s_{2\Delta \beta}] 
-\frac{c_{\beta} c^2_{\beta_B} t_{\Delta \beta} }{v_s c_{\Delta
    \beta}} [t_{h_u} + t_{h_d} t_{\beta} t^2_{\beta_B}]
\\& \quad\
+ \frac{\lambda M_W s_W s_{\Delta \beta}}{e^3 v_s} [2\lambda M_W^2
s_W^2 s_{2\beta} - 3\kappa e^2 v_s^2]
\displaybreak[3]\\ \nonumber
M^2_{P_{33}} &= M_{H^\pm}^2 \tan ^2{\Delta \beta}  + \frac{M_W^2 \sin
  ^2{\Delta \beta}}{e^2} [2 \lambda^2 s_W^2 - e^2]
\\& \quad\
+\frac{e }{2 M_W s_W c^2_{\Delta \beta}} [t_{h_d} c_{\beta - 2
  \beta_B} -t_{h_u} s_{\beta - 2 \beta_B}] \quad .
\end{align}
\s

\section*{B Fermionic Self-Energies}
In this Appendix we give the renormalised fermionic self-energies
which are needed in the determination of the counterterms $\delta v_s$
and $\delta \kappa$ from the chargino and the neutralino sector,
respectively. We start from the general structure of a fermionic
self-energy given by Eq.~\eqref{eq:strucself}, which also applies to
the corresponding renormalised self-energies. The left- and
right-chiral chargino fields $\psi^\pm_{L,R}$ in the interaction basis are given in
terms of the Weyl spinors for the gaugino fields $\tilde{W}_1,
\tilde{W}_2$ and the charged components of the higgsino fields
$\tilde{H}^\pm_d, \tilde{H}^\pm_u$, 
\beq
\psi_R^- = \left( \begin{array}{c} \tilde{W}^- \\
    \tilde{H}^-_{d} \end{array} \right) \; , \quad
\psi_L^+ = \left( \begin{array}{c} \tilde{W}^+ \\
    \tilde{H}^+_{u} \end{array} \right) \;,
\eeq
with $\tilde{W}^\pm = (\tilde{W}_1 \mp i \tilde{W}_2)/\sqrt{2} $. The
chargino fields in the interaction basis are replaced 
by the renormalised fields and the corresponding field renormalisation
constants, 
\beq
\psi^\pm_{L,R} \to \Big(\id + \frac{1}{2} \delta Z_{L,R} \Big)
\psi^\pm_{L,R} \; , \quad 
\mbox{with} \quad
\delta Z_{L,R} = \left( \begin{array}{cc} \delta Z_{L_1,R_1} & 0 \\
0 & \delta Z_{L_2,R_2} \end{array} \right) \;,
\eeq
and the $2\times 2$ chargino mass matrix by
\beq
M_C \to M_C + \delta (M_C) \;.
\eeq
Performing the rotation of $\psi^\pm_{L,R}$ to the fields $\chi^\pm_{L,R}$ in the mass
eigenbasis with the unitary matrices $U$ and $V$,
\beq
\chi^+_L = V \psi^+_L \qquad \; , \qquad \chi^-_R = U \psi^-_R
\eeq
yields for the renormalised self-energies $\hat{\Sigma}$ in the mass
eigenbasis in terms of the unrenormalised self-energies, field
renormalisation constants and mass matrix counterterm
\beq
\hat{\Sigma}^R_{\chi^\pm} (k^2) &=& \Sigma^R _{\chi^\pm} (k^2) +
\frac{1}{2} U^* 
(\delta Z_R + \delta Z_R^*) U^T \label{eq:cren1} \\
 \hat{\Sigma}^L_{\chi^\pm} (k^2) &=& \Sigma^L _{\chi^\pm} (k^2) +
 \frac{1}{2} V 
 (\delta Z_L + \delta Z_L^*) V^\dagger \label{eq:cren2} \\
\hat{\Sigma}^{Ls}_{\chi^\pm} (k^2) &=& \Sigma^{Ls}_{\chi^\pm} (k^2) -
\frac{1}{2} U^* (\delta Z_R 
M_C + M_C \delta Z_L) V^\dagger - U^* \delta (M_C) V^\dagger 
\label{eq:cren3} \\
\hat{\Sigma}^{Rs}_{\chi^\pm} (k^2) &=& \Sigma^{Rs}_{\chi^\pm} (k^2) -
\frac{1}{2} V (\delta Z_L^* M_C^\dagger + M_C^\dagger \delta Z_R^*)
U^T - V \delta (M_C^\dagger) U^T \;.
\label{eq:cren4} 
\eeq
The self-energies are $2\times 2$ matrices. The corresponding
renormalised self-energies in the neutralino sector, which are
$5\times 5$ matrices, are obtained analogously and read
\beq
\hat{\Sigma}^R_{\chi^0} (k^2) &=& \Sigma^R_{\chi^0} (k^2) +
\frac{1}{2} {\cal N}^* (\delta Z_R 
+ \delta Z_R^*) {\cal N}^T \label{eq:nren1} \\
 \hat{\Sigma}^L_{\chi^0} (k^2) &=& \Sigma^L_{\chi^0} (k^2) +
 \frac{1}{2} {\cal N} (\delta Z_L 
+ \delta Z_L^*) {\cal N}^\dagger \label{eq:nren2} \\
\hat{\Sigma}^{Ls}_{\chi^0} (k^2) &=& \Sigma^{Ls}_{\chi^0} (k^2) -
\frac{1}{2} {\cal N}^* (\delta Z_R  M_N + M_N \delta Z_L) {\cal
  N}^\dagger - {\cal N}^* \delta (M_N) 
{\cal N}^\dagger  
\label{eq:nren3} \\
\hat{\Sigma}^{Rs}_{\chi^0} (k^2) &=& \Sigma^{Rs}_{\chi^0} (k^2) -
\frac{1}{2} {\cal N} (\delta Z_L^* M_N^\dagger + M_N^\dagger \delta
Z_R^*) {\cal N}^T - {\cal N} \delta (M_N^\dagger) {\cal N}^T \;,
\label{eq:nren4} 
\eeq
where ${\cal N}$ is the unitary $5\times 5$ matrix which performs the
rotation from the interaction basis to the neutralino mass
eigenbasis. Allowing also for negative
neutralino mass values, as we do, the matrix is real. 
The neutralino mass matrix $M_N$ is given in Eq.~\eqref{eq:neutmat}
and $\delta Z_{L,R}$ denote diagonal $5\times 5$ matrices with the
field renormalisation constants in the interaction
eigenbasis\footnote{For better readability we did not choose a
  different notation for the field renormalisation constants in the
  chargino and the neutralino case. They are understood to be different,
  though.}, $\delta Z_{L_i,R_i}$ ($i=1,...,5$), as entries. Note that $\delta Z_L$ and
$\delta Z_R$ are related due to the Majorana character of the neutralinos. 

\section*{Acknowledgments}

\noindent
This research was supported in part by the Deutsche
Forschungsgemeinschaft via the Sonderforschungsbereich/Transregio
SFB/TR-9 Computational Particle Phy\-sics. Part of this work has been
performed while TG has been at the Institut f\"ur Theoretische Physik,
Karlsruhe Institute of Technology, 
as well as while HR has been at Physikalisches Institut,
Albert-Ludwigs-Universit\"at Freiburg. \s


\end{document}

%% file: Fig1a.tex
\begingroup
  \fontfamily{cmr}%
  \fontsize{9}{15}
\selectfont
  \makeatletter
  \providecommand\color[2][]{%
    \GenericError{(gnuplot) \space\space\space\@spaces}{%
      Package color not loaded in conjunction with
      terminal option `colourtext'%
    }{See the gnuplot documentation for explanation.%
    }{Either use 'blacktext' in gnuplot or load the package
      color.sty in LaTeX.}%
    \renewcommand\color[2][]{}%
  }%
  \providecommand\includegraphics[2][]{%
    \GenericError{(gnuplot) \space\space\space\@spaces}{%
      Package graphicx or graphics not loaded%
    }{See the gnuplot documentation for explanation.%
    }{The gnuplot epslatex terminal needs graphicx.sty or graphics.sty.}%
    \renewcommand\includegraphics[2][]{}%
  }%
  \providecommand\rotatebox[2]{#2}%
  \@ifundefined{ifGPcolor}{%
    \newif\ifGPcolor
    \GPcolortrue
  }{}%
  \@ifundefined{ifGPblacktext}{%
    \newif\ifGPblacktext
    \GPblacktexttrue
  }{}%
  \let\gplgaddtomacro\g@addto@macro
  \gdef\gplbacktext{}%
  \gdef\gplfronttext{}%
  \makeatother
  \ifGPblacktext
    \def\colorrgb#1{}%
    \def\colorgray#1{}%
  \else
    \ifGPcolor
      \def\colorrgb#1{\color[rgb]{#1}}%
      \def\colorgray#1{\color[gray]{#1}}%
      \expandafter\def\csname LTw\endcsname{\color{white}}%
      \expandafter\def\csname LTb\endcsname{\color{black}}%
      \expandafter\def\csname LTa\endcsname{\color{black}}%
      \expandafter\def\csname LT0\endcsname{\color[rgb]{1,0,0}}%
      \expandafter\def\csname LT1\endcsname{\color[rgb]{0,1,0}}%
      \expandafter\def\csname LT2\endcsname{\color[rgb]{0,0,1}}%
      \expandafter\def\csname LT3\endcsname{\color[rgb]{1,0,1}}%
      \expandafter\def\csname LT4\endcsname{\color[rgb]{0,1,1}}%
      \expandafter\def\csname LT5\endcsname{\color[rgb]{1,1,0}}%
      \expandafter\def\csname LT6\endcsname{\color[rgb]{0,0,0}}%
      \expandafter\def\csname LT7\endcsname{\color[rgb]{1,0.3,0}}%
      \expandafter\def\csname LT8\endcsname{\color[rgb]{0.5,0.5,0.5}}%
    \else
      \def\colorrgb#1{\color{black}}%
      \def\colorgray#1{\color[gray]{#1}}%
      \expandafter\def\csname LTw\endcsname{\color{white}}%
      \expandafter\def\csname LTb\endcsname{\color{black}}%
      \expandafter\def\csname LTa\endcsname{\color{black}}%
      \expandafter\def\csname LT0\endcsname{\color{black}}%
      \expandafter\def\csname LT1\endcsname{\color{black}}%
      \expandafter\def\csname LT2\endcsname{\color{black}}%
      \expandafter\def\csname LT3\endcsname{\color{black}}%
      \expandafter\def\csname LT4\endcsname{\color{black}}%
      \expandafter\def\csname LT5\endcsname{\color{black}}%
      \expandafter\def\csname LT6\endcsname{\color{black}}%
      \expandafter\def\csname LT7\endcsname{\color{black}}%
      \expandafter\def\csname LT8\endcsname{\color{black}}%
    \fi
  \fi
  \setlength{\unitlength}{0.0500bp}%
  \begin{picture}(4680.00,3276.00)%
    \gplgaddtomacro\gplbacktext{%
      \csname LTb\endcsname%
      \put(128,1797){\rotatebox{-270}{\makebox(0,0){\strut{}$(\mathcal{R}^S_{13})^2$}}}%
      \put(2539,112){\makebox(0,0){\strut{}$\lambda$}}%
    }%
    \gplgaddtomacro\gplfronttext{%
      \csname LTb\endcsname%
      \put(3464,2858){\makebox(0,0)[l]{\strut{}tree-level}}%
      \csname LTb\endcsname%
      \put(3464,2666){\makebox(0,0)[l]{\strut{}mixed}}%
      \csname LTb\endcsname%
      \put(3464,2474){\makebox(0,0)[l]{\strut{}$\overline{\text{DR}}$}}%
      \csname LTb\endcsname%
      \put(3464,2282){\makebox(0,0)[l]{\strut{}OS}}%
      \csname LTb\endcsname%
      \put(592,512){\makebox(0,0)[r]{\strut{} 0}}%
      \put(592,1026){\makebox(0,0)[r]{\strut{} 0.2}}%
      \put(592,1540){\makebox(0,0)[r]{\strut{} 0.4}}%
      \put(592,2055){\makebox(0,0)[r]{\strut{} 0.6}}%
      \put(592,2569){\makebox(0,0)[r]{\strut{} 0.8}}%
      \put(592,3083){\makebox(0,0)[r]{\strut{} 1}}%
      \put(688,352){\makebox(0,0){\strut{} 0}}%
      \put(1058,352){\makebox(0,0){\strut{} 0.1}}%
      \put(1429,352){\makebox(0,0){\strut{} 0.2}}%
      \put(1799,352){\makebox(0,0){\strut{} 0.3}}%
      \put(2169,352){\makebox(0,0){\strut{} 0.4}}%
      \put(2540,352){\makebox(0,0){\strut{} 0.5}}%
      \put(2910,352){\makebox(0,0){\strut{} 0.6}}%
      \put(3280,352){\makebox(0,0){\strut{} 0.7}}%
      \put(3650,352){\makebox(0,0){\strut{} 0.8}}%
      \put(4021,352){\makebox(0,0){\strut{} 0.9}}%
      \put(4391,352){\makebox(0,0){\strut{} 1}}%
    }%
    \gplbacktext
    \put(0,0){\includegraphics{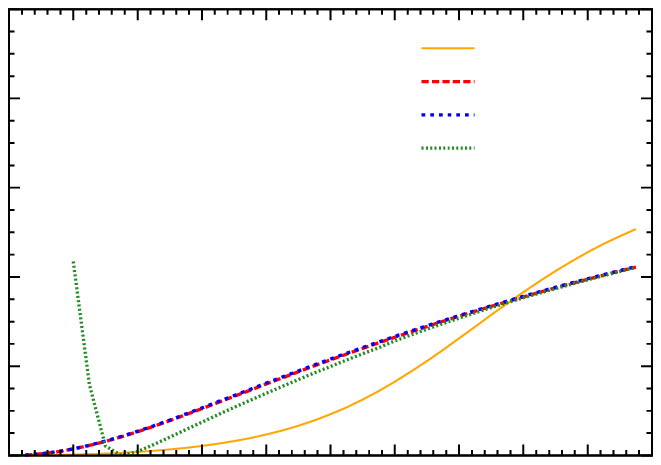}}%
    \gplfronttext
  \end{picture}%
\endgroup

%% file: Fig1b.tex
\begingroup
  \fontfamily{cmr}%
  \fontsize{9}{15}
\selectfont
  \makeatletter
  \providecommand\color[2][]{%
    \GenericError{(gnuplot) \space\space\space\@spaces}{%
      Package color not loaded in conjunction with
      terminal option `colourtext'%
    }{See the gnuplot documentation for explanation.%
    }{Either use 'blacktext' in gnuplot or load the package
      color.sty in LaTeX.}%
    \renewcommand\color[2][]{}%
  }%
  \providecommand\includegraphics[2][]{%
    \GenericError{(gnuplot) \space\space\space\@spaces}{%
      Package graphicx or graphics not loaded%
    }{See the gnuplot documentation for explanation.%
    }{The gnuplot epslatex terminal needs graphicx.sty or graphics.sty.}%
    \renewcommand\includegraphics[2][]{}%
  }%
  \providecommand\rotatebox[2]{#2}%
  \@ifundefined{ifGPcolor}{%
    \newif\ifGPcolor
    \GPcolortrue
  }{}%
  \@ifundefined{ifGPblacktext}{%
    \newif\ifGPblacktext
    \GPblacktexttrue
  }{}%
  \let\gplgaddtomacro\g@addto@macro
  \gdef\gplbacktext{}%
  \gdef\gplfronttext{}%
  \makeatother
  \ifGPblacktext
    \def\colorrgb#1{}%
    \def\colorgray#1{}%
  \else
    \ifGPcolor
      \def\colorrgb#1{\color[rgb]{#1}}%
      \def\colorgray#1{\color[gray]{#1}}%
      \expandafter\def\csname LTw\endcsname{\color{white}}%
      \expandafter\def\csname LTb\endcsname{\color{black}}%
      \expandafter\def\csname LTa\endcsname{\color{black}}%
      \expandafter\def\csname LT0\endcsname{\color[rgb]{1,0,0}}%
      \expandafter\def\csname LT1\endcsname{\color[rgb]{0,1,0}}%
      \expandafter\def\csname LT2\endcsname{\color[rgb]{0,0,1}}%
      \expandafter\def\csname LT3\endcsname{\color[rgb]{1,0,1}}%
      \expandafter\def\csname LT4\endcsname{\color[rgb]{0,1,1}}%
      \expandafter\def\csname LT5\endcsname{\color[rgb]{1,1,0}}%
      \expandafter\def\csname LT6\endcsname{\color[rgb]{0,0,0}}%
      \expandafter\def\csname LT7\endcsname{\color[rgb]{1,0.3,0}}%
      \expandafter\def\csname LT8\endcsname{\color[rgb]{0.5,0.5,0.5}}%
    \else
      \def\colorrgb#1{\color{black}}%
      \def\colorgray#1{\color[gray]{#1}}%
      \expandafter\def\csname LTw\endcsname{\color{white}}%
      \expandafter\def\csname LTb\endcsname{\color{black}}%
      \expandafter\def\csname LTa\endcsname{\color{black}}%
      \expandafter\def\csname LT0\endcsname{\color{black}}%
      \expandafter\def\csname LT1\endcsname{\color{black}}%
      \expandafter\def\csname LT2\endcsname{\color{black}}%
      \expandafter\def\csname LT3\endcsname{\color{black}}%
      \expandafter\def\csname LT4\endcsname{\color{black}}%
      \expandafter\def\csname LT5\endcsname{\color{black}}%
      \expandafter\def\csname LT6\endcsname{\color{black}}%
      \expandafter\def\csname LT7\endcsname{\color{black}}%
      \expandafter\def\csname LT8\endcsname{\color{black}}%
    \fi
  \fi
  \setlength{\unitlength}{0.0500bp}%
  \begin{picture}(4680.00,3276.00)%
    \gplgaddtomacro\gplbacktext{%
      \csname LTb\endcsname%
      \put(128,1797){\rotatebox{-270}{\makebox(0,0){\strut{}$(\mathcal{R}^S_{23})^2$}}}%
      \put(2539,112){\makebox(0,0){\strut{}$\lambda$}}%
    }%
    \gplgaddtomacro\gplfronttext{%
      \csname LTb\endcsname%
      \put(1242,1316){\makebox(0,0)[l]{\strut{}tree-level}}%
      \csname LTb\endcsname%
      \put(1242,1124){\makebox(0,0)[l]{\strut{}mixed}}%
      \csname LTb\endcsname%
      \put(1242,932){\makebox(0,0)[l]{\strut{}$\overline{\text{DR}}$}}%
      \csname LTb\endcsname%
      \put(1242,740){\makebox(0,0)[l]{\strut{}OS}}%
      \csname LTb\endcsname%
      \put(592,512){\makebox(0,0)[r]{\strut{} 0}}%
      \put(592,1026){\makebox(0,0)[r]{\strut{} 0.2}}%
      \put(592,1540){\makebox(0,0)[r]{\strut{} 0.4}}%
      \put(592,2055){\makebox(0,0)[r]{\strut{} 0.6}}%
      \put(592,2569){\makebox(0,0)[r]{\strut{} 0.8}}%
      \put(592,3083){\makebox(0,0)[r]{\strut{} 1}}%
      \put(688,352){\makebox(0,0){\strut{} 0}}%
      \put(1058,352){\makebox(0,0){\strut{} 0.1}}%
      \put(1429,352){\makebox(0,0){\strut{} 0.2}}%
      \put(1799,352){\makebox(0,0){\strut{} 0.3}}%
      \put(2169,352){\makebox(0,0){\strut{} 0.4}}%
      \put(2540,352){\makebox(0,0){\strut{} 0.5}}%
      \put(2910,352){\makebox(0,0){\strut{} 0.6}}%
      \put(3280,352){\makebox(0,0){\strut{} 0.7}}%
      \put(3650,352){\makebox(0,0){\strut{} 0.8}}%
      \put(4021,352){\makebox(0,0){\strut{} 0.9}}%
      \put(4391,352){\makebox(0,0){\strut{} 1}}%
    }%
    \gplbacktext
    \put(0,0){\includegraphics{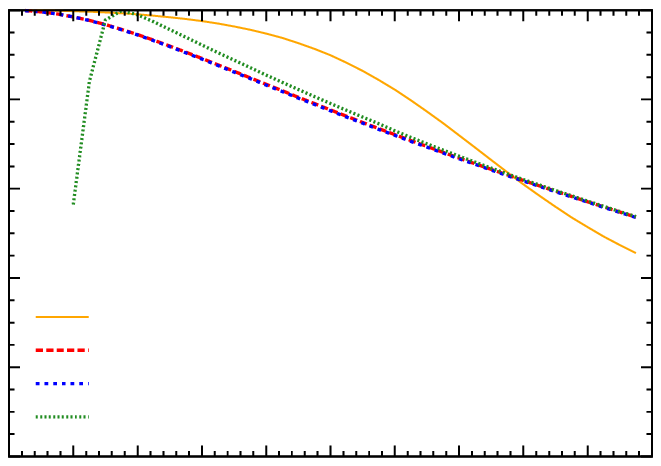}}%
    \gplfronttext
  \end{picture}%
\endgroup

%% file: Fig2a.tex
\begingroup
  \fontfamily{cmr}%
  \fontsize{9}{15}
\selectfont
  \makeatletter
  \providecommand\color[2][]{%
    \GenericError{(gnuplot) \space\space\space\@spaces}{%
      Package color not loaded in conjunction with
      terminal option `colourtext'%
    }{See the gnuplot documentation for explanation.%
    }{Either use 'blacktext' in gnuplot or load the package
      color.sty in LaTeX.}%
    \renewcommand\color[2][]{}%
  }%
  \providecommand\includegraphics[2][]{%
    \GenericError{(gnuplot) \space\space\space\@spaces}{%
      Package graphicx or graphics not loaded%
    }{See the gnuplot documentation for explanation.%
    }{The gnuplot epslatex terminal needs graphicx.sty or graphics.sty.}%
    \renewcommand\includegraphics[2][]{}%
  }%
  \providecommand\rotatebox[2]{#2}%
  \@ifundefined{ifGPcolor}{%
    \newif\ifGPcolor
    \GPcolortrue
  }{}%
  \@ifundefined{ifGPblacktext}{%
    \newif\ifGPblacktext
    \GPblacktexttrue
  }{}%
  \let\gplgaddtomacro\g@addto@macro
  \gdef\gplbacktext{}%
  \gdef\gplfronttext{}%
  \makeatother
  \ifGPblacktext
    \def\colorrgb#1{}%
    \def\colorgray#1{}%
  \else
    \ifGPcolor
      \def\colorrgb#1{\color[rgb]{#1}}%
      \def\colorgray#1{\color[gray]{#1}}%
      \expandafter\def\csname LTw\endcsname{\color{white}}%
      \expandafter\def\csname LTb\endcsname{\color{black}}%
      \expandafter\def\csname LTa\endcsname{\color{black}}%
      \expandafter\def\csname LT0\endcsname{\color[rgb]{1,0,0}}%
      \expandafter\def\csname LT1\endcsname{\color[rgb]{0,1,0}}%
      \expandafter\def\csname LT2\endcsname{\color[rgb]{0,0,1}}%
      \expandafter\def\csname LT3\endcsname{\color[rgb]{1,0,1}}%
      \expandafter\def\csname LT4\endcsname{\color[rgb]{0,1,1}}%
      \expandafter\def\csname LT5\endcsname{\color[rgb]{1,1,0}}%
      \expandafter\def\csname LT6\endcsname{\color[rgb]{0,0,0}}%
      \expandafter\def\csname LT7\endcsname{\color[rgb]{1,0.3,0}}%
      \expandafter\def\csname LT8\endcsname{\color[rgb]{0.5,0.5,0.5}}%
    \else
      \def\colorrgb#1{\color{black}}%
      \def\colorgray#1{\color[gray]{#1}}%
      \expandafter\def\csname LTw\endcsname{\color{white}}%
      \expandafter\def\csname LTb\endcsname{\color{black}}%
      \expandafter\def\csname LTa\endcsname{\color{black}}%
      \expandafter\def\csname LT0\endcsname{\color{black}}%
      \expandafter\def\csname LT1\endcsname{\color{black}}%
      \expandafter\def\csname LT2\endcsname{\color{black}}%
      \expandafter\def\csname LT3\endcsname{\color{black}}%
      \expandafter\def\csname LT4\endcsname{\color{black}}%
      \expandafter\def\csname LT5\endcsname{\color{black}}%
      \expandafter\def\csname LT6\endcsname{\color{black}}%
      \expandafter\def\csname LT7\endcsname{\color{black}}%
      \expandafter\def\csname LT8\endcsname{\color{black}}%
    \fi
  \fi
  \setlength{\unitlength}{0.0500bp}%
  \begin{picture}(4680.00,3276.00)%
    \gplgaddtomacro\gplbacktext{%
      \csname LTb\endcsname%
      \put(128,1797){\rotatebox{-270}{\makebox(0,0){\strut{}$M_{H_1} \quad [\text{GeV}]$}}}%
      \put(2539,112){\makebox(0,0){\strut{}$\lambda$}}%
    }%
    \gplgaddtomacro\gplfronttext{%
      \csname LTb\endcsname%
      \put(1242,2858){\makebox(0,0)[l]{\strut{}tree-level}}%
      \csname LTb\endcsname%
      \put(1242,2666){\makebox(0,0)[l]{\strut{}mixed}}%
      \csname LTb\endcsname%
      \put(1242,2474){\makebox(0,0)[l]{\strut{}$\overline{\text{DR}}$}}%
      \csname LTb\endcsname%
      \put(1242,2282){\makebox(0,0)[l]{\strut{}OS}}%
      \csname LTb\endcsname%
      \put(592,512){\makebox(0,0)[r]{\strut{} 50}}%
      \put(592,769){\makebox(0,0)[r]{\strut{} 60}}%
      \put(592,1026){\makebox(0,0)[r]{\strut{} 70}}%
      \put(592,1283){\makebox(0,0)[r]{\strut{} 80}}%
      \put(592,1540){\makebox(0,0)[r]{\strut{} 90}}%
      \put(592,1798){\makebox(0,0)[r]{\strut{} 100}}%
      \put(592,2055){\makebox(0,0)[r]{\strut{} 110}}%
      \put(592,2312){\makebox(0,0)[r]{\strut{} 120}}%
      \put(592,2569){\makebox(0,0)[r]{\strut{} 130}}%
      \put(592,2826){\makebox(0,0)[r]{\strut{} 140}}%
      \put(592,3083){\makebox(0,0)[r]{\strut{} 150}}%
      \put(688,352){\makebox(0,0){\strut{} 0}}%
      \put(1058,352){\makebox(0,0){\strut{} 0.1}}%
      \put(1429,352){\makebox(0,0){\strut{} 0.2}}%
      \put(1799,352){\makebox(0,0){\strut{} 0.3}}%
      \put(2169,352){\makebox(0,0){\strut{} 0.4}}%
      \put(2540,352){\makebox(0,0){\strut{} 0.5}}%
      \put(2910,352){\makebox(0,0){\strut{} 0.6}}%
      \put(3280,352){\makebox(0,0){\strut{} 0.7}}%
      \put(3650,352){\makebox(0,0){\strut{} 0.8}}%
      \put(4021,352){\makebox(0,0){\strut{} 0.9}}%
      \put(4391,352){\makebox(0,0){\strut{} 1}}%
    }%
    \gplbacktext
    \put(0,0){\includegraphics{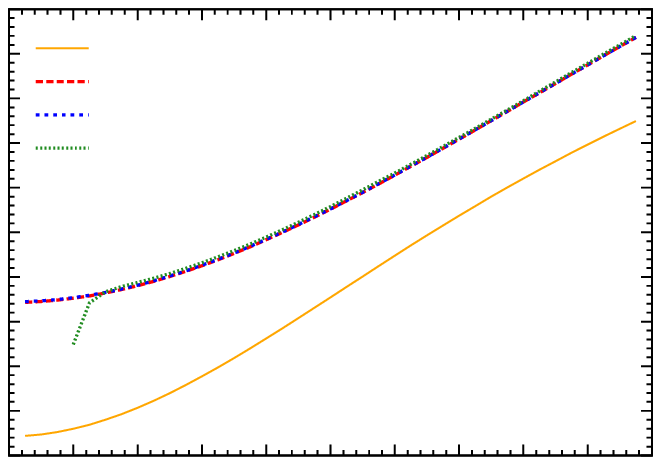}}%
    \gplfronttext
  \end{picture}%
\endgroup

%% file: Fig2b.tex
\begingroup
  \fontfamily{cmr}%
  \fontsize{9}{15}
\selectfont
  \makeatletter
  \providecommand\color[2][]{%
    \GenericError{(gnuplot) \space\space\space\@spaces}{%
      Package color not loaded in conjunction with
      terminal option `colourtext'%
    }{See the gnuplot documentation for explanation.%
    }{Either use 'blacktext' in gnuplot or load the package
      color.sty in LaTeX.}%
    \renewcommand\color[2][]{}%
  }%
  \providecommand\includegraphics[2][]{%
    \GenericError{(gnuplot) \space\space\space\@spaces}{%
      Package graphicx or graphics not loaded%
    }{See the gnuplot documentation for explanation.%
    }{The gnuplot epslatex terminal needs graphicx.sty or graphics.sty.}%
    \renewcommand\includegraphics[2][]{}%
  }%
  \providecommand\rotatebox[2]{#2}%
  \@ifundefined{ifGPcolor}{%
    \newif\ifGPcolor
    \GPcolortrue
  }{}%
  \@ifundefined{ifGPblacktext}{%
    \newif\ifGPblacktext
    \GPblacktexttrue
  }{}%
  \let\gplgaddtomacro\g@addto@macro
  \gdef\gplbacktext{}%
  \gdef\gplfronttext{}%
  \makeatother
  \ifGPblacktext
    \def\colorrgb#1{}%
    \def\colorgray#1{}%
  \else
    \ifGPcolor
      \def\colorrgb#1{\color[rgb]{#1}}%
      \def\colorgray#1{\color[gray]{#1}}%
      \expandafter\def\csname LTw\endcsname{\color{white}}%
      \expandafter\def\csname LTb\endcsname{\color{black}}%
      \expandafter\def\csname LTa\endcsname{\color{black}}%
      \expandafter\def\csname LT0\endcsname{\color[rgb]{1,0,0}}%
      \expandafter\def\csname LT1\endcsname{\color[rgb]{0,1,0}}%
      \expandafter\def\csname LT2\endcsname{\color[rgb]{0,0,1}}%
      \expandafter\def\csname LT3\endcsname{\color[rgb]{1,0,1}}%
      \expandafter\def\csname LT4\endcsname{\color[rgb]{0,1,1}}%
      \expandafter\def\csname LT5\endcsname{\color[rgb]{1,1,0}}%
      \expandafter\def\csname LT6\endcsname{\color[rgb]{0,0,0}}%
      \expandafter\def\csname LT7\endcsname{\color[rgb]{1,0.3,0}}%
      \expandafter\def\csname LT8\endcsname{\color[rgb]{0.5,0.5,0.5}}%
    \else
      \def\colorrgb#1{\color{black}}%
      \def\colorgray#1{\color[gray]{#1}}%
      \expandafter\def\csname LTw\endcsname{\color{white}}%
      \expandafter\def\csname LTb\endcsname{\color{black}}%
      \expandafter\def\csname LTa\endcsname{\color{black}}%
      \expandafter\def\csname LT0\endcsname{\color{black}}%
      \expandafter\def\csname LT1\endcsname{\color{black}}%
      \expandafter\def\csname LT2\endcsname{\color{black}}%
      \expandafter\def\csname LT3\endcsname{\color{black}}%
      \expandafter\def\csname LT4\endcsname{\color{black}}%
      \expandafter\def\csname LT5\endcsname{\color{black}}%
      \expandafter\def\csname LT6\endcsname{\color{black}}%
      \expandafter\def\csname LT7\endcsname{\color{black}}%
      \expandafter\def\csname LT8\endcsname{\color{black}}%
    \fi
  \fi
  \setlength{\unitlength}{0.0500bp}%
  \begin{picture}(4680.00,3276.00)%
    \gplgaddtomacro\gplbacktext{%
      \csname LTb\endcsname%
      \put(128,1797){\rotatebox{-270}{\makebox(0,0){\strut{}$M_{H_2} \quad [\text{GeV}]$}}}%
      \put(2539,112){\makebox(0,0){\strut{}$\lambda$}}%
    }%
    \gplgaddtomacro\gplfronttext{%
      \csname LTb\endcsname%
      \put(1242,2844){\makebox(0,0)[l]{\strut{}tree-level}}%
      \csname LTb\endcsname%
      \put(1242,2652){\makebox(0,0)[l]{\strut{}mixed}}%
      \csname LTb\endcsname%
      \put(1242,2460){\makebox(0,0)[l]{\strut{}$\overline{\text{DR}}$}}%
      \csname LTb\endcsname%
      \put(1242,2268){\makebox(0,0)[l]{\strut{}OS}}%
      \csname LTb\endcsname%
      \put(592,512){\makebox(0,0)[r]{\strut{} 90}}%
      \put(592,798){\makebox(0,0)[r]{\strut{} 100}}%
      \put(592,1083){\makebox(0,0)[r]{\strut{} 110}}%
      \put(592,1369){\makebox(0,0)[r]{\strut{} 120}}%
      \put(592,1655){\makebox(0,0)[r]{\strut{} 130}}%
      \put(592,1940){\makebox(0,0)[r]{\strut{} 140}}%
      \put(592,2226){\makebox(0,0)[r]{\strut{} 150}}%
      \put(592,2512){\makebox(0,0)[r]{\strut{} 160}}%
      \put(592,2797){\makebox(0,0)[r]{\strut{} 170}}%
      \put(592,3083){\makebox(0,0)[r]{\strut{} 180}}%
      \put(688,352){\makebox(0,0){\strut{} 0}}%
      \put(1058,352){\makebox(0,0){\strut{} 0.1}}%
      \put(1429,352){\makebox(0,0){\strut{} 0.2}}%
      \put(1799,352){\makebox(0,0){\strut{} 0.3}}%
      \put(2169,352){\makebox(0,0){\strut{} 0.4}}%
      \put(2540,352){\makebox(0,0){\strut{} 0.5}}%
      \put(2910,352){\makebox(0,0){\strut{} 0.6}}%
      \put(3280,352){\makebox(0,0){\strut{} 0.7}}%
      \put(3650,352){\makebox(0,0){\strut{} 0.8}}%
      \put(4021,352){\makebox(0,0){\strut{} 0.9}}%
      \put(4391,352){\makebox(0,0){\strut{} 1}}%
    }%
    \gplbacktext
    \put(0,0){\includegraphics{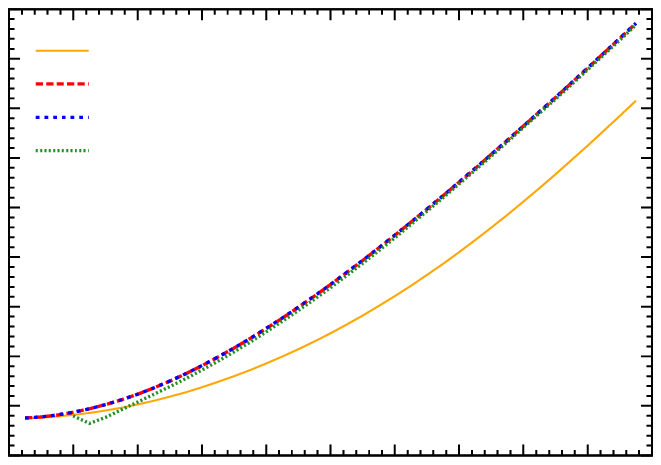}}%
    \gplfronttext
  \end{picture}%
\endgroup

%% file: Fig3a.tex
\begingroup
  \fontfamily{cmr}%
  \fontsize{9}{15}
\selectfont
  \makeatletter
  \providecommand\color[2][]{%
    \GenericError{(gnuplot) \space\space\space\@spaces}{%
      Package color not loaded in conjunction with
      terminal option `colourtext'%
    }{See the gnuplot documentation for explanation.%
    }{Either use 'blacktext' in gnuplot or load the package
      color.sty in LaTeX.}%
    \renewcommand\color[2][]{}%
  }%
  \providecommand\includegraphics[2][]{%
    \GenericError{(gnuplot) \space\space\space\@spaces}{%
      Package graphicx or graphics not loaded%
    }{See the gnuplot documentation for explanation.%
    }{The gnuplot epslatex terminal needs graphicx.sty or graphics.sty.}%
    \renewcommand\includegraphics[2][]{}%
  }%
  \providecommand\rotatebox[2]{#2}%
  \@ifundefined{ifGPcolor}{%
    \newif\ifGPcolor
    \GPcolortrue
  }{}%
  \@ifundefined{ifGPblacktext}{%
    \newif\ifGPblacktext
    \GPblacktexttrue
  }{}%
  \let\gplgaddtomacro\g@addto@macro
  \gdef\gplbacktext{}%
  \gdef\gplfronttext{}%
  \makeatother
  \ifGPblacktext
    \def\colorrgb#1{}%
    \def\colorgray#1{}%
  \else
    \ifGPcolor
      \def\colorrgb#1{\color[rgb]{#1}}%
      \def\colorgray#1{\color[gray]{#1}}%
      \expandafter\def\csname LTw\endcsname{\color{white}}%
      \expandafter\def\csname LTb\endcsname{\color{black}}%
      \expandafter\def\csname LTa\endcsname{\color{black}}%
      \expandafter\def\csname LT0\endcsname{\color[rgb]{1,0,0}}%
      \expandafter\def\csname LT1\endcsname{\color[rgb]{0,1,0}}%
      \expandafter\def\csname LT2\endcsname{\color[rgb]{0,0,1}}%
      \expandafter\def\csname LT3\endcsname{\color[rgb]{1,0,1}}%
      \expandafter\def\csname LT4\endcsname{\color[rgb]{0,1,1}}%
      \expandafter\def\csname LT5\endcsname{\color[rgb]{1,1,0}}%
      \expandafter\def\csname LT6\endcsname{\color[rgb]{0,0,0}}%
      \expandafter\def\csname LT7\endcsname{\color[rgb]{1,0.3,0}}%
      \expandafter\def\csname LT8\endcsname{\color[rgb]{0.5,0.5,0.5}}%
    \else
      \def\colorrgb#1{\color{black}}%
      \def\colorgray#1{\color[gray]{#1}}%
      \expandafter\def\csname LTw\endcsname{\color{white}}%
      \expandafter\def\csname LTb\endcsname{\color{black}}%
      \expandafter\def\csname LTa\endcsname{\color{black}}%
      \expandafter\def\csname LT0\endcsname{\color{black}}%
      \expandafter\def\csname LT1\endcsname{\color{black}}%
      \expandafter\def\csname LT2\endcsname{\color{black}}%
      \expandafter\def\csname LT3\endcsname{\color{black}}%
      \expandafter\def\csname LT4\endcsname{\color{black}}%
      \expandafter\def\csname LT5\endcsname{\color{black}}%
      \expandafter\def\csname LT6\endcsname{\color{black}}%
      \expandafter\def\csname LT7\endcsname{\color{black}}%
      \expandafter\def\csname LT8\endcsname{\color{black}}%
    \fi
  \fi
  \setlength{\unitlength}{0.0500bp}%
  \begin{picture}(4680.00,3276.00)%
    \gplgaddtomacro\gplbacktext{%
      \csname LTb\endcsname%
      \put(128,1797){\rotatebox{-270}{\makebox(0,0){\strut{}$(\mathcal{R}^S_{13})^2$}}}%
      \put(2539,112){\makebox(0,0){\strut{}$\lambda$}}%
    }%
    \gplgaddtomacro\gplfronttext{%
      \csname LTb\endcsname%
      \put(2146,2858){\makebox(0,0)[l]{\strut{}tree-level}}%
      \csname LTb\endcsname%
      \put(2146,2666){\makebox(0,0)[l]{\strut{}one-loop (pole mass)}}%
      \csname LTb\endcsname%
      \put(2146,2474){\makebox(0,0)[l]{\strut{}one-loop (running mass)}}%
      \csname LTb\endcsname%
      \put(592,512){\makebox(0,0)[r]{\strut{} 0}}%
      \put(592,1026){\makebox(0,0)[r]{\strut{} 0.2}}%
      \put(592,1540){\makebox(0,0)[r]{\strut{} 0.4}}%
      \put(592,2055){\makebox(0,0)[r]{\strut{} 0.6}}%
      \put(592,2569){\makebox(0,0)[r]{\strut{} 0.8}}%
      \put(592,3083){\makebox(0,0)[r]{\strut{} 1}}%
      \put(688,352){\makebox(0,0){\strut{} 0}}%
      \put(1058,352){\makebox(0,0){\strut{} 0.1}}%
      \put(1429,352){\makebox(0,0){\strut{} 0.2}}%
      \put(1799,352){\makebox(0,0){\strut{} 0.3}}%
      \put(2169,352){\makebox(0,0){\strut{} 0.4}}%
      \put(2540,352){\makebox(0,0){\strut{} 0.5}}%
      \put(2910,352){\makebox(0,0){\strut{} 0.6}}%
      \put(3280,352){\makebox(0,0){\strut{} 0.7}}%
      \put(3650,352){\makebox(0,0){\strut{} 0.8}}%
      \put(4021,352){\makebox(0,0){\strut{} 0.9}}%
      \put(4391,352){\makebox(0,0){\strut{} 1}}%
    }%
    \gplbacktext
    \put(0,0){\includegraphics{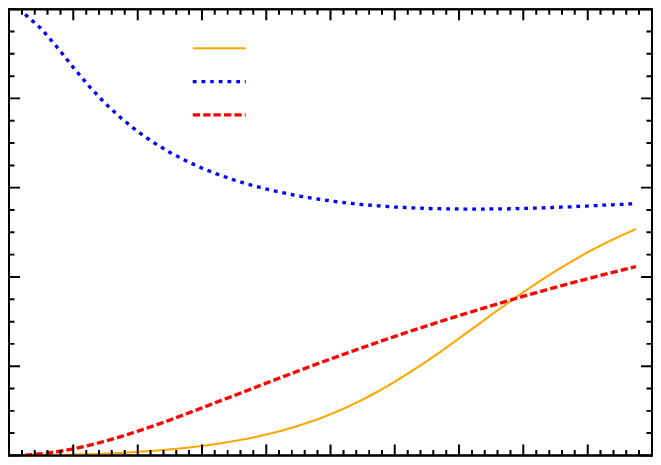}}%
    \gplfronttext
  \end{picture}%
\endgroup

%% file: Fig3b.tex
\begingroup
  \fontfamily{cmr}%
  \fontsize{9}{15}
\selectfont
  \makeatletter
  \providecommand\color[2][]{%
    \GenericError{(gnuplot) \space\space\space\@spaces}{%
      Package color not loaded in conjunction with
      terminal option `colourtext'%
    }{See the gnuplot documentation for explanation.%
    }{Either use 'blacktext' in gnuplot or load the package
      color.sty in LaTeX.}%
    \renewcommand\color[2][]{}%
  }%
  \providecommand\includegraphics[2][]{%
    \GenericError{(gnuplot) \space\space\space\@spaces}{%
      Package graphicx or graphics not loaded%
    }{See the gnuplot documentation for explanation.%
    }{The gnuplot epslatex terminal needs graphicx.sty or graphics.sty.}%
    \renewcommand\includegraphics[2][]{}%
  }%
  \providecommand\rotatebox[2]{#2}%
  \@ifundefined{ifGPcolor}{%
    \newif\ifGPcolor
    \GPcolortrue
  }{}%
  \@ifundefined{ifGPblacktext}{%
    \newif\ifGPblacktext
    \GPblacktexttrue
  }{}%
  \let\gplgaddtomacro\g@addto@macro
  \gdef\gplbacktext{}%
  \gdef\gplfronttext{}%
  \makeatother
  \ifGPblacktext
    \def\colorrgb#1{}%
    \def\colorgray#1{}%
  \else
    \ifGPcolor
      \def\colorrgb#1{\color[rgb]{#1}}%
      \def\colorgray#1{\color[gray]{#1}}%
      \expandafter\def\csname LTw\endcsname{\color{white}}%
      \expandafter\def\csname LTb\endcsname{\color{black}}%
      \expandafter\def\csname LTa\endcsname{\color{black}}%
      \expandafter\def\csname LT0\endcsname{\color[rgb]{1,0,0}}%
      \expandafter\def\csname LT1\endcsname{\color[rgb]{0,1,0}}%
      \expandafter\def\csname LT2\endcsname{\color[rgb]{0,0,1}}%
      \expandafter\def\csname LT3\endcsname{\color[rgb]{1,0,1}}%
      \expandafter\def\csname LT4\endcsname{\color[rgb]{0,1,1}}%
      \expandafter\def\csname LT5\endcsname{\color[rgb]{1,1,0}}%
      \expandafter\def\csname LT6\endcsname{\color[rgb]{0,0,0}}%
      \expandafter\def\csname LT7\endcsname{\color[rgb]{1,0.3,0}}%
      \expandafter\def\csname LT8\endcsname{\color[rgb]{0.5,0.5,0.5}}%
    \else
      \def\colorrgb#1{\color{black}}%
      \def\colorgray#1{\color[gray]{#1}}%
      \expandafter\def\csname LTw\endcsname{\color{white}}%
      \expandafter\def\csname LTb\endcsname{\color{black}}%
      \expandafter\def\csname LTa\endcsname{\color{black}}%
      \expandafter\def\csname LT0\endcsname{\color{black}}%
      \expandafter\def\csname LT1\endcsname{\color{black}}%
      \expandafter\def\csname LT2\endcsname{\color{black}}%
      \expandafter\def\csname LT3\endcsname{\color{black}}%
      \expandafter\def\csname LT4\endcsname{\color{black}}%
      \expandafter\def\csname LT5\endcsname{\color{black}}%
      \expandafter\def\csname LT6\endcsname{\color{black}}%
      \expandafter\def\csname LT7\endcsname{\color{black}}%
      \expandafter\def\csname LT8\endcsname{\color{black}}%
    \fi
  \fi
  \setlength{\unitlength}{0.0500bp}%
  \begin{picture}(4680.00,3276.00)%
    \gplgaddtomacro\gplbacktext{%
      \csname LTb\endcsname%
      \put(128,1797){\rotatebox{-270}{\makebox(0,0){\strut{}$(\mathcal{R}^S_{23})^2$}}}%
      \put(2539,112){\makebox(0,0){\strut{}$\lambda$}}%
    }%
    \gplgaddtomacro\gplfronttext{%
      \csname LTb\endcsname%
      \put(2146,1084){\makebox(0,0)[l]{\strut{}tree-level}}%
      \csname LTb\endcsname%
      \put(2146,892){\makebox(0,0)[l]{\strut{}one-loop (pole mass)}}%
      \csname LTb\endcsname%
      \put(2146,700){\makebox(0,0)[l]{\strut{}one-loop (running mass)}}%
      \csname LTb\endcsname%
      \put(592,512){\makebox(0,0)[r]{\strut{} 0}}%
      \put(592,1026){\makebox(0,0)[r]{\strut{} 0.2}}%
      \put(592,1540){\makebox(0,0)[r]{\strut{} 0.4}}%
      \put(592,2055){\makebox(0,0)[r]{\strut{} 0.6}}%
      \put(592,2569){\makebox(0,0)[r]{\strut{} 0.8}}%
      \put(592,3083){\makebox(0,0)[r]{\strut{} 1}}%
      \put(688,352){\makebox(0,0){\strut{} 0}}%
      \put(1058,352){\makebox(0,0){\strut{} 0.1}}%
      \put(1429,352){\makebox(0,0){\strut{} 0.2}}%
      \put(1799,352){\makebox(0,0){\strut{} 0.3}}%
      \put(2169,352){\makebox(0,0){\strut{} 0.4}}%
      \put(2540,352){\makebox(0,0){\strut{} 0.5}}%
      \put(2910,352){\makebox(0,0){\strut{} 0.6}}%
      \put(3280,352){\makebox(0,0){\strut{} 0.7}}%
      \put(3650,352){\makebox(0,0){\strut{} 0.8}}%
      \put(4021,352){\makebox(0,0){\strut{} 0.9}}%
      \put(4391,352){\makebox(0,0){\strut{} 1}}%
    }%
    \gplbacktext
    \put(0,0){\includegraphics{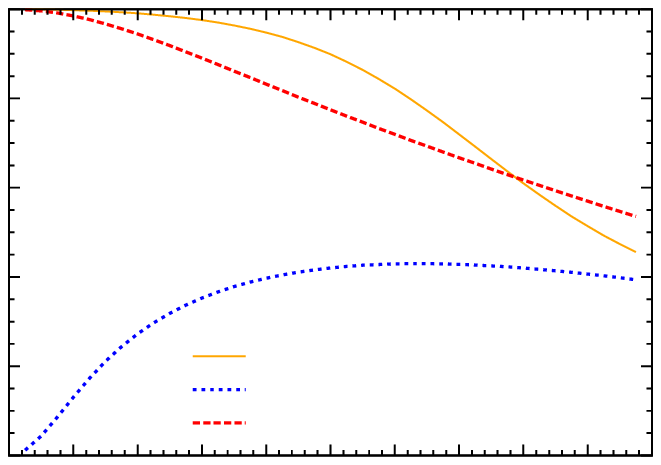}}%
    \gplfronttext
  \end{picture}%
\endgroup

%% file: Fig4a.tex
\begingroup
  \fontfamily{cmr}%
  \fontsize{9}{15}
\selectfont
  \makeatletter
  \providecommand\color[2][]{%
    \GenericError{(gnuplot) \space\space\space\@spaces}{%
      Package color not loaded in conjunction with
      terminal option `colourtext'%
    }{See the gnuplot documentation for explanation.%
    }{Either use 'blacktext' in gnuplot or load the package
      color.sty in LaTeX.}%
    \renewcommand\color[2][]{}%
  }%
  \providecommand\includegraphics[2][]{%
    \GenericError{(gnuplot) \space\space\space\@spaces}{%
      Package graphicx or graphics not loaded%
    }{See the gnuplot documentation for explanation.%
    }{The gnuplot epslatex terminal needs graphicx.sty or graphics.sty.}%
    \renewcommand\includegraphics[2][]{}%
  }%
  \providecommand\rotatebox[2]{#2}%
  \@ifundefined{ifGPcolor}{%
    \newif\ifGPcolor
    \GPcolortrue
  }{}%
  \@ifundefined{ifGPblacktext}{%
    \newif\ifGPblacktext
    \GPblacktexttrue
  }{}%
  \let\gplgaddtomacro\g@addto@macro
  \gdef\gplbacktext{}%
  \gdef\gplfronttext{}%
  \makeatother
  \ifGPblacktext
    \def\colorrgb#1{}%
    \def\colorgray#1{}%
  \else
    \ifGPcolor
      \def\colorrgb#1{\color[rgb]{#1}}%
      \def\colorgray#1{\color[gray]{#1}}%
      \expandafter\def\csname LTw\endcsname{\color{white}}%
      \expandafter\def\csname LTb\endcsname{\color{black}}%
      \expandafter\def\csname LTa\endcsname{\color{black}}%
      \expandafter\def\csname LT0\endcsname{\color[rgb]{1,0,0}}%
      \expandafter\def\csname LT1\endcsname{\color[rgb]{0,1,0}}%
      \expandafter\def\csname LT2\endcsname{\color[rgb]{0,0,1}}%
      \expandafter\def\csname LT3\endcsname{\color[rgb]{1,0,1}}%
      \expandafter\def\csname LT4\endcsname{\color[rgb]{0,1,1}}%
      \expandafter\def\csname LT5\endcsname{\color[rgb]{1,1,0}}%
      \expandafter\def\csname LT6\endcsname{\color[rgb]{0,0,0}}%
      \expandafter\def\csname LT7\endcsname{\color[rgb]{1,0.3,0}}%
      \expandafter\def\csname LT8\endcsname{\color[rgb]{0.5,0.5,0.5}}%
    \else
      \def\colorrgb#1{\color{black}}%
      \def\colorgray#1{\color[gray]{#1}}%
      \expandafter\def\csname LTw\endcsname{\color{white}}%
      \expandafter\def\csname LTb\endcsname{\color{black}}%
      \expandafter\def\csname LTa\endcsname{\color{black}}%
      \expandafter\def\csname LT0\endcsname{\color{black}}%
      \expandafter\def\csname LT1\endcsname{\color{black}}%
      \expandafter\def\csname LT2\endcsname{\color{black}}%
      \expandafter\def\csname LT3\endcsname{\color{black}}%
      \expandafter\def\csname LT4\endcsname{\color{black}}%
      \expandafter\def\csname LT5\endcsname{\color{black}}%
      \expandafter\def\csname LT6\endcsname{\color{black}}%
      \expandafter\def\csname LT7\endcsname{\color{black}}%
      \expandafter\def\csname LT8\endcsname{\color{black}}%
    \fi
  \fi
  \setlength{\unitlength}{0.0500bp}%
  \begin{picture}(4680.00,3276.00)%
    \gplgaddtomacro\gplbacktext{%
      \csname LTb\endcsname%
      \put(128,1797){\rotatebox{-270}{\makebox(0,0){\strut{}$M_{H_1} \quad [\text{GeV}]$}}}%
      \put(2539,112){\makebox(0,0){\strut{}$\lambda$}}%
    }%
    \gplgaddtomacro\gplfronttext{%
      \csname LTb\endcsname%
      \put(1182,2858){\makebox(0,0)[l]{\strut{}tree-level}}%
      \csname LTb\endcsname%
      \put(1182,2666){\makebox(0,0)[l]{\strut{}one-loop (pole)}}%
      \csname LTb\endcsname%
      \put(1182,2474){\makebox(0,0)[l]{\strut{}one-loop (running)}}%
      \csname LTb\endcsname%
      \put(592,512){\makebox(0,0)[r]{\strut{} 50}}%
      \put(592,769){\makebox(0,0)[r]{\strut{} 60}}%
      \put(592,1026){\makebox(0,0)[r]{\strut{} 70}}%
      \put(592,1283){\makebox(0,0)[r]{\strut{} 80}}%
      \put(592,1540){\makebox(0,0)[r]{\strut{} 90}}%
      \put(592,1798){\makebox(0,0)[r]{\strut{} 100}}%
      \put(592,2055){\makebox(0,0)[r]{\strut{} 110}}%
      \put(592,2312){\makebox(0,0)[r]{\strut{} 120}}%
      \put(592,2569){\makebox(0,0)[r]{\strut{} 130}}%
      \put(592,2826){\makebox(0,0)[r]{\strut{} 140}}%
      \put(592,3083){\makebox(0,0)[r]{\strut{} 150}}%
      \put(688,352){\makebox(0,0){\strut{} 0}}%
      \put(1058,352){\makebox(0,0){\strut{} 0.1}}%
      \put(1429,352){\makebox(0,0){\strut{} 0.2}}%
      \put(1799,352){\makebox(0,0){\strut{} 0.3}}%
      \put(2169,352){\makebox(0,0){\strut{} 0.4}}%
      \put(2540,352){\makebox(0,0){\strut{} 0.5}}%
      \put(2910,352){\makebox(0,0){\strut{} 0.6}}%
      \put(3280,352){\makebox(0,0){\strut{} 0.7}}%
      \put(3650,352){\makebox(0,0){\strut{} 0.8}}%
      \put(4021,352){\makebox(0,0){\strut{} 0.9}}%
      \put(4391,352){\makebox(0,0){\strut{} 1}}%
    }%
    \gplbacktext
    \put(0,0){\includegraphics{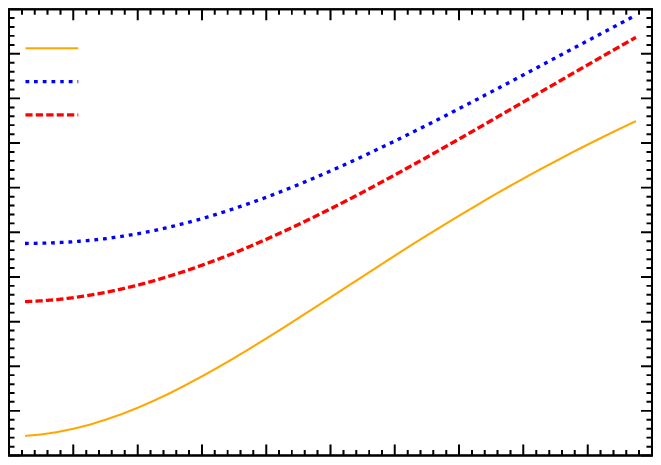}}%
    \gplfronttext
  \end{picture}%
\endgroup

%% file: Fig4b.tex
\begingroup
  \fontfamily{cmr}%
  \fontsize{9}{15}
\selectfont
  \makeatletter
  \providecommand\color[2][]{%
    \GenericError{(gnuplot) \space\space\space\@spaces}{%
      Package color not loaded in conjunction with
      terminal option `colourtext'%
    }{See the gnuplot documentation for explanation.%
    }{Either use 'blacktext' in gnuplot or load the package
      color.sty in LaTeX.}%
    \renewcommand\color[2][]{}%
  }%
  \providecommand\includegraphics[2][]{%
    \GenericError{(gnuplot) \space\space\space\@spaces}{%
      Package graphicx or graphics not loaded%
    }{See the gnuplot documentation for explanation.%
    }{The gnuplot epslatex terminal needs graphicx.sty or graphics.sty.}%
    \renewcommand\includegraphics[2][]{}%
  }%
  \providecommand\rotatebox[2]{#2}%
  \@ifundefined{ifGPcolor}{%
    \newif\ifGPcolor
    \GPcolortrue
  }{}%
  \@ifundefined{ifGPblacktext}{%
    \newif\ifGPblacktext
    \GPblacktexttrue
  }{}%
  \let\gplgaddtomacro\g@addto@macro
  \gdef\gplbacktext{}%
  \gdef\gplfronttext{}%
  \makeatother
  \ifGPblacktext
    \def\colorrgb#1{}%
    \def\colorgray#1{}%
  \else
    \ifGPcolor
      \def\colorrgb#1{\color[rgb]{#1}}%
      \def\colorgray#1{\color[gray]{#1}}%
      \expandafter\def\csname LTw\endcsname{\color{white}}%
      \expandafter\def\csname LTb\endcsname{\color{black}}%
      \expandafter\def\csname LTa\endcsname{\color{black}}%
      \expandafter\def\csname LT0\endcsname{\color[rgb]{1,0,0}}%
      \expandafter\def\csname LT1\endcsname{\color[rgb]{0,1,0}}%
      \expandafter\def\csname LT2\endcsname{\color[rgb]{0,0,1}}%
      \expandafter\def\csname LT3\endcsname{\color[rgb]{1,0,1}}%
      \expandafter\def\csname LT4\endcsname{\color[rgb]{0,1,1}}%
      \expandafter\def\csname LT5\endcsname{\color[rgb]{1,1,0}}%
      \expandafter\def\csname LT6\endcsname{\color[rgb]{0,0,0}}%
      \expandafter\def\csname LT7\endcsname{\color[rgb]{1,0.3,0}}%
      \expandafter\def\csname LT8\endcsname{\color[rgb]{0.5,0.5,0.5}}%
    \else
      \def\colorrgb#1{\color{black}}%
      \def\colorgray#1{\color[gray]{#1}}%
      \expandafter\def\csname LTw\endcsname{\color{white}}%
      \expandafter\def\csname LTb\endcsname{\color{black}}%
      \expandafter\def\csname LTa\endcsname{\color{black}}%
      \expandafter\def\csname LT0\endcsname{\color{black}}%
      \expandafter\def\csname LT1\endcsname{\color{black}}%
      \expandafter\def\csname LT2\endcsname{\color{black}}%
      \expandafter\def\csname LT3\endcsname{\color{black}}%
      \expandafter\def\csname LT4\endcsname{\color{black}}%
      \expandafter\def\csname LT5\endcsname{\color{black}}%
      \expandafter\def\csname LT6\endcsname{\color{black}}%
      \expandafter\def\csname LT7\endcsname{\color{black}}%
      \expandafter\def\csname LT8\endcsname{\color{black}}%
    \fi
  \fi
  \setlength{\unitlength}{0.0500bp}%
  \begin{picture}(4680.00,3276.00)%
    \gplgaddtomacro\gplbacktext{%
      \csname LTb\endcsname%
      \put(128,1797){\rotatebox{-270}{\makebox(0,0){\strut{}$M_{H_2} \quad [\text{GeV}]$}}}%
      \put(2539,112){\makebox(0,0){\strut{}$\lambda$}}%
    }%
    \gplgaddtomacro\gplfronttext{%
      \csname LTb\endcsname%
      \put(1182,2858){\makebox(0,0)[l]{\strut{}tree-level}}%
      \csname LTb\endcsname%
      \put(1182,2666){\makebox(0,0)[l]{\strut{}one-loop (pole)}}%
      \csname LTb\endcsname%
      \put(1182,2474){\makebox(0,0)[l]{\strut{}one-loop (running)}}%
      \csname LTb\endcsname%
      \put(592,512){\makebox(0,0)[r]{\strut{} 90}}%
      \put(592,769){\makebox(0,0)[r]{\strut{} 100}}%
      \put(592,1026){\makebox(0,0)[r]{\strut{} 110}}%
      \put(592,1283){\makebox(0,0)[r]{\strut{} 120}}%
      \put(592,1540){\makebox(0,0)[r]{\strut{} 130}}%
      \put(592,1798){\makebox(0,0)[r]{\strut{} 140}}%
      \put(592,2055){\makebox(0,0)[r]{\strut{} 150}}%
      \put(592,2312){\makebox(0,0)[r]{\strut{} 160}}%
      \put(592,2569){\makebox(0,0)[r]{\strut{} 170}}%
      \put(592,2826){\makebox(0,0)[r]{\strut{} 180}}%
      \put(592,3083){\makebox(0,0)[r]{\strut{} 190}}%
      \put(688,352){\makebox(0,0){\strut{} 0}}%
      \put(1058,352){\makebox(0,0){\strut{} 0.1}}%
      \put(1429,352){\makebox(0,0){\strut{} 0.2}}%
      \put(1799,352){\makebox(0,0){\strut{} 0.3}}%
      \put(2169,352){\makebox(0,0){\strut{} 0.4}}%
      \put(2540,352){\makebox(0,0){\strut{} 0.5}}%
      \put(2910,352){\makebox(0,0){\strut{} 0.6}}%
      \put(3280,352){\makebox(0,0){\strut{} 0.7}}%
      \put(3650,352){\makebox(0,0){\strut{} 0.8}}%
      \put(4021,352){\makebox(0,0){\strut{} 0.9}}%
      \put(4391,352){\makebox(0,0){\strut{} 1}}%
    }%
    \gplbacktext
    \put(0,0){\includegraphics{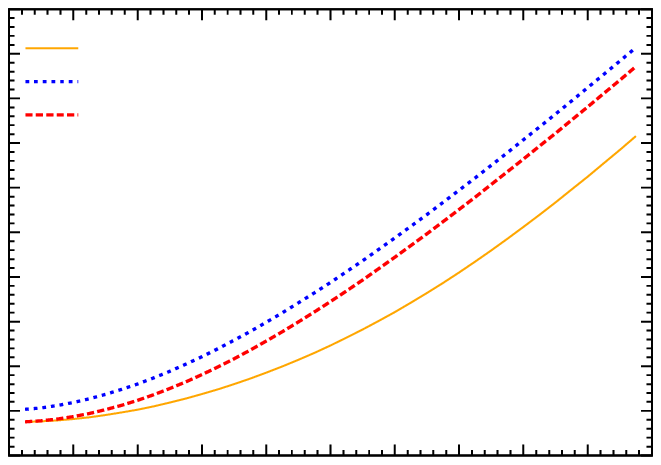}}%
    \gplfronttext
  \end{picture}%
\endgroup

%% file: Fig5a.tex
\begingroup
  \fontfamily{cmr}%
  \fontsize{9}{15}
\selectfont
  \makeatletter
  \providecommand\color[2][]{%
    \GenericError{(gnuplot) \space\space\space\@spaces}{%
      Package color not loaded in conjunction with
      terminal option `colourtext'%
    }{See the gnuplot documentation for explanation.%
    }{Either use 'blacktext' in gnuplot or load the package
      color.sty in LaTeX.}%
    \renewcommand\color[2][]{}%
  }%
  \providecommand\includegraphics[2][]{%
    \GenericError{(gnuplot) \space\space\space\@spaces}{%
      Package graphicx or graphics not loaded%
    }{See the gnuplot documentation for explanation.%
    }{The gnuplot epslatex terminal needs graphicx.sty or graphics.sty.}%
    \renewcommand\includegraphics[2][]{}%
  }%
  \providecommand\rotatebox[2]{#2}%
  \@ifundefined{ifGPcolor}{%
    \newif\ifGPcolor
    \GPcolortrue
  }{}%
  \@ifundefined{ifGPblacktext}{%
    \newif\ifGPblacktext
    \GPblacktexttrue
  }{}%
  \let\gplgaddtomacro\g@addto@macro
  \gdef\gplbacktext{}%
  \gdef\gplfronttext{}%
  \makeatother
  \ifGPblacktext
    \def\colorrgb#1{}%
    \def\colorgray#1{}%
  \else
    \ifGPcolor
      \def\colorrgb#1{\color[rgb]{#1}}%
      \def\colorgray#1{\color[gray]{#1}}%
      \expandafter\def\csname LTw\endcsname{\color{white}}%
      \expandafter\def\csname LTb\endcsname{\color{black}}%
      \expandafter\def\csname LTa\endcsname{\color{black}}%
      \expandafter\def\csname LT0\endcsname{\color[rgb]{1,0,0}}%
      \expandafter\def\csname LT1\endcsname{\color[rgb]{0,1,0}}%
      \expandafter\def\csname LT2\endcsname{\color[rgb]{0,0,1}}%
      \expandafter\def\csname LT3\endcsname{\color[rgb]{1,0,1}}%
      \expandafter\def\csname LT4\endcsname{\color[rgb]{0,1,1}}%
      \expandafter\def\csname LT5\endcsname{\color[rgb]{1,1,0}}%
      \expandafter\def\csname LT6\endcsname{\color[rgb]{0,0,0}}%
      \expandafter\def\csname LT7\endcsname{\color[rgb]{1,0.3,0}}%
      \expandafter\def\csname LT8\endcsname{\color[rgb]{0.5,0.5,0.5}}%
    \else
      \def\colorrgb#1{\color{black}}%
      \def\colorgray#1{\color[gray]{#1}}%
      \expandafter\def\csname LTw\endcsname{\color{white}}%
      \expandafter\def\csname LTb\endcsname{\color{black}}%
      \expandafter\def\csname LTa\endcsname{\color{black}}%
      \expandafter\def\csname LT0\endcsname{\color{black}}%
      \expandafter\def\csname LT1\endcsname{\color{black}}%
      \expandafter\def\csname LT2\endcsname{\color{black}}%
      \expandafter\def\csname LT3\endcsname{\color{black}}%
      \expandafter\def\csname LT4\endcsname{\color{black}}%
      \expandafter\def\csname LT5\endcsname{\color{black}}%
      \expandafter\def\csname LT6\endcsname{\color{black}}%
      \expandafter\def\csname LT7\endcsname{\color{black}}%
      \expandafter\def\csname LT8\endcsname{\color{black}}%
    \fi
  \fi
  \setlength{\unitlength}{0.0500bp}%
  \begin{picture}(4680.00,3276.00)%
    \gplgaddtomacro\gplbacktext{%
      \csname LTb\endcsname%
      \put(128,1797){\rotatebox{-270}{\makebox(0,0){\strut{}$(\mathcal{R}^S_{13})^2$}}}%
      \put(2539,112){\makebox(0,0){\strut{}$\lambda$}}%
    }%
    \gplgaddtomacro\gplfronttext{%
      \csname LTb\endcsname%
      \put(1271,2858){\makebox(0,0)[l]{\strut{}tree-level}}%
      \csname LTb\endcsname%
      \put(1271,2666){\makebox(0,0)[l]{\strut{}$Q_0=150\text{GeV}$}}%
      \csname LTb\endcsname%
      \put(1271,2474){\makebox(0,0)[l]{\strut{}$Q_0=300\text{GeV}$}}%
      \csname LTb\endcsname%
      \put(1271,2282){\makebox(0,0)[l]{\strut{}$Q_0=600\text{GeV}$}}%
      \csname LTb\endcsname%
      \put(592,512){\makebox(0,0)[r]{\strut{} 0}}%
      \put(592,1026){\makebox(0,0)[r]{\strut{} 0.2}}%
      \put(592,1540){\makebox(0,0)[r]{\strut{} 0.4}}%
      \put(592,2055){\makebox(0,0)[r]{\strut{} 0.6}}%
      \put(592,2569){\makebox(0,0)[r]{\strut{} 0.8}}%
      \put(592,3083){\makebox(0,0)[r]{\strut{} 1}}%
      \put(688,352){\makebox(0,0){\strut{} 0}}%
      \put(1058,352){\makebox(0,0){\strut{} 0.1}}%
      \put(1429,352){\makebox(0,0){\strut{} 0.2}}%
      \put(1799,352){\makebox(0,0){\strut{} 0.3}}%
      \put(2169,352){\makebox(0,0){\strut{} 0.4}}%
      \put(2540,352){\makebox(0,0){\strut{} 0.5}}%
      \put(2910,352){\makebox(0,0){\strut{} 0.6}}%
      \put(3280,352){\makebox(0,0){\strut{} 0.7}}%
      \put(3650,352){\makebox(0,0){\strut{} 0.8}}%
      \put(4021,352){\makebox(0,0){\strut{} 0.9}}%
      \put(4391,352){\makebox(0,0){\strut{} 1}}%
    }%
    \gplbacktext
    \put(0,0){\includegraphics{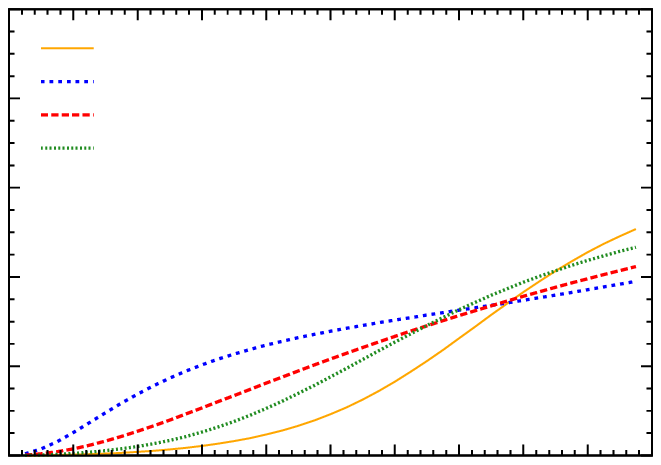}}%
    \gplfronttext
  \end{picture}%
\endgroup

%% file: Fig5b.tex
\begingroup
  \fontfamily{cmr}%
  \fontsize{9}{15}
\selectfont
  \makeatletter
  \providecommand\color[2][]{%
    \GenericError{(gnuplot) \space\space\space\@spaces}{%
      Package color not loaded in conjunction with
      terminal option `colourtext'%
    }{See the gnuplot documentation for explanation.%
    }{Either use 'blacktext' in gnuplot or load the package
      color.sty in LaTeX.}%
    \renewcommand\color[2][]{}%
  }%
  \providecommand\includegraphics[2][]{%
    \GenericError{(gnuplot) \space\space\space\@spaces}{%
      Package graphicx or graphics not loaded%
    }{See the gnuplot documentation for explanation.%
    }{The gnuplot epslatex terminal needs graphicx.sty or graphics.sty.}%
    \renewcommand\includegraphics[2][]{}%
  }%
  \providecommand\rotatebox[2]{#2}%
  \@ifundefined{ifGPcolor}{%
    \newif\ifGPcolor
    \GPcolortrue
  }{}%
  \@ifundefined{ifGPblacktext}{%
    \newif\ifGPblacktext
    \GPblacktexttrue
  }{}%
  \let\gplgaddtomacro\g@addto@macro
  \gdef\gplbacktext{}%
  \gdef\gplfronttext{}%
  \makeatother
  \ifGPblacktext
    \def\colorrgb#1{}%
    \def\colorgray#1{}%
  \else
    \ifGPcolor
      \def\colorrgb#1{\color[rgb]{#1}}%
      \def\colorgray#1{\color[gray]{#1}}%
      \expandafter\def\csname LTw\endcsname{\color{white}}%
      \expandafter\def\csname LTb\endcsname{\color{black}}%
      \expandafter\def\csname LTa\endcsname{\color{black}}%
      \expandafter\def\csname LT0\endcsname{\color[rgb]{1,0,0}}%
      \expandafter\def\csname LT1\endcsname{\color[rgb]{0,1,0}}%
      \expandafter\def\csname LT2\endcsname{\color[rgb]{0,0,1}}%
      \expandafter\def\csname LT3\endcsname{\color[rgb]{1,0,1}}%
      \expandafter\def\csname LT4\endcsname{\color[rgb]{0,1,1}}%
      \expandafter\def\csname LT5\endcsname{\color[rgb]{1,1,0}}%
      \expandafter\def\csname LT6\endcsname{\color[rgb]{0,0,0}}%
      \expandafter\def\csname LT7\endcsname{\color[rgb]{1,0.3,0}}%
      \expandafter\def\csname LT8\endcsname{\color[rgb]{0.5,0.5,0.5}}%
    \else
      \def\colorrgb#1{\color{black}}%
      \def\colorgray#1{\color[gray]{#1}}%
      \expandafter\def\csname LTw\endcsname{\color{white}}%
      \expandafter\def\csname LTb\endcsname{\color{black}}%
      \expandafter\def\csname LTa\endcsname{\color{black}}%
      \expandafter\def\csname LT0\endcsname{\color{black}}%
      \expandafter\def\csname LT1\endcsname{\color{black}}%
      \expandafter\def\csname LT2\endcsname{\color{black}}%
      \expandafter\def\csname LT3\endcsname{\color{black}}%
      \expandafter\def\csname LT4\endcsname{\color{black}}%
      \expandafter\def\csname LT5\endcsname{\color{black}}%
      \expandafter\def\csname LT6\endcsname{\color{black}}%
      \expandafter\def\csname LT7\endcsname{\color{black}}%
      \expandafter\def\csname LT8\endcsname{\color{black}}%
    \fi
  \fi
  \setlength{\unitlength}{0.0500bp}%
  \begin{picture}(4680.00,3276.00)%
    \gplgaddtomacro\gplbacktext{%
      \csname LTb\endcsname%
      \put(128,1797){\rotatebox{-270}{\makebox(0,0){\strut{}$(\mathcal{R}^S_{23})^2$}}}%
      \put(2539,112){\makebox(0,0){\strut{}$\lambda$}}%
    }%
    \gplgaddtomacro\gplfronttext{%
      \csname LTb\endcsname%
      \put(1271,1316){\makebox(0,0)[l]{\strut{}tree-level}}%
      \csname LTb\endcsname%
      \put(1271,1124){\makebox(0,0)[l]{\strut{}$Q_0=150\text{GeV}$}}%
      \csname LTb\endcsname%
      \put(1271,932){\makebox(0,0)[l]{\strut{}$Q_0=300\text{GeV}$}}%
      \csname LTb\endcsname%
      \put(1271,740){\makebox(0,0)[l]{\strut{}$Q_0=600\text{GeV}$}}%
      \csname LTb\endcsname%
      \put(592,512){\makebox(0,0)[r]{\strut{} 0}}%
      \put(592,1026){\makebox(0,0)[r]{\strut{} 0.2}}%
      \put(592,1540){\makebox(0,0)[r]{\strut{} 0.4}}%
      \put(592,2055){\makebox(0,0)[r]{\strut{} 0.6}}%
      \put(592,2569){\makebox(0,0)[r]{\strut{} 0.8}}%
      \put(592,3083){\makebox(0,0)[r]{\strut{} 1}}%
      \put(688,352){\makebox(0,0){\strut{} 0}}%
      \put(1058,352){\makebox(0,0){\strut{} 0.1}}%
      \put(1429,352){\makebox(0,0){\strut{} 0.2}}%
      \put(1799,352){\makebox(0,0){\strut{} 0.3}}%
      \put(2169,352){\makebox(0,0){\strut{} 0.4}}%
      \put(2540,352){\makebox(0,0){\strut{} 0.5}}%
      \put(2910,352){\makebox(0,0){\strut{} 0.6}}%
      \put(3280,352){\makebox(0,0){\strut{} 0.7}}%
      \put(3650,352){\makebox(0,0){\strut{} 0.8}}%
      \put(4021,352){\makebox(0,0){\strut{} 0.9}}%
      \put(4391,352){\makebox(0,0){\strut{} 1}}%
    }%
    \gplbacktext
    \put(0,0){\includegraphics{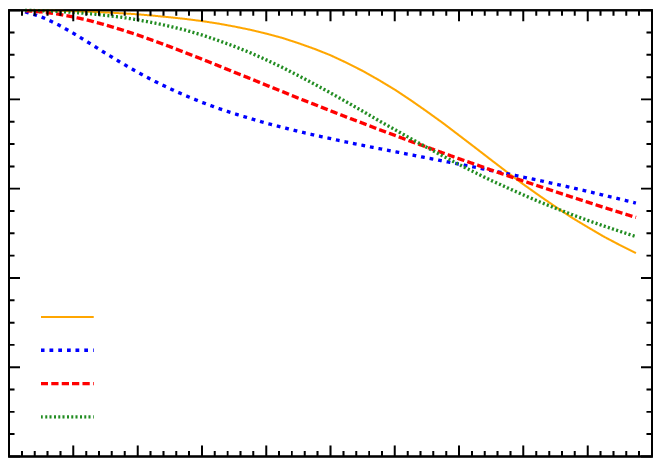}}%
    \gplfronttext
  \end{picture}%
\endgroup

%% file: Fig6a.tex
\begingroup
  \fontfamily{cmr}%
  \fontsize{9}{15}
\selectfont
  \makeatletter
  \providecommand\color[2][]{%
    \GenericError{(gnuplot) \space\space\space\@spaces}{%
      Package color not loaded in conjunction with
      terminal option `colourtext'%
    }{See the gnuplot documentation for explanation.%
    }{Either use 'blacktext' in gnuplot or load the package
      color.sty in LaTeX.}%
    \renewcommand\color[2][]{}%
  }%
  \providecommand\includegraphics[2][]{%
    \GenericError{(gnuplot) \space\space\space\@spaces}{%
      Package graphicx or graphics not loaded%
    }{See the gnuplot documentation for explanation.%
    }{The gnuplot epslatex terminal needs graphicx.sty or graphics.sty.}%
    \renewcommand\includegraphics[2][]{}%
  }%
  \providecommand\rotatebox[2]{#2}%
  \@ifundefined{ifGPcolor}{%
    \newif\ifGPcolor
    \GPcolortrue
  }{}%
  \@ifundefined{ifGPblacktext}{%
    \newif\ifGPblacktext
    \GPblacktexttrue
  }{}%
  \let\gplgaddtomacro\g@addto@macro
  \gdef\gplbacktext{}%
  \gdef\gplfronttext{}%
  \makeatother
  \ifGPblacktext
    \def\colorrgb#1{}%
    \def\colorgray#1{}%
  \else
    \ifGPcolor
      \def\colorrgb#1{\color[rgb]{#1}}%
      \def\colorgray#1{\color[gray]{#1}}%
      \expandafter\def\csname LTw\endcsname{\color{white}}%
      \expandafter\def\csname LTb\endcsname{\color{black}}%
      \expandafter\def\csname LTa\endcsname{\color{black}}%
      \expandafter\def\csname LT0\endcsname{\color[rgb]{1,0,0}}%
      \expandafter\def\csname LT1\endcsname{\color[rgb]{0,1,0}}%
      \expandafter\def\csname LT2\endcsname{\color[rgb]{0,0,1}}%
      \expandafter\def\csname LT3\endcsname{\color[rgb]{1,0,1}}%
      \expandafter\def\csname LT4\endcsname{\color[rgb]{0,1,1}}%
      \expandafter\def\csname LT5\endcsname{\color[rgb]{1,1,0}}%
      \expandafter\def\csname LT6\endcsname{\color[rgb]{0,0,0}}%
      \expandafter\def\csname LT7\endcsname{\color[rgb]{1,0.3,0}}%
      \expandafter\def\csname LT8\endcsname{\color[rgb]{0.5,0.5,0.5}}%
    \else
      \def\colorrgb#1{\color{black}}%
      \def\colorgray#1{\color[gray]{#1}}%
      \expandafter\def\csname LTw\endcsname{\color{white}}%
      \expandafter\def\csname LTb\endcsname{\color{black}}%
      \expandafter\def\csname LTa\endcsname{\color{black}}%
      \expandafter\def\csname LT0\endcsname{\color{black}}%
      \expandafter\def\csname LT1\endcsname{\color{black}}%
      \expandafter\def\csname LT2\endcsname{\color{black}}%
      \expandafter\def\csname LT3\endcsname{\color{black}}%
      \expandafter\def\csname LT4\endcsname{\color{black}}%
      \expandafter\def\csname LT5\endcsname{\color{black}}%
      \expandafter\def\csname LT6\endcsname{\color{black}}%
      \expandafter\def\csname LT7\endcsname{\color{black}}%
      \expandafter\def\csname LT8\endcsname{\color{black}}%
    \fi
  \fi
  \setlength{\unitlength}{0.0500bp}%
  \begin{picture}(4680.00,3276.00)%
    \gplgaddtomacro\gplbacktext{%
      \csname LTb\endcsname%
      \put(128,1797){\rotatebox{-270}{\makebox(0,0){\strut{}$M_{H_1} \quad [\text{GeV}]$}}}%
      \put(2539,112){\makebox(0,0){\strut{}$\lambda$}}%
    }%
    \gplgaddtomacro\gplfronttext{%
      \csname LTb\endcsname%
      \put(1271,2858){\makebox(0,0)[l]{\strut{}tree-level}}%
      \csname LTb\endcsname%
      \put(1271,2666){\makebox(0,0)[l]{\strut{}$Q_0=150\text{GeV}$}}%
      \csname LTb\endcsname%
      \put(1271,2474){\makebox(0,0)[l]{\strut{}$Q_0=300\text{GeV}$}}%
      \csname LTb\endcsname%
      \put(1271,2282){\makebox(0,0)[l]{\strut{}$Q_0=600\text{GeV}$}}%
      \csname LTb\endcsname%
      \put(592,512){\makebox(0,0)[r]{\strut{} 50}}%
      \put(592,769){\makebox(0,0)[r]{\strut{} 60}}%
      \put(592,1026){\makebox(0,0)[r]{\strut{} 70}}%
      \put(592,1283){\makebox(0,0)[r]{\strut{} 80}}%
      \put(592,1540){\makebox(0,0)[r]{\strut{} 90}}%
      \put(592,1798){\makebox(0,0)[r]{\strut{} 100}}%
      \put(592,2055){\makebox(0,0)[r]{\strut{} 110}}%
      \put(592,2312){\makebox(0,0)[r]{\strut{} 120}}%
      \put(592,2569){\makebox(0,0)[r]{\strut{} 130}}%
      \put(592,2826){\makebox(0,0)[r]{\strut{} 140}}%
      \put(592,3083){\makebox(0,0)[r]{\strut{} 150}}%
      \put(688,352){\makebox(0,0){\strut{} 0}}%
      \put(1058,352){\makebox(0,0){\strut{} 0.1}}%
      \put(1429,352){\makebox(0,0){\strut{} 0.2}}%
      \put(1799,352){\makebox(0,0){\strut{} 0.3}}%
      \put(2169,352){\makebox(0,0){\strut{} 0.4}}%
      \put(2540,352){\makebox(0,0){\strut{} 0.5}}%
      \put(2910,352){\makebox(0,0){\strut{} 0.6}}%
      \put(3280,352){\makebox(0,0){\strut{} 0.7}}%
      \put(3650,352){\makebox(0,0){\strut{} 0.8}}%
      \put(4021,352){\makebox(0,0){\strut{} 0.9}}%
      \put(4391,352){\makebox(0,0){\strut{} 1}}%
    }%
    \gplbacktext
    \put(0,0){\includegraphics{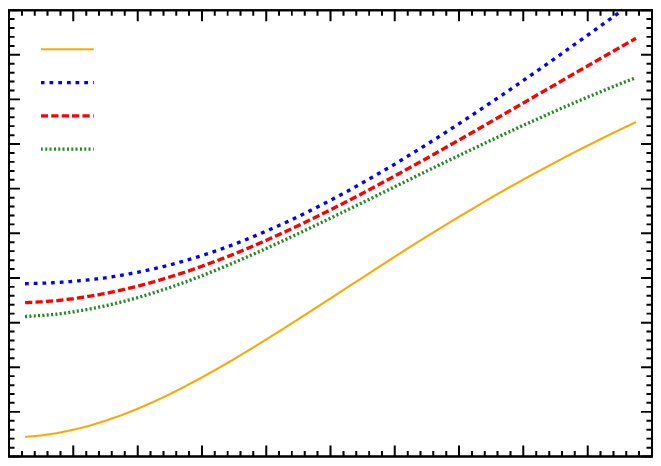}}%
    \gplfronttext
  \end{picture}%
\endgroup

%% file: Fig6b.tex
\begingroup
  \fontfamily{cmr}%
  \fontsize{9}{15}
\selectfont
  \makeatletter
  \providecommand\color[2][]{%
    \GenericError{(gnuplot) \space\space\space\@spaces}{%
      Package color not loaded in conjunction with
      terminal option `colourtext'%
    }{See the gnuplot documentation for explanation.%
    }{Either use 'blacktext' in gnuplot or load the package
      color.sty in LaTeX.}%
    \renewcommand\color[2][]{}%
  }%
  \providecommand\includegraphics[2][]{%
    \GenericError{(gnuplot) \space\space\space\@spaces}{%
      Package graphicx or graphics not loaded%
    }{See the gnuplot documentation for explanation.%
    }{The gnuplot epslatex terminal needs graphicx.sty or graphics.sty.}%
    \renewcommand\includegraphics[2][]{}%
  }%
  \providecommand\rotatebox[2]{#2}%
  \@ifundefined{ifGPcolor}{%
    \newif\ifGPcolor
    \GPcolortrue
  }{}%
  \@ifundefined{ifGPblacktext}{%
    \newif\ifGPblacktext
    \GPblacktexttrue
  }{}%
  \let\gplgaddtomacro\g@addto@macro
  \gdef\gplbacktext{}%
  \gdef\gplfronttext{}%
  \makeatother
  \ifGPblacktext
    \def\colorrgb#1{}%
    \def\colorgray#1{}%
  \else
    \ifGPcolor
      \def\colorrgb#1{\color[rgb]{#1}}%
      \def\colorgray#1{\color[gray]{#1}}%
      \expandafter\def\csname LTw\endcsname{\color{white}}%
      \expandafter\def\csname LTb\endcsname{\color{black}}%
      \expandafter\def\csname LTa\endcsname{\color{black}}%
      \expandafter\def\csname LT0\endcsname{\color[rgb]{1,0,0}}%
      \expandafter\def\csname LT1\endcsname{\color[rgb]{0,1,0}}%
      \expandafter\def\csname LT2\endcsname{\color[rgb]{0,0,1}}%
      \expandafter\def\csname LT3\endcsname{\color[rgb]{1,0,1}}%
      \expandafter\def\csname LT4\endcsname{\color[rgb]{0,1,1}}%
      \expandafter\def\csname LT5\endcsname{\color[rgb]{1,1,0}}%
      \expandafter\def\csname LT6\endcsname{\color[rgb]{0,0,0}}%
      \expandafter\def\csname LT7\endcsname{\color[rgb]{1,0.3,0}}%
      \expandafter\def\csname LT8\endcsname{\color[rgb]{0.5,0.5,0.5}}%
    \else
      \def\colorrgb#1{\color{black}}%
      \def\colorgray#1{\color[gray]{#1}}%
      \expandafter\def\csname LTw\endcsname{\color{white}}%
      \expandafter\def\csname LTb\endcsname{\color{black}}%
      \expandafter\def\csname LTa\endcsname{\color{black}}%
      \expandafter\def\csname LT0\endcsname{\color{black}}%
      \expandafter\def\csname LT1\endcsname{\color{black}}%
      \expandafter\def\csname LT2\endcsname{\color{black}}%
      \expandafter\def\csname LT3\endcsname{\color{black}}%
      \expandafter\def\csname LT4\endcsname{\color{black}}%
      \expandafter\def\csname LT5\endcsname{\color{black}}%
      \expandafter\def\csname LT6\endcsname{\color{black}}%
      \expandafter\def\csname LT7\endcsname{\color{black}}%
      \expandafter\def\csname LT8\endcsname{\color{black}}%
    \fi
  \fi
  \setlength{\unitlength}{0.0500bp}%
  \begin{picture}(4680.00,3276.00)%
    \gplgaddtomacro\gplbacktext{%
      \csname LTb\endcsname%
      \put(128,1797){\rotatebox{-270}{\makebox(0,0){\strut{}$M_{H_2} \quad [\text{GeV}]$}}}%
      \put(2539,112){\makebox(0,0){\strut{}$\lambda$}}%
    }%
    \gplgaddtomacro\gplfronttext{%
      \csname LTb\endcsname%
      \put(1271,2858){\makebox(0,0)[l]{\strut{}tree-level}}%
      \csname LTb\endcsname%
      \put(1271,2666){\makebox(0,0)[l]{\strut{}$Q_0=150\text{GeV}$}}%
      \csname LTb\endcsname%
      \put(1271,2474){\makebox(0,0)[l]{\strut{}$Q_0=300\text{GeV}$}}%
      \csname LTb\endcsname%
      \put(1271,2282){\makebox(0,0)[l]{\strut{}$Q_0=600\text{GeV}$}}%
      \csname LTb\endcsname%
      \put(592,512){\makebox(0,0)[r]{\strut{} 90}}%
      \put(592,769){\makebox(0,0)[r]{\strut{} 100}}%
      \put(592,1026){\makebox(0,0)[r]{\strut{} 110}}%
      \put(592,1283){\makebox(0,0)[r]{\strut{} 120}}%
      \put(592,1540){\makebox(0,0)[r]{\strut{} 130}}%
      \put(592,1798){\makebox(0,0)[r]{\strut{} 140}}%
      \put(592,2055){\makebox(0,0)[r]{\strut{} 150}}%
      \put(592,2312){\makebox(0,0)[r]{\strut{} 160}}%
      \put(592,2569){\makebox(0,0)[r]{\strut{} 170}}%
      \put(592,2826){\makebox(0,0)[r]{\strut{} 180}}%
      \put(592,3083){\makebox(0,0)[r]{\strut{} 190}}%
      \put(688,352){\makebox(0,0){\strut{} 0}}%
      \put(1058,352){\makebox(0,0){\strut{} 0.1}}%
      \put(1429,352){\makebox(0,0){\strut{} 0.2}}%
      \put(1799,352){\makebox(0,0){\strut{} 0.3}}%
      \put(2169,352){\makebox(0,0){\strut{} 0.4}}%
      \put(2540,352){\makebox(0,0){\strut{} 0.5}}%
      \put(2910,352){\makebox(0,0){\strut{} 0.6}}%
      \put(3280,352){\makebox(0,0){\strut{} 0.7}}%
      \put(3650,352){\makebox(0,0){\strut{} 0.8}}%
      \put(4021,352){\makebox(0,0){\strut{} 0.9}}%
      \put(4391,352){\makebox(0,0){\strut{} 1}}%
    }%
    \gplbacktext
    \put(0,0){\includegraphics{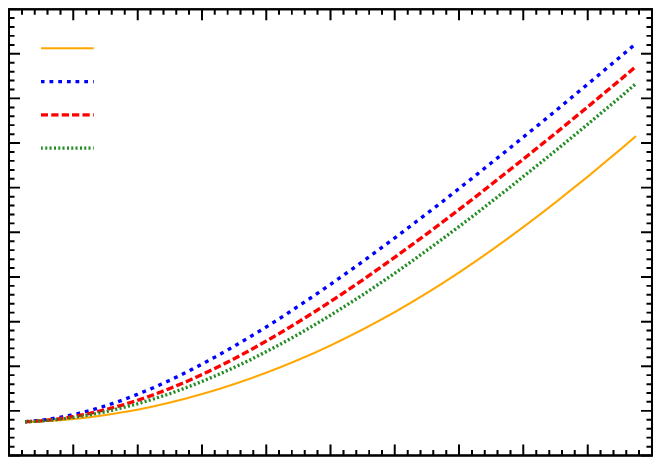}}%
    \gplfronttext
  \end{picture}%
\endgroup

%% file: Fig7.tex
\begingroup
  \fontfamily{cmr}%
  \fontsize{9}{15}
\selectfont
  \makeatletter
  \providecommand\color[2][]{%
    \GenericError{(gnuplot) \space\space\space\@spaces}{%
      Package color not loaded in conjunction with
      terminal option `colourtext'%
    }{See the gnuplot documentation for explanation.%
    }{Either use 'blacktext' in gnuplot or load the package
      color.sty in LaTeX.}%
    \renewcommand\color[2][]{}%
  }%
  \providecommand\includegraphics[2][]{%
    \GenericError{(gnuplot) \space\space\space\@spaces}{%
      Package graphicx or graphics not loaded%
    }{See the gnuplot documentation for explanation.%
    }{The gnuplot epslatex terminal needs graphicx.sty or graphics.sty.}%
    \renewcommand\includegraphics[2][]{}%
  }%
  \providecommand\rotatebox[2]{#2}%
  \@ifundefined{ifGPcolor}{%
    \newif\ifGPcolor
    \GPcolortrue
  }{}%
  \@ifundefined{ifGPblacktext}{%
    \newif\ifGPblacktext
    \GPblacktexttrue
  }{}%
  \let\gplgaddtomacro\g@addto@macro
  \gdef\gplbacktext{}%
  \gdef\gplfronttext{}%
  \makeatother
  \ifGPblacktext
    \def\colorrgb#1{}%
    \def\colorgray#1{}%
  \else
    \ifGPcolor
      \def\colorrgb#1{\color[rgb]{#1}}%
      \def\colorgray#1{\color[gray]{#1}}%
      \expandafter\def\csname LTw\endcsname{\color{white}}%
      \expandafter\def\csname LTb\endcsname{\color{black}}%
      \expandafter\def\csname LTa\endcsname{\color{black}}%
      \expandafter\def\csname LT0\endcsname{\color[rgb]{1,0,0}}%
      \expandafter\def\csname LT1\endcsname{\color[rgb]{0,1,0}}%
      \expandafter\def\csname LT2\endcsname{\color[rgb]{0,0,1}}%
      \expandafter\def\csname LT3\endcsname{\color[rgb]{1,0,1}}%
      \expandafter\def\csname LT4\endcsname{\color[rgb]{0,1,1}}%
      \expandafter\def\csname LT5\endcsname{\color[rgb]{1,1,0}}%
      \expandafter\def\csname LT6\endcsname{\color[rgb]{0,0,0}}%
      \expandafter\def\csname LT7\endcsname{\color[rgb]{1,0.3,0}}%
      \expandafter\def\csname LT8\endcsname{\color[rgb]{0.5,0.5,0.5}}%
    \else
      \def\colorrgb#1{\color{black}}%
      \def\colorgray#1{\color[gray]{#1}}%
      \expandafter\def\csname LTw\endcsname{\color{white}}%
      \expandafter\def\csname LTb\endcsname{\color{black}}%
      \expandafter\def\csname LTa\endcsname{\color{black}}%
      \expandafter\def\csname LT0\endcsname{\color{black}}%
      \expandafter\def\csname LT1\endcsname{\color{black}}%
      \expandafter\def\csname LT2\endcsname{\color{black}}%
      \expandafter\def\csname LT3\endcsname{\color{black}}%
      \expandafter\def\csname LT4\endcsname{\color{black}}%
      \expandafter\def\csname LT5\endcsname{\color{black}}%
      \expandafter\def\csname LT6\endcsname{\color{black}}%
      \expandafter\def\csname LT7\endcsname{\color{black}}%
      \expandafter\def\csname LT8\endcsname{\color{black}}%
    \fi
  \fi
  \setlength{\unitlength}{0.0500bp}%
  \begin{picture}(5400.00,3780.00)%
    \gplgaddtomacro\gplbacktext{%
      \csname LTb\endcsname%
      \put(128,2049){\rotatebox{-270}{\makebox(0,0){\strut{}$M_{H_i} \quad [\text{GeV}]$}}}%
      \put(2899,112){\makebox(0,0){\strut{}$\lambda$}}%
    }%
    \gplgaddtomacro\gplfronttext{%
      \csname LTb\endcsname%
      \put(1268,3381){\makebox(0,0)[l]{\strut{}$H_1$  tree-level}}%
      \csname LTb\endcsname%
      \put(1268,3189){\makebox(0,0)[l]{\strut{}$H_1$  one-loop}}%
      \csname LTb\endcsname%
      \put(1268,2997){\makebox(0,0)[l]{\strut{}$H_2$  tree-level}}%
      \csname LTb\endcsname%
      \put(1268,2805){\makebox(0,0)[l]{\strut{}$H_2$  one-loop}}%
      \csname LTb\endcsname%
      \put(592,512){\makebox(0,0)[r]{\strut{} 40}}%
      \put(592,951){\makebox(0,0)[r]{\strut{} 60}}%
      \put(592,1391){\makebox(0,0)[r]{\strut{} 80}}%
      \put(592,1830){\makebox(0,0)[r]{\strut{} 100}}%
      \put(592,2269){\makebox(0,0)[r]{\strut{} 120}}%
      \put(592,2708){\makebox(0,0)[r]{\strut{} 140}}%
      \put(592,3148){\makebox(0,0)[r]{\strut{} 160}}%
      \put(592,3587){\makebox(0,0)[r]{\strut{} 180}}%
      \put(688,352){\makebox(0,0){\strut{} 0}}%
      \put(1130,352){\makebox(0,0){\strut{} 0.1}}%
      \put(1573,352){\makebox(0,0){\strut{} 0.2}}%
      \put(2015,352){\makebox(0,0){\strut{} 0.3}}%
      \put(2457,352){\makebox(0,0){\strut{} 0.4}}%
      \put(2900,352){\makebox(0,0){\strut{} 0.5}}%
      \put(3342,352){\makebox(0,0){\strut{} 0.6}}%
      \put(3784,352){\makebox(0,0){\strut{} 0.7}}%
      \put(4226,352){\makebox(0,0){\strut{} 0.8}}%
      \put(4669,352){\makebox(0,0){\strut{} 0.9}}%
      \put(5111,352){\makebox(0,0){\strut{} 1}}%
      \put(2457,1061){\makebox(0,0)[l]{\strut{}LEP}}%
      \put(2457,907){\makebox(0,0)[l]{\strut{}excluded}}%
    }%
    \gplbacktext
    \put(0,0){\includegraphics{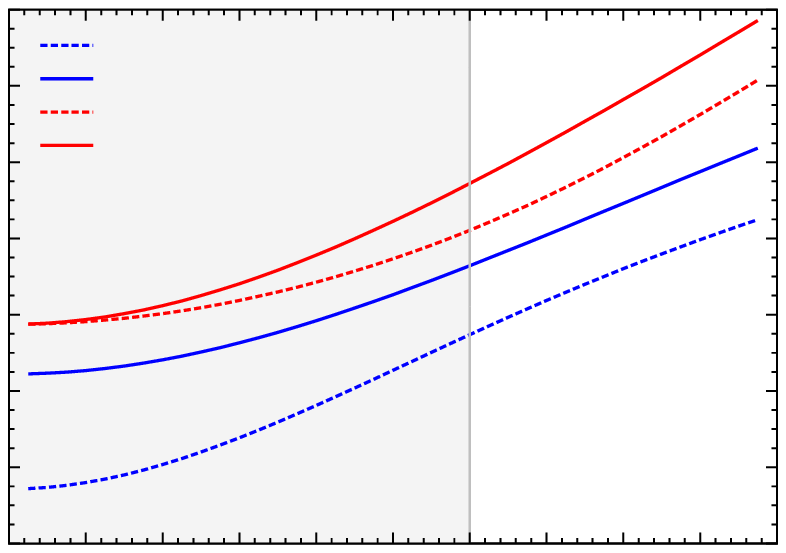}}%
    \gplfronttext
  \end{picture}%
\endgroup

%% file: Fig8a.tex
\begingroup
  \fontfamily{cmr}%
  \fontsize{9}{15}
\selectfont
  \makeatletter
  \providecommand\color[2][]{%
    \GenericError{(gnuplot) \space\space\space\@spaces}{%
      Package color not loaded in conjunction with
      terminal option `colourtext'%
    }{See the gnuplot documentation for explanation.%
    }{Either use 'blacktext' in gnuplot or load the package
      color.sty in LaTeX.}%
    \renewcommand\color[2][]{}%
  }%
  \providecommand\includegraphics[2][]{%
    \GenericError{(gnuplot) \space\space\space\@spaces}{%
      Package graphicx or graphics not loaded%
    }{See the gnuplot documentation for explanation.%
    }{The gnuplot epslatex terminal needs graphicx.sty or graphics.sty.}%
    \renewcommand\includegraphics[2][]{}%
  }%
  \providecommand\rotatebox[2]{#2}%
  \@ifundefined{ifGPcolor}{%
    \newif\ifGPcolor
    \GPcolortrue
  }{}%
  \@ifundefined{ifGPblacktext}{%
    \newif\ifGPblacktext
    \GPblacktexttrue
  }{}%
  \let\gplgaddtomacro\g@addto@macro
  \gdef\gplbacktext{}%
  \gdef\gplfronttext{}%
  \makeatother
  \ifGPblacktext
    \def\colorrgb#1{}%
    \def\colorgray#1{}%
  \else
    \ifGPcolor
      \def\colorrgb#1{\color[rgb]{#1}}%
      \def\colorgray#1{\color[gray]{#1}}%
      \expandafter\def\csname LTw\endcsname{\color{white}}%
      \expandafter\def\csname LTb\endcsname{\color{black}}%
      \expandafter\def\csname LTa\endcsname{\color{black}}%
      \expandafter\def\csname LT0\endcsname{\color[rgb]{1,0,0}}%
      \expandafter\def\csname LT1\endcsname{\color[rgb]{0,1,0}}%
      \expandafter\def\csname LT2\endcsname{\color[rgb]{0,0,1}}%
      \expandafter\def\csname LT3\endcsname{\color[rgb]{1,0,1}}%
      \expandafter\def\csname LT4\endcsname{\color[rgb]{0,1,1}}%
      \expandafter\def\csname LT5\endcsname{\color[rgb]{1,1,0}}%
      \expandafter\def\csname LT6\endcsname{\color[rgb]{0,0,0}}%
      \expandafter\def\csname LT7\endcsname{\color[rgb]{1,0.3,0}}%
      \expandafter\def\csname LT8\endcsname{\color[rgb]{0.5,0.5,0.5}}%
    \else
      \def\colorrgb#1{\color{black}}%
      \def\colorgray#1{\color[gray]{#1}}%
      \expandafter\def\csname LTw\endcsname{\color{white}}%
      \expandafter\def\csname LTb\endcsname{\color{black}}%
      \expandafter\def\csname LTa\endcsname{\color{black}}%
      \expandafter\def\csname LT0\endcsname{\color{black}}%
      \expandafter\def\csname LT1\endcsname{\color{black}}%
      \expandafter\def\csname LT2\endcsname{\color{black}}%
      \expandafter\def\csname LT3\endcsname{\color{black}}%
      \expandafter\def\csname LT4\endcsname{\color{black}}%
      \expandafter\def\csname LT5\endcsname{\color{black}}%
      \expandafter\def\csname LT6\endcsname{\color{black}}%
      \expandafter\def\csname LT7\endcsname{\color{black}}%
      \expandafter\def\csname LT8\endcsname{\color{black}}%
    \fi
  \fi
  \setlength{\unitlength}{0.0500bp}%
  \begin{picture}(4680.00,3276.00)%
    \gplgaddtomacro\gplbacktext{%
      \csname LTb\endcsname%
      \put(128,1797){\rotatebox{-270}{\makebox(0,0){\strut{}$(\mathcal{R}^S_{i3})^2$}}}%
      \put(2539,112){\makebox(0,0){\strut{}$A_{\kappa}\quad [\text{GeV}]$}}%
    }%
    \gplgaddtomacro\gplfronttext{%
      \csname LTb\endcsname%
      \put(3126,1959){\makebox(0,0)[l]{\strut{}$H_1$ tree-level}}%
      \csname LTb\endcsname%
      \put(3126,1767){\makebox(0,0)[l]{\strut{}$H_1$ one-loop}}%
      \csname LTb\endcsname%
      \put(3126,1575){\makebox(0,0)[l]{\strut{}$H_2$ tree-level}}%
      \csname LTb\endcsname%
      \put(3126,1383){\makebox(0,0)[l]{\strut{}$H_2$ one-loop}}%
      \csname LTb\endcsname%
      \put(592,512){\makebox(0,0)[r]{\strut{} 0}}%
      \put(592,1026){\makebox(0,0)[r]{\strut{} 0.2}}%
      \put(592,1540){\makebox(0,0)[r]{\strut{} 0.4}}%
      \put(592,2055){\makebox(0,0)[r]{\strut{} 0.6}}%
      \put(592,2569){\makebox(0,0)[r]{\strut{} 0.8}}%
      \put(592,3083){\makebox(0,0)[r]{\strut{} 1}}%
      \put(778,352){\makebox(0,0){\strut{}-400}}%
      \put(1230,352){\makebox(0,0){\strut{}-350}}%
      \put(1681,352){\makebox(0,0){\strut{}-300}}%
      \put(2133,352){\makebox(0,0){\strut{}-250}}%
      \put(2585,352){\makebox(0,0){\strut{}-200}}%
      \put(3036,352){\makebox(0,0){\strut{}-150}}%
      \put(3488,352){\makebox(0,0){\strut{}-100}}%
      \put(3939,352){\makebox(0,0){\strut{}-50}}%
      \put(4391,352){\makebox(0,0){\strut{} 0}}%
    }%
    \gplbacktext
    \put(0,0){\includegraphics{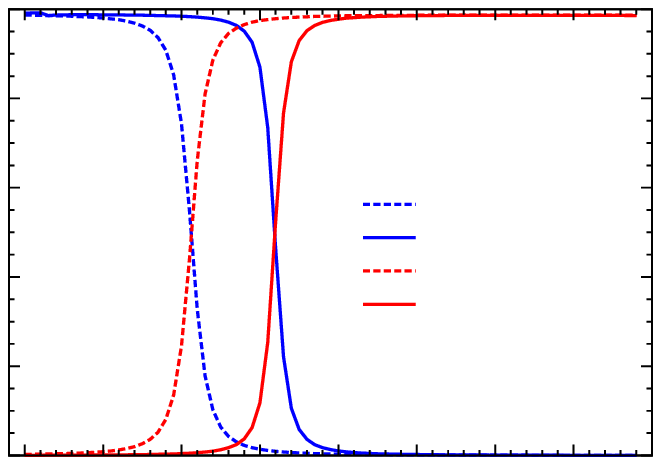}}%
    \gplfronttext
  \end{picture}%
\endgroup

%% file: Fig8b.tex
\begingroup
  \fontfamily{cmr}%
  \fontsize{9}{15}
\selectfont
  \makeatletter
  \providecommand\color[2][]{%
    \GenericError{(gnuplot) \space\space\space\@spaces}{%
      Package color not loaded in conjunction with
      terminal option `colourtext'%
    }{See the gnuplot documentation for explanation.%
    }{Either use 'blacktext' in gnuplot or load the package
      color.sty in LaTeX.}%
    \renewcommand\color[2][]{}%
  }%
  \providecommand\includegraphics[2][]{%
    \GenericError{(gnuplot) \space\space\space\@spaces}{%
      Package graphicx or graphics not loaded%
    }{See the gnuplot documentation for explanation.%
    }{The gnuplot epslatex terminal needs graphicx.sty or graphics.sty.}%
    \renewcommand\includegraphics[2][]{}%
  }%
  \providecommand\rotatebox[2]{#2}%
  \@ifundefined{ifGPcolor}{%
    \newif\ifGPcolor
    \GPcolortrue
  }{}%
  \@ifundefined{ifGPblacktext}{%
    \newif\ifGPblacktext
    \GPblacktexttrue
  }{}%
  \let\gplgaddtomacro\g@addto@macro
  \gdef\gplbacktext{}%
  \gdef\gplfronttext{}%
  \makeatother
  \ifGPblacktext
    \def\colorrgb#1{}%
    \def\colorgray#1{}%
  \else
    \ifGPcolor
      \def\colorrgb#1{\color[rgb]{#1}}%
      \def\colorgray#1{\color[gray]{#1}}%
      \expandafter\def\csname LTw\endcsname{\color{white}}%
      \expandafter\def\csname LTb\endcsname{\color{black}}%
      \expandafter\def\csname LTa\endcsname{\color{black}}%
      \expandafter\def\csname LT0\endcsname{\color[rgb]{1,0,0}}%
      \expandafter\def\csname LT1\endcsname{\color[rgb]{0,1,0}}%
      \expandafter\def\csname LT2\endcsname{\color[rgb]{0,0,1}}%
      \expandafter\def\csname LT3\endcsname{\color[rgb]{1,0,1}}%
      \expandafter\def\csname LT4\endcsname{\color[rgb]{0,1,1}}%
      \expandafter\def\csname LT5\endcsname{\color[rgb]{1,1,0}}%
      \expandafter\def\csname LT6\endcsname{\color[rgb]{0,0,0}}%
      \expandafter\def\csname LT7\endcsname{\color[rgb]{1,0.3,0}}%
      \expandafter\def\csname LT8\endcsname{\color[rgb]{0.5,0.5,0.5}}%
    \else
      \def\colorrgb#1{\color{black}}%
      \def\colorgray#1{\color[gray]{#1}}%
      \expandafter\def\csname LTw\endcsname{\color{white}}%
      \expandafter\def\csname LTb\endcsname{\color{black}}%
      \expandafter\def\csname LTa\endcsname{\color{black}}%
      \expandafter\def\csname LT0\endcsname{\color{black}}%
      \expandafter\def\csname LT1\endcsname{\color{black}}%
      \expandafter\def\csname LT2\endcsname{\color{black}}%
      \expandafter\def\csname LT3\endcsname{\color{black}}%
      \expandafter\def\csname LT4\endcsname{\color{black}}%
      \expandafter\def\csname LT5\endcsname{\color{black}}%
      \expandafter\def\csname LT6\endcsname{\color{black}}%
      \expandafter\def\csname LT7\endcsname{\color{black}}%
      \expandafter\def\csname LT8\endcsname{\color{black}}%
    \fi
  \fi
  \setlength{\unitlength}{0.0500bp}%
  \begin{picture}(4680.00,3276.00)%
    \gplgaddtomacro\gplbacktext{%
      \csname LTb\endcsname%
      \put(128,1797){\rotatebox{-270}{\makebox(0,0){\strut{}$M_{H_i} \quad [\text{GeV}]$}}}%
      \put(2539,112){\makebox(0,0){\strut{}$A_{\kappa}\quad [\text{GeV}]$}}%
    }%
    \gplgaddtomacro\gplfronttext{%
      \csname LTb\endcsname%
      \put(3081,1380){\makebox(0,0)[l]{\strut{}$H_1$ tree-level}}%
      \csname LTb\endcsname%
      \put(3081,1188){\makebox(0,0)[l]{\strut{}$H_1$ one-loop}}%
      \csname LTb\endcsname%
      \put(3081,996){\makebox(0,0)[l]{\strut{}$H_2$ tree-level}}%
      \csname LTb\endcsname%
      \put(3081,804){\makebox(0,0)[l]{\strut{}$H_2$ one-loop}}%
      \csname LTb\endcsname%
      \put(592,576){\makebox(0,0)[r]{\strut{} 20}}%
      \put(592,833){\makebox(0,0)[r]{\strut{} 40}}%
      \put(592,1090){\makebox(0,0)[r]{\strut{} 60}}%
      \put(592,1348){\makebox(0,0)[r]{\strut{} 80}}%
      \put(592,1605){\makebox(0,0)[r]{\strut{} 100}}%
      \put(592,1862){\makebox(0,0)[r]{\strut{} 120}}%
      \put(592,2119){\makebox(0,0)[r]{\strut{} 140}}%
      \put(592,2376){\makebox(0,0)[r]{\strut{} 160}}%
      \put(592,2633){\makebox(0,0)[r]{\strut{} 180}}%
      \put(592,2890){\makebox(0,0)[r]{\strut{} 200}}%
      \put(778,352){\makebox(0,0){\strut{}-400}}%
      \put(1230,352){\makebox(0,0){\strut{}-350}}%
      \put(1681,352){\makebox(0,0){\strut{}-300}}%
      \put(2133,352){\makebox(0,0){\strut{}-250}}%
      \put(2585,352){\makebox(0,0){\strut{}-200}}%
      \put(3036,352){\makebox(0,0){\strut{}-150}}%
      \put(3488,352){\makebox(0,0){\strut{}-100}}%
      \put(3939,352){\makebox(0,0){\strut{}-50}}%
      \put(4391,352){\makebox(0,0){\strut{} 0}}%
    }%
    \gplbacktext
    \put(0,0){\includegraphics{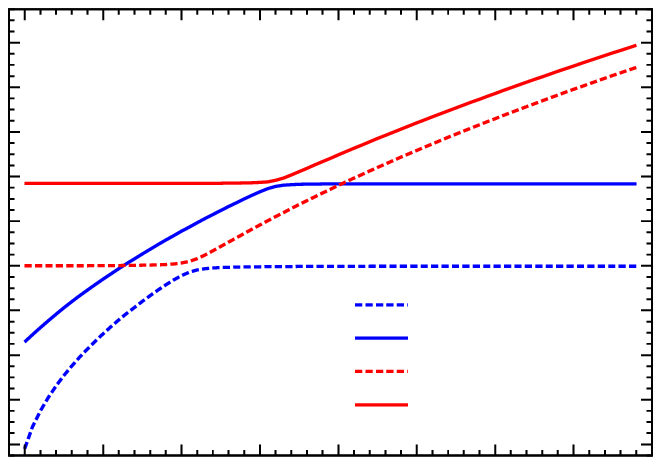}}%
    \gplfronttext
  \end{picture}%
\endgroup

%% file: Fig9a.tex
\begingroup
  \fontfamily{cmr}%
  \fontsize{9}{15}
\selectfont
  \makeatletter
  \providecommand\color[2][]{%
    \GenericError{(gnuplot) \space\space\space\@spaces}{%
      Package color not loaded in conjunction with
      terminal option `colourtext'%
    }{See the gnuplot documentation for explanation.%
    }{Either use 'blacktext' in gnuplot or load the package
      color.sty in LaTeX.}%
    \renewcommand\color[2][]{}%
  }%
  \providecommand\includegraphics[2][]{%
    \GenericError{(gnuplot) \space\space\space\@spaces}{%
      Package graphicx or graphics not loaded%
    }{See the gnuplot documentation for explanation.%
    }{The gnuplot epslatex terminal needs graphicx.sty or graphics.sty.}%
    \renewcommand\includegraphics[2][]{}%
  }%
  \providecommand\rotatebox[2]{#2}%
  \@ifundefined{ifGPcolor}{%
    \newif\ifGPcolor
    \GPcolortrue
  }{}%
  \@ifundefined{ifGPblacktext}{%
    \newif\ifGPblacktext
    \GPblacktexttrue
  }{}%
  \let\gplgaddtomacro\g@addto@macro
  \gdef\gplbacktext{}%
  \gdef\gplfronttext{}%
  \makeatother
  \ifGPblacktext
    \def\colorrgb#1{}%
    \def\colorgray#1{}%
  \else
    \ifGPcolor
      \def\colorrgb#1{\color[rgb]{#1}}%
      \def\colorgray#1{\color[gray]{#1}}%
      \expandafter\def\csname LTw\endcsname{\color{white}}%
      \expandafter\def\csname LTb\endcsname{\color{black}}%
      \expandafter\def\csname LTa\endcsname{\color{black}}%
      \expandafter\def\csname LT0\endcsname{\color[rgb]{1,0,0}}%
      \expandafter\def\csname LT1\endcsname{\color[rgb]{0,1,0}}%
      \expandafter\def\csname LT2\endcsname{\color[rgb]{0,0,1}}%
      \expandafter\def\csname LT3\endcsname{\color[rgb]{1,0,1}}%
      \expandafter\def\csname LT4\endcsname{\color[rgb]{0,1,1}}%
      \expandafter\def\csname LT5\endcsname{\color[rgb]{1,1,0}}%
      \expandafter\def\csname LT6\endcsname{\color[rgb]{0,0,0}}%
      \expandafter\def\csname LT7\endcsname{\color[rgb]{1,0.3,0}}%
      \expandafter\def\csname LT8\endcsname{\color[rgb]{0.5,0.5,0.5}}%
    \else
      \def\colorrgb#1{\color{black}}%
      \def\colorgray#1{\color[gray]{#1}}%
      \expandafter\def\csname LTw\endcsname{\color{white}}%
      \expandafter\def\csname LTb\endcsname{\color{black}}%
      \expandafter\def\csname LTa\endcsname{\color{black}}%
      \expandafter\def\csname LT0\endcsname{\color{black}}%
      \expandafter\def\csname LT1\endcsname{\color{black}}%
      \expandafter\def\csname LT2\endcsname{\color{black}}%
      \expandafter\def\csname LT3\endcsname{\color{black}}%
      \expandafter\def\csname LT4\endcsname{\color{black}}%
      \expandafter\def\csname LT5\endcsname{\color{black}}%
      \expandafter\def\csname LT6\endcsname{\color{black}}%
      \expandafter\def\csname LT7\endcsname{\color{black}}%
      \expandafter\def\csname LT8\endcsname{\color{black}}%
    \fi
  \fi
  \setlength{\unitlength}{0.0500bp}%
  \begin{picture}(4680.00,3276.00)%
    \gplgaddtomacro\gplbacktext{%
      \csname LTb\endcsname%
      \put(128,1797){\rotatebox{-270}{\makebox(0,0){\strut{}$(\mathcal{R}^S_{i3})^2$}}}%
      \put(2539,112){\makebox(0,0){\strut{}$M_{H^\pm}\quad [\text{GeV}]$}}%
    }%
    \gplgaddtomacro\gplfronttext{%
      \csname LTb\endcsname%
      \put(3225,2023){\makebox(0,0)[l]{\strut{}$H_1$ tree-level}}%
      \csname LTb\endcsname%
      \put(3225,1831){\makebox(0,0)[l]{\strut{}$H_1$ one-loop}}%
      \csname LTb\endcsname%
      \put(3225,1639){\makebox(0,0)[l]{\strut{}$H_2$ tree-level}}%
      \csname LTb\endcsname%
      \put(3225,1447){\makebox(0,0)[l]{\strut{}$H_2$ one-loop}}%
      \csname LTb\endcsname%
      \put(592,512){\makebox(0,0)[r]{\strut{} 0}}%
      \put(592,1026){\makebox(0,0)[r]{\strut{} 0.2}}%
      \put(592,1540){\makebox(0,0)[r]{\strut{} 0.4}}%
      \put(592,2055){\makebox(0,0)[r]{\strut{} 0.6}}%
      \put(592,2569){\makebox(0,0)[r]{\strut{} 0.8}}%
      \put(592,3083){\makebox(0,0)[r]{\strut{} 1}}%
      \put(873,352){\makebox(0,0){\strut{} 425}}%
      \put(1336,352){\makebox(0,0){\strut{} 450}}%
      \put(1799,352){\makebox(0,0){\strut{} 475}}%
      \put(2262,352){\makebox(0,0){\strut{} 500}}%
      \put(2725,352){\makebox(0,0){\strut{} 525}}%
      \put(3188,352){\makebox(0,0){\strut{} 550}}%
      \put(3650,352){\makebox(0,0){\strut{} 575}}%
      \put(4113,352){\makebox(0,0){\strut{} 600}}%
    }%
    \gplbacktext
    \put(0,0){\includegraphics{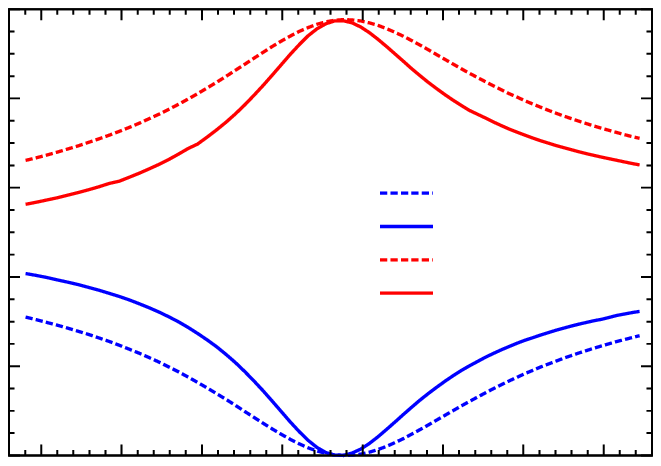}}%
    \gplfronttext
  \end{picture}%
\endgroup

%% file: Fig9b.tex
\begingroup
  \fontfamily{cmr}%
  \fontsize{9}{15}
\selectfont
  \makeatletter
  \providecommand\color[2][]{%
    \GenericError{(gnuplot) \space\space\space\@spaces}{%
      Package color not loaded in conjunction with
      terminal option `colourtext'%
    }{See the gnuplot documentation for explanation.%
    }{Either use 'blacktext' in gnuplot or load the package
      color.sty in LaTeX.}%
    \renewcommand\color[2][]{}%
  }%
  \providecommand\includegraphics[2][]{%
    \GenericError{(gnuplot) \space\space\space\@spaces}{%
      Package graphicx or graphics not loaded%
    }{See the gnuplot documentation for explanation.%
    }{The gnuplot epslatex terminal needs graphicx.sty or graphics.sty.}%
    \renewcommand\includegraphics[2][]{}%
  }%
  \providecommand\rotatebox[2]{#2}%
  \@ifundefined{ifGPcolor}{%
    \newif\ifGPcolor
    \GPcolortrue
  }{}%
  \@ifundefined{ifGPblacktext}{%
    \newif\ifGPblacktext
    \GPblacktexttrue
  }{}%
  \let\gplgaddtomacro\g@addto@macro
  \gdef\gplbacktext{}%
  \gdef\gplfronttext{}%
  \makeatother
  \ifGPblacktext
    \def\colorrgb#1{}%
    \def\colorgray#1{}%
  \else
    \ifGPcolor
      \def\colorrgb#1{\color[rgb]{#1}}%
      \def\colorgray#1{\color[gray]{#1}}%
      \expandafter\def\csname LTw\endcsname{\color{white}}%
      \expandafter\def\csname LTb\endcsname{\color{black}}%
      \expandafter\def\csname LTa\endcsname{\color{black}}%
      \expandafter\def\csname LT0\endcsname{\color[rgb]{1,0,0}}%
      \expandafter\def\csname LT1\endcsname{\color[rgb]{0,1,0}}%
      \expandafter\def\csname LT2\endcsname{\color[rgb]{0,0,1}}%
      \expandafter\def\csname LT3\endcsname{\color[rgb]{1,0,1}}%
      \expandafter\def\csname LT4\endcsname{\color[rgb]{0,1,1}}%
      \expandafter\def\csname LT5\endcsname{\color[rgb]{1,1,0}}%
      \expandafter\def\csname LT6\endcsname{\color[rgb]{0,0,0}}%
      \expandafter\def\csname LT7\endcsname{\color[rgb]{1,0.3,0}}%
      \expandafter\def\csname LT8\endcsname{\color[rgb]{0.5,0.5,0.5}}%
    \else
      \def\colorrgb#1{\color{black}}%
      \def\colorgray#1{\color[gray]{#1}}%
      \expandafter\def\csname LTw\endcsname{\color{white}}%
      \expandafter\def\csname LTb\endcsname{\color{black}}%
      \expandafter\def\csname LTa\endcsname{\color{black}}%
      \expandafter\def\csname LT0\endcsname{\color{black}}%
      \expandafter\def\csname LT1\endcsname{\color{black}}%
      \expandafter\def\csname LT2\endcsname{\color{black}}%
      \expandafter\def\csname LT3\endcsname{\color{black}}%
      \expandafter\def\csname LT4\endcsname{\color{black}}%
      \expandafter\def\csname LT5\endcsname{\color{black}}%
      \expandafter\def\csname LT6\endcsname{\color{black}}%
      \expandafter\def\csname LT7\endcsname{\color{black}}%
      \expandafter\def\csname LT8\endcsname{\color{black}}%
    \fi
  \fi
  \setlength{\unitlength}{0.0500bp}%
  \begin{picture}(4680.00,3276.00)%
    \gplgaddtomacro\gplbacktext{%
      \csname LTb\endcsname%
      \put(128,1797){\rotatebox{-270}{\makebox(0,0){\strut{}$M_{H_i} \quad [\text{GeV}]$}}}%
      \put(2539,112){\makebox(0,0){\strut{}$M_{H^\pm}\quad [\text{GeV}]$}}%
    }%
    \gplgaddtomacro\gplfronttext{%
      \csname LTb\endcsname%
      \put(2207,1273){\makebox(0,0)[l]{\strut{}$H_1$ tree-level}}%
      \csname LTb\endcsname%
      \put(2207,1081){\makebox(0,0)[l]{\strut{}$H_1$ one-loop}}%
      \csname LTb\endcsname%
      \put(2207,889){\makebox(0,0)[l]{\strut{}$H_2$ tree-level}}%
      \csname LTb\endcsname%
      \put(2207,697){\makebox(0,0)[l]{\strut{}$H_2$ one-loop}}%
      \csname LTb\endcsname%
      \put(592,512){\makebox(0,0)[r]{\strut{} 0}}%
      \put(592,798){\makebox(0,0)[r]{\strut{} 25}}%
      \put(592,1083){\makebox(0,0)[r]{\strut{} 50}}%
      \put(592,1369){\makebox(0,0)[r]{\strut{} 75}}%
      \put(592,1655){\makebox(0,0)[r]{\strut{} 100}}%
      \put(592,1940){\makebox(0,0)[r]{\strut{} 125}}%
      \put(592,2226){\makebox(0,0)[r]{\strut{} 150}}%
      \put(592,2512){\makebox(0,0)[r]{\strut{} 175}}%
      \put(592,2797){\makebox(0,0)[r]{\strut{} 200}}%
      \put(592,3083){\makebox(0,0)[r]{\strut{} 225}}%
      \put(873,352){\makebox(0,0){\strut{} 425}}%
      \put(1336,352){\makebox(0,0){\strut{} 450}}%
      \put(1799,352){\makebox(0,0){\strut{} 475}}%
      \put(2262,352){\makebox(0,0){\strut{} 500}}%
      \put(2725,352){\makebox(0,0){\strut{} 525}}%
      \put(3188,352){\makebox(0,0){\strut{} 550}}%
      \put(3650,352){\makebox(0,0){\strut{} 575}}%
      \put(4113,352){\makebox(0,0){\strut{} 600}}%
      \put(1058,1198){\rotatebox{90}{\makebox(0,0)[l]{\strut{}LEP excluded}}}%
      \put(4021,1198){\rotatebox{90}{\makebox(0,0)[l]{\strut{}LEP excluded}}}%
    }%
    \gplbacktext
    \put(0,0){\includegraphics{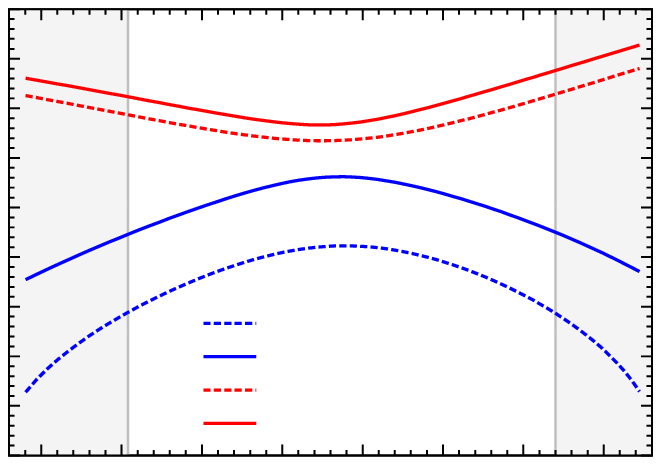}}%
    \gplfronttext
  \end{picture}%
\endgroup

%% file: Fig10a.tex
\begingroup
  \fontfamily{cmr}%
  \fontsize{9}{15}
\selectfont
  \makeatletter
  \providecommand\color[2][]{%
    \GenericError{(gnuplot) \space\space\space\@spaces}{%
      Package color not loaded in conjunction with
      terminal option `colourtext'%
    }{See the gnuplot documentation for explanation.%
    }{Either use 'blacktext' in gnuplot or load the package
      color.sty in LaTeX.}%
    \renewcommand\color[2][]{}%
  }%
  \providecommand\includegraphics[2][]{%
    \GenericError{(gnuplot) \space\space\space\@spaces}{%
      Package graphicx or graphics not loaded%
    }{See the gnuplot documentation for explanation.%
    }{The gnuplot epslatex terminal needs graphicx.sty or graphics.sty.}%
    \renewcommand\includegraphics[2][]{}%
  }%
  \providecommand\rotatebox[2]{#2}%
  \@ifundefined{ifGPcolor}{%
    \newif\ifGPcolor
    \GPcolortrue
  }{}%
  \@ifundefined{ifGPblacktext}{%
    \newif\ifGPblacktext
    \GPblacktexttrue
  }{}%
  \let\gplgaddtomacro\g@addto@macro
  \gdef\gplbacktext{}%
  \gdef\gplfronttext{}%
  \makeatother
  \ifGPblacktext
    \def\colorrgb#1{}%
    \def\colorgray#1{}%
  \else
    \ifGPcolor
      \def\colorrgb#1{\color[rgb]{#1}}%
      \def\colorgray#1{\color[gray]{#1}}%
      \expandafter\def\csname LTw\endcsname{\color{white}}%
      \expandafter\def\csname LTb\endcsname{\color{black}}%
      \expandafter\def\csname LTa\endcsname{\color{black}}%
      \expandafter\def\csname LT0\endcsname{\color[rgb]{1,0,0}}%
      \expandafter\def\csname LT1\endcsname{\color[rgb]{0,1,0}}%
      \expandafter\def\csname LT2\endcsname{\color[rgb]{0,0,1}}%
      \expandafter\def\csname LT3\endcsname{\color[rgb]{1,0,1}}%
      \expandafter\def\csname LT4\endcsname{\color[rgb]{0,1,1}}%
      \expandafter\def\csname LT5\endcsname{\color[rgb]{1,1,0}}%
      \expandafter\def\csname LT6\endcsname{\color[rgb]{0,0,0}}%
      \expandafter\def\csname LT7\endcsname{\color[rgb]{1,0.3,0}}%
      \expandafter\def\csname LT8\endcsname{\color[rgb]{0.5,0.5,0.5}}%
    \else
      \def\colorrgb#1{\color{black}}%
      \def\colorgray#1{\color[gray]{#1}}%
      \expandafter\def\csname LTw\endcsname{\color{white}}%
      \expandafter\def\csname LTb\endcsname{\color{black}}%
      \expandafter\def\csname LTa\endcsname{\color{black}}%
      \expandafter\def\csname LT0\endcsname{\color{black}}%
      \expandafter\def\csname LT1\endcsname{\color{black}}%
      \expandafter\def\csname LT2\endcsname{\color{black}}%
      \expandafter\def\csname LT3\endcsname{\color{black}}%
      \expandafter\def\csname LT4\endcsname{\color{black}}%
      \expandafter\def\csname LT5\endcsname{\color{black}}%
      \expandafter\def\csname LT6\endcsname{\color{black}}%
      \expandafter\def\csname LT7\endcsname{\color{black}}%
      \expandafter\def\csname LT8\endcsname{\color{black}}%
    \fi
  \fi
  \setlength{\unitlength}{0.0500bp}%
  \begin{picture}(4680.00,3276.00)%
    \gplgaddtomacro\gplbacktext{%
      \csname LTb\endcsname%
      \put(128,1797){\rotatebox{-270}{\makebox(0,0){\strut{}$M_{H_1} \quad [\text{GeV}]$}}}%
      \put(2539,112){\makebox(0,0){\strut{}$M_{H^\pm}\quad [\text{GeV}]$}}%
    }%
    \gplgaddtomacro\gplfronttext{%
      \csname LTb\endcsname%
      \put(2557,1207){\makebox(0,0)[l]{\strut{}tree-level}}%
      \csname LTb\endcsname%
      \put(2557,1015){\makebox(0,0)[l]{\strut{}mixed}}%
      \csname LTb\endcsname%
      \put(2557,823){\makebox(0,0)[l]{\strut{}$\overline{\text{DR}}$}}%
      \csname LTb\endcsname%
      \put(2557,631){\makebox(0,0)[l]{\strut{}OS}}%
      \csname LTb\endcsname%
      \put(592,512){\makebox(0,0)[r]{\strut{} 20}}%
      \put(592,908){\makebox(0,0)[r]{\strut{} 40}}%
      \put(592,1303){\makebox(0,0)[r]{\strut{} 60}}%
      \put(592,1699){\makebox(0,0)[r]{\strut{} 80}}%
      \put(592,2094){\makebox(0,0)[r]{\strut{} 100}}%
      \put(592,2490){\makebox(0,0)[r]{\strut{} 120}}%
      \put(592,2885){\makebox(0,0)[r]{\strut{} 140}}%
      \put(873,352){\makebox(0,0){\strut{} 425}}%
      \put(1336,352){\makebox(0,0){\strut{} 450}}%
      \put(1799,352){\makebox(0,0){\strut{} 475}}%
      \put(2262,352){\makebox(0,0){\strut{} 500}}%
      \put(2725,352){\makebox(0,0){\strut{} 525}}%
      \put(3188,352){\makebox(0,0){\strut{} 550}}%
      \put(3650,352){\makebox(0,0){\strut{} 575}}%
      \put(4113,352){\makebox(0,0){\strut{} 600}}%
    }%
    \gplbacktext
    \put(0,0){\includegraphics{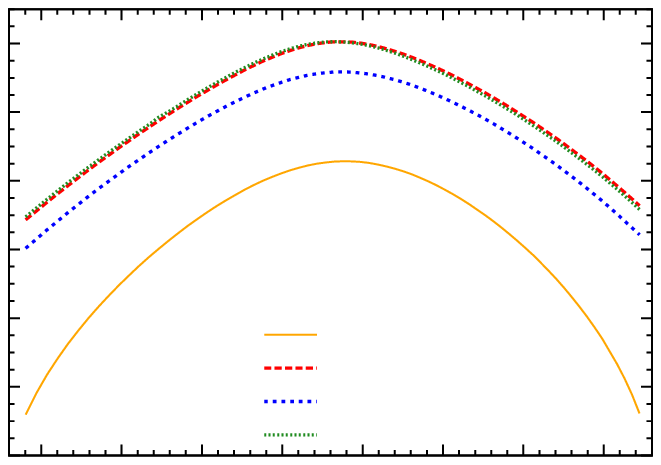}}%
    \gplfronttext
  \end{picture}%
\endgroup

%% file: Fig10b.tex
\begingroup
  \fontfamily{cmr}%
  \fontsize{9}{15}
\selectfont
  \makeatletter
  \providecommand\color[2][]{%
    \GenericError{(gnuplot) \space\space\space\@spaces}{%
      Package color not loaded in conjunction with
      terminal option `colourtext'%
    }{See the gnuplot documentation for explanation.%
    }{Either use 'blacktext' in gnuplot or load the package
      color.sty in LaTeX.}%
    \renewcommand\color[2][]{}%
  }%
  \providecommand\includegraphics[2][]{%
    \GenericError{(gnuplot) \space\space\space\@spaces}{%
      Package graphicx or graphics not loaded%
    }{See the gnuplot documentation for explanation.%
    }{The gnuplot epslatex terminal needs graphicx.sty or graphics.sty.}%
    \renewcommand\includegraphics[2][]{}%
  }%
  \providecommand\rotatebox[2]{#2}%
  \@ifundefined{ifGPcolor}{%
    \newif\ifGPcolor
    \GPcolortrue
  }{}%
  \@ifundefined{ifGPblacktext}{%
    \newif\ifGPblacktext
    \GPblacktexttrue
  }{}%
  \let\gplgaddtomacro\g@addto@macro
  \gdef\gplbacktext{}%
  \gdef\gplfronttext{}%
  \makeatother
  \ifGPblacktext
    \def\colorrgb#1{}%
    \def\colorgray#1{}%
  \else
    \ifGPcolor
      \def\colorrgb#1{\color[rgb]{#1}}%
      \def\colorgray#1{\color[gray]{#1}}%
      \expandafter\def\csname LTw\endcsname{\color{white}}%
      \expandafter\def\csname LTb\endcsname{\color{black}}%
      \expandafter\def\csname LTa\endcsname{\color{black}}%
      \expandafter\def\csname LT0\endcsname{\color[rgb]{1,0,0}}%
      \expandafter\def\csname LT1\endcsname{\color[rgb]{0,1,0}}%
      \expandafter\def\csname LT2\endcsname{\color[rgb]{0,0,1}}%
      \expandafter\def\csname LT3\endcsname{\color[rgb]{1,0,1}}%
      \expandafter\def\csname LT4\endcsname{\color[rgb]{0,1,1}}%
      \expandafter\def\csname LT5\endcsname{\color[rgb]{1,1,0}}%
      \expandafter\def\csname LT6\endcsname{\color[rgb]{0,0,0}}%
      \expandafter\def\csname LT7\endcsname{\color[rgb]{1,0.3,0}}%
      \expandafter\def\csname LT8\endcsname{\color[rgb]{0.5,0.5,0.5}}%
    \else
      \def\colorrgb#1{\color{black}}%
      \def\colorgray#1{\color[gray]{#1}}%
      \expandafter\def\csname LTw\endcsname{\color{white}}%
      \expandafter\def\csname LTb\endcsname{\color{black}}%
      \expandafter\def\csname LTa\endcsname{\color{black}}%
      \expandafter\def\csname LT0\endcsname{\color{black}}%
      \expandafter\def\csname LT1\endcsname{\color{black}}%
      \expandafter\def\csname LT2\endcsname{\color{black}}%
      \expandafter\def\csname LT3\endcsname{\color{black}}%
      \expandafter\def\csname LT4\endcsname{\color{black}}%
      \expandafter\def\csname LT5\endcsname{\color{black}}%
      \expandafter\def\csname LT6\endcsname{\color{black}}%
      \expandafter\def\csname LT7\endcsname{\color{black}}%
      \expandafter\def\csname LT8\endcsname{\color{black}}%
    \fi
  \fi
  \setlength{\unitlength}{0.0500bp}%
  \begin{picture}(4680.00,3276.00)%
    \gplgaddtomacro\gplbacktext{%
      \csname LTb\endcsname%
      \put(128,1797){\rotatebox{-270}{\makebox(0,0){\strut{}$M_{H_2} \quad [\text{GeV}]$}}}%
      \put(2539,112){\makebox(0,0){\strut{}$M_{H^\pm}\quad [\text{GeV}]$}}%
    }%
    \gplgaddtomacro\gplfronttext{%
      \csname LTb\endcsname%
      \put(2557,2847){\makebox(0,0)[l]{\strut{}tree-level}}%
      \csname LTb\endcsname%
      \put(2557,2655){\makebox(0,0)[l]{\strut{}mixed}}%
      \csname LTb\endcsname%
      \put(2557,2463){\makebox(0,0)[l]{\strut{}$\overline{\text{DR}}$}}%
      \csname LTb\endcsname%
      \put(2557,2271){\makebox(0,0)[l]{\strut{}OS}}%
      \csname LTb\endcsname%
      \put(592,746){\makebox(0,0)[r]{\strut{} 160}}%
      \put(592,1213){\makebox(0,0)[r]{\strut{} 170}}%
      \put(592,1681){\makebox(0,0)[r]{\strut{} 180}}%
      \put(592,2148){\makebox(0,0)[r]{\strut{} 190}}%
      \put(592,2616){\makebox(0,0)[r]{\strut{} 200}}%
      \put(592,3083){\makebox(0,0)[r]{\strut{} 210}}%
      \put(873,352){\makebox(0,0){\strut{} 425}}%
      \put(1336,352){\makebox(0,0){\strut{} 450}}%
      \put(1799,352){\makebox(0,0){\strut{} 475}}%
      \put(2262,352){\makebox(0,0){\strut{} 500}}%
      \put(2725,352){\makebox(0,0){\strut{} 525}}%
      \put(3188,352){\makebox(0,0){\strut{} 550}}%
      \put(3650,352){\makebox(0,0){\strut{} 575}}%
      \put(4113,352){\makebox(0,0){\strut{} 600}}%
    }%
    \gplbacktext
    \put(0,0){\includegraphics{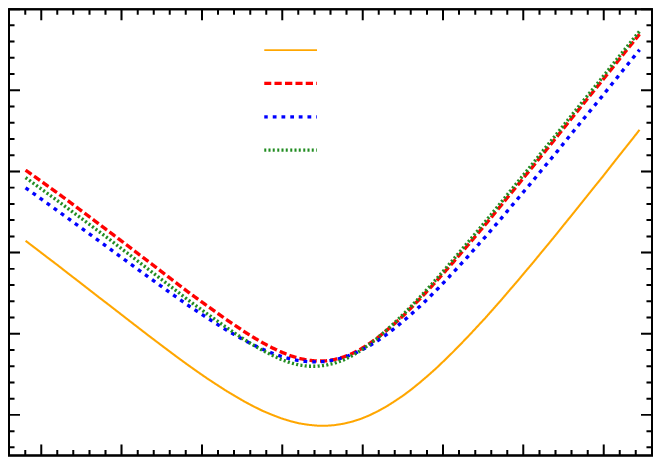}}%
    \gplfronttext
  \end{picture}%
\endgroup

%% file: Fig16a.tex
\begingroup
  \fontfamily{cmr}%
  \fontsize{9}{15}
\selectfont
  \makeatletter
  \providecommand\color[2][]{%
    \GenericError{(gnuplot) \space\space\space\@spaces}{%
      Package color not loaded in conjunction with
      terminal option `colourtext'%
    }{See the gnuplot documentation for explanation.%
    }{Either use 'blacktext' in gnuplot or load the package
      color.sty in LaTeX.}%
    \renewcommand\color[2][]{}%
  }%
  \providecommand\includegraphics[2][]{%
    \GenericError{(gnuplot) \space\space\space\@spaces}{%
      Package graphicx or graphics not loaded%
    }{See the gnuplot documentation for explanation.%
    }{The gnuplot epslatex terminal needs graphicx.sty or graphics.sty.}%
    \renewcommand\includegraphics[2][]{}%
  }%
  \providecommand\rotatebox[2]{#2}%
  \@ifundefined{ifGPcolor}{%
    \newif\ifGPcolor
    \GPcolortrue
  }{}%
  \@ifundefined{ifGPblacktext}{%
    \newif\ifGPblacktext
    \GPblacktexttrue
  }{}%
  \let\gplgaddtomacro\g@addto@macro
  \gdef\gplbacktext{}%
  \gdef\gplfronttext{}%
  \makeatother
  \ifGPblacktext
    \def\colorrgb#1{}%
    \def\colorgray#1{}%
  \else
    \ifGPcolor
      \def\colorrgb#1{\color[rgb]{#1}}%
      \def\colorgray#1{\color[gray]{#1}}%
      \expandafter\def\csname LTw\endcsname{\color{white}}%
      \expandafter\def\csname LTb\endcsname{\color{black}}%
      \expandafter\def\csname LTa\endcsname{\color{black}}%
      \expandafter\def\csname LT0\endcsname{\color[rgb]{1,0,0}}%
      \expandafter\def\csname LT1\endcsname{\color[rgb]{0,1,0}}%
      \expandafter\def\csname LT2\endcsname{\color[rgb]{0,0,1}}%
      \expandafter\def\csname LT3\endcsname{\color[rgb]{1,0,1}}%
      \expandafter\def\csname LT4\endcsname{\color[rgb]{0,1,1}}%
      \expandafter\def\csname LT5\endcsname{\color[rgb]{1,1,0}}%
      \expandafter\def\csname LT6\endcsname{\color[rgb]{0,0,0}}%
      \expandafter\def\csname LT7\endcsname{\color[rgb]{1,0.3,0}}%
      \expandafter\def\csname LT8\endcsname{\color[rgb]{0.5,0.5,0.5}}%
    \else
      \def\colorrgb#1{\color{black}}%
      \def\colorgray#1{\color[gray]{#1}}%
      \expandafter\def\csname LTw\endcsname{\color{white}}%
      \expandafter\def\csname LTb\endcsname{\color{black}}%
      \expandafter\def\csname LTa\endcsname{\color{black}}%
      \expandafter\def\csname LT0\endcsname{\color{black}}%
      \expandafter\def\csname LT1\endcsname{\color{black}}%
      \expandafter\def\csname LT2\endcsname{\color{black}}%
      \expandafter\def\csname LT3\endcsname{\color{black}}%
      \expandafter\def\csname LT4\endcsname{\color{black}}%
      \expandafter\def\csname LT5\endcsname{\color{black}}%
      \expandafter\def\csname LT6\endcsname{\color{black}}%
      \expandafter\def\csname LT7\endcsname{\color{black}}%
      \expandafter\def\csname LT8\endcsname{\color{black}}%
    \fi
  \fi
  \setlength{\unitlength}{0.0500bp}%
  \begin{picture}(4680.00,3276.00)%
    \gplgaddtomacro\gplbacktext{%
      \csname LTb\endcsname%
      \put(128,1797){\rotatebox{-270}{\makebox(0,0){\strut{}$M_{A_1} \quad [\text{GeV}]$}}}%
      \put(2539,112){\makebox(0,0){\strut{}$M_{H^\pm}\quad [\text{GeV}]$}}%
    }%
    \gplgaddtomacro\gplfronttext{%
      \csname LTb\endcsname%
      \put(1302,2858){\makebox(0,0)[l]{\strut{}tree-level}}%
      \csname LTb\endcsname%
      \put(1302,2666){\makebox(0,0)[l]{\strut{}one-loop}}%
      \csname LTb\endcsname%
      \put(592,512){\makebox(0,0)[r]{\strut{} 116}}%
      \put(592,769){\makebox(0,0)[r]{\strut{} 117}}%
      \put(592,1026){\makebox(0,0)[r]{\strut{} 118}}%
      \put(592,1283){\makebox(0,0)[r]{\strut{} 119}}%
      \put(592,1540){\makebox(0,0)[r]{\strut{} 120}}%
      \put(592,1798){\makebox(0,0)[r]{\strut{} 121}}%
      \put(592,2055){\makebox(0,0)[r]{\strut{} 122}}%
      \put(592,2312){\makebox(0,0)[r]{\strut{} 123}}%
      \put(592,2569){\makebox(0,0)[r]{\strut{} 124}}%
      \put(592,2826){\makebox(0,0)[r]{\strut{} 125}}%
      \put(592,3083){\makebox(0,0)[r]{\strut{} 126}}%
      \put(873,352){\makebox(0,0){\strut{} 425}}%
      \put(1336,352){\makebox(0,0){\strut{} 450}}%
      \put(1799,352){\makebox(0,0){\strut{} 475}}%
      \put(2262,352){\makebox(0,0){\strut{} 500}}%
      \put(2725,352){\makebox(0,0){\strut{} 525}}%
      \put(3188,352){\makebox(0,0){\strut{} 550}}%
      \put(3650,352){\makebox(0,0){\strut{} 575}}%
      \put(4113,352){\makebox(0,0){\strut{} 600}}%
    }%
    \gplbacktext
    \put(0,0){\includegraphics{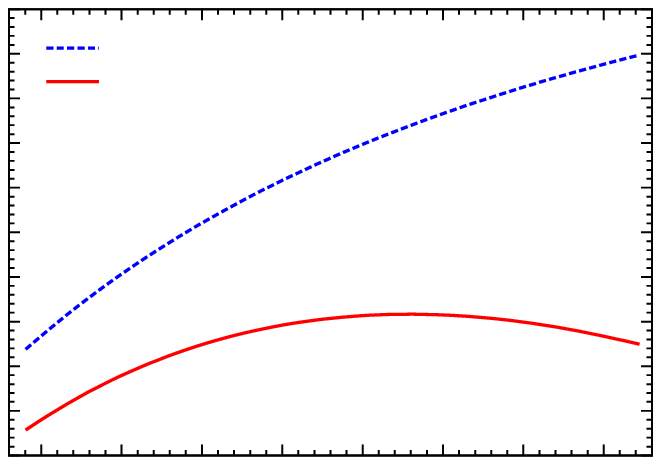}}%
    \gplfronttext
  \end{picture}%
\endgroup

%% file: Fig16b.tex
\begingroup
  \fontfamily{cmr}%
  \fontsize{9}{15}
\selectfont
  \makeatletter
  \providecommand\color[2][]{%
    \GenericError{(gnuplot) \space\space\space\@spaces}{%
      Package color not loaded in conjunction with
      terminal option `colourtext'%
    }{See the gnuplot documentation for explanation.%
    }{Either use 'blacktext' in gnuplot or load the package
      color.sty in LaTeX.}%
    \renewcommand\color[2][]{}%
  }%
  \providecommand\includegraphics[2][]{%
    \GenericError{(gnuplot) \space\space\space\@spaces}{%
      Package graphicx or graphics not loaded%
    }{See the gnuplot documentation for explanation.%
    }{The gnuplot epslatex terminal needs graphicx.sty or graphics.sty.}%
    \renewcommand\includegraphics[2][]{}%
  }%
  \providecommand\rotatebox[2]{#2}%
  \@ifundefined{ifGPcolor}{%
    \newif\ifGPcolor
    \GPcolortrue
  }{}%
  \@ifundefined{ifGPblacktext}{%
    \newif\ifGPblacktext
    \GPblacktexttrue
  }{}%
  \let\gplgaddtomacro\g@addto@macro
  \gdef\gplbacktext{}%
  \gdef\gplfronttext{}%
  \makeatother
  \ifGPblacktext
    \def\colorrgb#1{}%
    \def\colorgray#1{}%
  \else
    \ifGPcolor
      \def\colorrgb#1{\color[rgb]{#1}}%
      \def\colorgray#1{\color[gray]{#1}}%
      \expandafter\def\csname LTw\endcsname{\color{white}}%
      \expandafter\def\csname LTb\endcsname{\color{black}}%
      \expandafter\def\csname LTa\endcsname{\color{black}}%
      \expandafter\def\csname LT0\endcsname{\color[rgb]{1,0,0}}%
      \expandafter\def\csname LT1\endcsname{\color[rgb]{0,1,0}}%
      \expandafter\def\csname LT2\endcsname{\color[rgb]{0,0,1}}%
      \expandafter\def\csname LT3\endcsname{\color[rgb]{1,0,1}}%
      \expandafter\def\csname LT4\endcsname{\color[rgb]{0,1,1}}%
      \expandafter\def\csname LT5\endcsname{\color[rgb]{1,1,0}}%
      \expandafter\def\csname LT6\endcsname{\color[rgb]{0,0,0}}%
      \expandafter\def\csname LT7\endcsname{\color[rgb]{1,0.3,0}}%
      \expandafter\def\csname LT8\endcsname{\color[rgb]{0.5,0.5,0.5}}%
    \else
      \def\colorrgb#1{\color{black}}%
      \def\colorgray#1{\color[gray]{#1}}%
      \expandafter\def\csname LTw\endcsname{\color{white}}%
      \expandafter\def\csname LTb\endcsname{\color{black}}%
      \expandafter\def\csname LTa\endcsname{\color{black}}%
      \expandafter\def\csname LT0\endcsname{\color{black}}%
      \expandafter\def\csname LT1\endcsname{\color{black}}%
      \expandafter\def\csname LT2\endcsname{\color{black}}%
      \expandafter\def\csname LT3\endcsname{\color{black}}%
      \expandafter\def\csname LT4\endcsname{\color{black}}%
      \expandafter\def\csname LT5\endcsname{\color{black}}%
      \expandafter\def\csname LT6\endcsname{\color{black}}%
      \expandafter\def\csname LT7\endcsname{\color{black}}%
      \expandafter\def\csname LT8\endcsname{\color{black}}%
    \fi
  \fi
  \setlength{\unitlength}{0.0500bp}%
  \begin{picture}(4680.00,3276.00)%
    \gplgaddtomacro\gplbacktext{%
      \csname LTb\endcsname%
      \put(128,1797){\rotatebox{-270}{\makebox(0,0){\strut{}$M_{A_2} \quad [\text{GeV}]$}}}%
      \put(2539,112){\makebox(0,0){\strut{}$M_{H^\pm}\quad [\text{GeV}]$}}%
    }%
    \gplgaddtomacro\gplfronttext{%
      \csname LTb\endcsname%
      \put(1302,2865){\makebox(0,0)[l]{\strut{}tree-level}}%
      \csname LTb\endcsname%
      \put(1302,2673){\makebox(0,0)[l]{\strut{}one-loop}}%
      \csname LTb\endcsname%
      \put(592,512){\makebox(0,0)[r]{\strut{} 420}}%
      \put(592,757){\makebox(0,0)[r]{\strut{} 440}}%
      \put(592,1002){\makebox(0,0)[r]{\strut{} 460}}%
      \put(592,1247){\makebox(0,0)[r]{\strut{} 480}}%
      \put(592,1491){\makebox(0,0)[r]{\strut{} 500}}%
      \put(592,1736){\makebox(0,0)[r]{\strut{} 520}}%
      \put(592,1981){\makebox(0,0)[r]{\strut{} 540}}%
      \put(592,2226){\makebox(0,0)[r]{\strut{} 560}}%
      \put(592,2471){\makebox(0,0)[r]{\strut{} 580}}%
      \put(592,2716){\makebox(0,0)[r]{\strut{} 600}}%
      \put(592,2961){\makebox(0,0)[r]{\strut{} 620}}%
      \put(873,352){\makebox(0,0){\strut{} 425}}%
      \put(1336,352){\makebox(0,0){\strut{} 450}}%
      \put(1799,352){\makebox(0,0){\strut{} 475}}%
      \put(2262,352){\makebox(0,0){\strut{} 500}}%
      \put(2725,352){\makebox(0,0){\strut{} 525}}%
      \put(3188,352){\makebox(0,0){\strut{} 550}}%
      \put(3650,352){\makebox(0,0){\strut{} 575}}%
      \put(4113,352){\makebox(0,0){\strut{} 600}}%
    }%
    \gplbacktext
    \put(0,0){\includegraphics{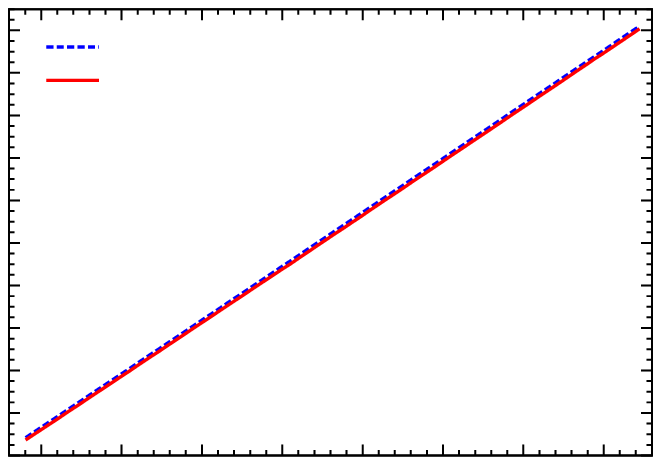}}%
    \gplfronttext
  \end{picture}%
\endgroup